\documentclass[aps,prx,twocolumn,longbibliography,showpacs,floatfix]{revtex4-1}

\usepackage{amsfonts,amssymb,amsmath,amsthm}
\usepackage{hyperref}
\usepackage{graphicx}
\usepackage{dcolumn}
\usepackage{bm}
\usepackage{verbatim}
\usepackage{mathrsfs}
\usepackage{color}
\usepackage{sidecap}
\usepackage{soul}
\usepackage{dsfont}
\usepackage{subfigure}
\usepackage{float}
\usepackage[T1]{fontenc}
\usepackage{booktabs}

\allowdisplaybreaks

\hypersetup{colorlinks,
linkcolor=blue, citecolor=blue, filecolor=blue, urlcolor=blue }

\def\Tr{{\text{Tr}}\,}

\newcommand{\ket}[1]{|#1\rangle}
\newcommand{\bra}[1]{\langle #1 |}
\newcommand{\braket}[2]{\left< #1 \vphantom{#2} \right| \left. #2 \vphantom{#1} \right>} 

\begin{document}

\title{Higher-order topological phases in crystalline and non-crystalline systems: a review}
\author{Yan-Bin Yang$^{1}$}
\email{yangyb@ust.hk}
\author{Jiong-Hao Wang$^{2}$}
\author{Kai Li$^{2}$}
\author{Yong Xu$^{2,3}$}
\email{yongxuphy@tsinghua.edu.cn}
\affiliation{$^{1}$Department of Physics, Hong Kong University of Science and Technology, Clear Water Bay, Hong Kong, People's Republic of China}
\affiliation{$^{2}$Center for Quantum Information, IIIS, Tsinghua University, Beijing 100084, People's Republic of China}
\affiliation{$^{3}$Hefei National Laboratory, Hefei 230088, People's Republic of China}


\begin{abstract}
In recent years, higher-order topological phases have attracted great interest in various fields of physics.
These phases have protected boundary states at lower-dimensional boundaries than
the conventional first-order topological phases due to the higher-order bulk-boundary correspondence.
In this review, we summarize current research progress on higher-order topological phases in both crystalline and non-crystalline systems.
We firstly introduce prototypical models of higher-order topological phases in crystals and their topological characterizations.
We then discuss effects of quenched disorder on higher-order topology and demonstrate disorder-induced higher-order topological insulators.
We also review the theoretical studies on higher-order topological insulators in amorphous systems without any crystalline symmetry
and higher-order topological phases in non-periodic lattices including quasicrystals, hyperbolic lattices, and fractals,
which have no crystalline counterparts.
We conclude the review by a summary of experimental realizations of higher-order topological phases
and discussions on potential directions for future study.
\end{abstract}

\maketitle

\tableofcontents


\section{Introduction} \label{Sec1}

Quantum phases of matter, according to Landau's paradigm, were believed to be classified based on the symmetry of the ground state for a long time.
However, in 1980, the discovery of quantum Hall effect showcased the example of topological phases of matter,
for which different phases cannot be distinguished by spontaneously symmetry breaking but are characterized by the bulk topology~\cite{Klitzing1980PRL,TKNN1982PRL}.
In 1988, the famous Haldane model was proposed to describe a topological insulator (TI) with quantum Hall conductance resulted from chiral edge states,
which is characterized by a nonzero Chern number of bulk wave functions~\cite{Haldane1988PRL}.
The connection between nontrivial bulk topology and topologically protected boundary states
is referred to as bulk-boundary correspondence of TIs.
Later, inspired by the discovery of time-reversal-invariant TIs~\cite{Kane2005PRL1,Kane2005PRL2,Fu2007PRL,Moore2007PRB},
the concept of TIs has been extended to symmetry-protected topological states which are protected by either internal symmetries or spatial symmetries~\cite{ryu2010topological,Hasan2010RMP,Qi2011RMP,Fu2011PRL,ChenXie2013PRB,slager2013space,
Sato2014PRB,Fu2015Review}.
The symmetry-protected TIs cannot be adiabatically deformed into trivial insulators without closing the bulk gap if the symmetry is not broken,
and they can be classified based on homotopy theory~\cite{Ludwig2008PRB,schnyder2009AIP,kitaev2009periodic} and symmetry indicators based on band representations developed for topological crystalline insulators~\cite{Slager2017PRX,Po2017NC,Bradlyn2017Nature}.
Besides TIs, people have also discovered topological superconductors described by gapped mean-field Hamiltonian~\cite{Read2000PRB,Kitaev2001Majorana,Sato2017RPP} and
topological semimetals with topological gapless points and protected surface states~\cite{Weng2016JPCM,Vishwanath2018RMP,xu2019topological,lv2021experimental}.
During the past decades, the topological phases of matter have attracted broad research interest and become a very active field in condensed matter physics~\cite{Senthil2015Review,Das2016RMP,Chiu2016RMP,WenXG2017RMP,Haldane2017RMP}.

The conventional topological phases in $d$ dimensions host $(d-1)$-dimensional boundary states according to the bulk-boundary correspondence~\cite{Hasan2010RMP,Qi2011RMP}.
In recent years, a new type of topological phases dubbed higher-order topological insulators (HOTIs) were discovered, for which
a $d$-dimensional $n$th-order topological phase supports $(d-n)$-dimensional boundary states with
the codimension $n>1$~\cite{Benalcazar2017Science,Benalcazar2017PRB,Langbehn2017PRL,Song2017PRL,Schindler2018SA,
slager2015impurity,Ezawa2018PRL,Ezawa2018PRB,kunst2018lattice,Fulga2018PRB,Brouwer2018PRB,Khalaf2018PRB,VanMiert2018PRB,Wieder2018Fragile,Khalaf2018PRX,
Fang2019SA,Brouwer2019PRX,YangBJ2019PRB,hwang2019fragile,YangBJ2019PRX,YangBJ2019CPB,Liu2019PRL,XuYF2019PRL,Park2019PRL,YangSA2019PRL,
Benalcazar2019PRB,
okuma2019classification,YangBJ2020npjQM,
SunKai2020PRL,Tiwari2020PRL,ZhangRX2020Mobius,Niu2020PRL,Ghorashi2020PRL,Yang2020PRR,Li2020PRB,Tao2020NJP,
Khalaf2021PRR,Benalcazar2022PRL,saha2023multiple,kim2023replica,lin2022spin}.
In 2017, Benalcazar et al. generalized the concept of the dipole moment to the multipole moment and
proposed a model for multipole TIs protected by mirror symmetries, now known as the Benalcazar-Bernevig-Hughes (BBH) model,
which exhibits quantized multipole moment and in-gap corner modes as the defining signature of higher-order topology~\cite{Benalcazar2017Science,Benalcazar2017PRB}.
Specifically, a two-dimensional (2D) quadrupole topological insulator is featured by quantized edge polarizations and fractional corner charges resulted from corner modes.
The corner modes arise as the manifestation of nontrivial topology of edge Hamiltonians
and cannot be removed by symmetry-preserving perturbations unless the bulk or edge energy gap closes~\cite{Benalcazar2017Science,Benalcazar2017PRB,Khalaf2021PRR,Yang2020PRR}.
Analogously, an octupole topological insulator is a three-dimensional (3D) third-order topological insulator with the quantized octupole moment and corner modes~\cite{Benalcazar2017Science}.
In fact, it is not always true that a quadrupole TI always has corner modes. For instance,
very recently, a quadrupole TI (with nonzero quadrupole moment) without corner modes in the energy spectrum
has been found~\cite{Tao2023Quadrupole,YangYi2023Quadrupole}.

Soon after the discovery of the multipole TIs, several works discovered 3D HOTIs
with gapped surface states but gapless hinge states, in contrast to
first-order TIs with gapless surface states~\cite{Langbehn2017PRL,Song2017PRL,Schindler2018SA}.
In the absence of time-reversal symmetry (TRS), a typical class of second-order topological insulators (SOTIs) is featured by chiral hinge states,
which can be protected by the product of time-reversal and rotational symmetry such as the $C_4 T$ symmetry~\cite{Schindler2018SA}.
In the presence of TRS, there exist Kramers pairs of hinge states, namely, helical hinge states,
which can be protected by additional crystalline symmetry such as mirror or rotational symmetry~\cite{Langbehn2017PRL,Song2017PRL,Schindler2018SA}.
These gapless hinge modes give rise to quantized conductance, which is the transport signature of the higher-order topology.
For many HOTIs with crystalline symmetry,
the anomalous boundary states can be understood as the domain walls between massive Dirac Hamiltonians of adjacent gapped surfaces
with opposite signs of masses due to the symmetry constraints~\cite{Brouwer2019PRX}.
In fact, in an early study on 3D topological superfluid $^3$He-B~\cite{volovik2010JETP},
it was found that in the presence of magnetic fields, the surfaces become gapped while
there exists one-dimensional (1D) gapless modes at the line between different domains with opposite signs of the mass term due to magnetic field.
Later, in two studies~\cite{Sitte2012PRL,ZhangFan2013PRL}, similar in-gap hinge modes were also found in 3D TIs,
where the surface states are gapped out by the magnetic field or magnetization breaking the TRS.
The chiral hinge modes emerge at the interface of adjacent surfaces with opposite signs of magnetic gaps.
However, according to higher-order bulk-boundary correspondence proposed in Ref.~\cite{Brouwer2019PRX},
these higher-order topological phases (HOTPs) depending on the surface orientations or lattice terminations
belong to extrinsic HOTPs,
in contrast to intrinsic HOTPs originating from the nontrivial bulk topology
protected by crystalline symmetry.
The concept of higher-order topology has also been extended to higher-order topological superconductors
hosting Majorana modes at corners or hinges~\cite{WangZhong2018PRL,ZhangFan2018PRL,Loss2018PRL,Shinsei2018PRB,ZhuXY2018PRB,WangYX2018PRB,Nori2018PRB,
	Volpez2019PRL,ZhuXY2019PRL,ZhangRX2019Helical,ZhangRX2019PRL,Zhongbo2019PRL,HuangWen2019PRB,Bultinck2019PRB,Fulga2019PRB,Zhongbo2019PRB,
	Hsu2020PRL,kheirkhah2020PRL,Roberts2020PRB,ZhangRX2020PRB,kheirkhah2020majorana,laubscher2020kramers,YangBJ2020PRR,Tiwari2020PRR,Vu2020PRR,WuXX2020PRX,
	Zhang2021Intrinsic,LiuXin2021PRB,kheirkhah2021vortex,ZhangRX2022arXiv,ZhangRX2023kitaev},
and higher-order topological semimetals hosting gapless points and
Fermi arcs at hinges~\cite{Lin2018PRB,WangZJ2019PRL,Roy2019PRB,Roy2019PRR,Okugawa2019PRB,Zilberberg2020PRR,Zeng2020PRB,
Hughes2020PRL,JiangJH2020PRL,Wieder2020NC,ZhaoYX2020PRL,YangSA2022PRL,Brouwer2022PRB}.

Since the pioneering works reporting HOTIs,
it has witnessed a rapid development of both theoretical and experimental research in this field,
and many models with higher-order topology were studied~\cite{Benalcazar2017Science,Benalcazar2017PRB,Langbehn2017PRL,Song2017PRL,Schindler2018SA,
slager2015impurity,Ezawa2018PRL,Ezawa2018PRB,kunst2018lattice,Fulga2018PRB,Brouwer2018PRB,Khalaf2018PRB,VanMiert2018PRB,Wieder2018Fragile,Khalaf2018PRX,
Fang2019SA,Brouwer2019PRX,YangBJ2019PRB,hwang2019fragile,YangBJ2019PRX,YangBJ2019CPB,Liu2019PRL,XuYF2019PRL,Park2019PRL,YangSA2019PRL,Benalcazar2019PRB,Roy2019PRR,
okuma2019classification,YangBJ2020npjQM,
SunKai2020PRL,Tiwari2020PRL,ZhangRX2020Mobius,Niu2020PRL,Ghorashi2020PRL,Yang2020PRR,Zilberberg2020PRR,Zeng2020PRB,Li2020PRB,Tao2020NJP,
Khalaf2021PRR,Benalcazar2022PRL,
Tao2023Quadrupole,YangYi2023Quadrupole,
Sitte2012PRL,ZhangFan2013PRL,saha2023multiple,kim2023replica,lin2022spin,
volovik2010JETP,
WangZhong2018PRL,ZhangFan2018PRL,Loss2018PRL,Shinsei2018PRB,ZhuXY2018PRB,WangYX2018PRB,Nori2018PRB,
Volpez2019PRL,ZhuXY2019PRL,ZhangRX2019Helical,ZhangRX2019PRL,Zhongbo2019PRL,HuangWen2019PRB,Bultinck2019PRB,Fulga2019PRB,Zhongbo2019PRB,
Hsu2020PRL,kheirkhah2020PRL,Roberts2020PRB,ZhangRX2020PRB,kheirkhah2020majorana,laubscher2020kramers,YangBJ2020PRR,Tiwari2020PRR,Vu2020PRR,WuXX2020PRX,
Zhang2021Intrinsic,LiuXin2021PRB,kheirkhah2021vortex,ZhangRX2022arXiv,ZhangRX2023kitaev,
Lin2018PRB,WangZJ2019PRL,Roy2019PRB,Okugawa2019PRB,
Zeng2020PRB,Hughes2020PRL,JiangJH2020PRL,Wieder2020NC,ZhaoYX2020PRL,YangSA2022PRL,Brouwer2022PRB}.
For example, in 2Ds, the SOTIs with corner modes were
found in breathing kagome and pyrochlore lattices,
which are characterized by the quantized bulk polarizations~\cite{Ezawa2018PRL}.
In Ref.~\cite{Benalcazar2019PRB}, the authors systematically investigated 2D SOTIs with
quantized fractional corner charges protected by rotational symmetries.
Moreover, several works studied fractional charge responses to topological defects such as dislocations and disclinations,
which are related to higher-order bulk topology~\cite{liu2019shift,li2020fractional,liu2021bulk,may2022crystalline,zhang2022fractional,schindler2022NC}.
In 3Ds, it was found that HOTIs with anomalous hinge states
can be protected by various crystalline symmetries including inversion symmetry, rotational symmetry,
and magnetic symmetry such as $C_2 T$~\cite{Khalaf2018PRB,Khalaf2018PRX,Brouwer2018PRB,VanMiert2018PRB,Fang2019SA,Brouwer2019PRX,Wieder2018Fragile,YangBJ2019PRB}.
In addition, the space-time inversion symmetry (spinless $PT$ symmetry) can protect 2D fragile topology with corner modes
and 3D nodal-line semimetals with hinge Fermi arcs~\cite{WangZJ2019PRL,YangBJ2019PRX,YangBJ2019CPB,ZhaoYX2020PRL}.
Very recently, a Klein-bottle HOTI with non-chiral hinge modes protected by a pair of momentum-space glide
reflection symmetries has been found~\cite{XuChen2023arXiv}.

In the presence of crystalline symmetry, HOTIs can usually be detected by symmetry indicators~\cite{Song2017PRL,Khalaf2018PRX,Benalcazar2019PRB}.
On the other hand, the higher-order topology
can also be characterized by some bulk topological invariants.
For instance, the 2D quadrupole TIs
are characterized by the quadrupole moment as a $\mathbb{Z}_2$ topological invariant~\cite{Kang2019PRB,Wheeler2019PRB}.
It was proved that the quadrupole moment is quantized by chiral symmetry~\cite{Yang2021PRB,Li2020PRL} or particle-hole symmetry~\cite{Li2020PRL}.
The $C_4 T$ symmetric SOTIs with chiral hinge modes
can be characterized by the Chern-Simons invariant~\cite{Schindler2018SA}.
Later, it was found that these chiral hinge modes can be also characterized by the winding number of the quadrupole moment~\cite{kang2021many,WangJH2021PRL}.
For the SOTIs with TRS,
the helical hinge states can be characterized by the mirror Chern number in the presence of mirror symmetries~\cite{Schindler2018SA}.
In the absence of mirror symmetries, the helical modes can be characterized by a $\mathbb{Z}_2$ invariant proposed in Ref.~\cite{WangJH2021PRL}.
Remarkably, either the quadrupole moment winding for chiral modes and the
$\mathbb{Z}_2$ invariant for helical modes do not require the presence of crystalline symmetries.
Therefore, both the 2D quadrupole TIs and 3D chiral or helical SOTIs can
exist in systems without any crystalline symmetry.

People have also studied the effects of quenched disorder on the higher-order topology,
and found that the higher-order topological states remain robust against weak disorder~\cite{Benalcazar2017Science,Hatsugai2019PRB,Fulga2019PRB,Li2020PRB,JiangH2019CPB,Yang2021PRB,Li2020PRL,WangXR2020PRR,Szabo2020PRR,
WangXR2021PRB,ChenCZ2021PRB,
JiangH2021PRB,Lu2023PRB,Castro2023arXiv,Qiao2023arXiv}.
More remarkably, the disorder is not always detrimental to topological phases.
In two works studying quadrupole TIs in the presence of quenched disorder,
it was found that disorder can induce a topological phase transition from a
trivial insulator to a quadrupole TI with
nonzero quadrupole moment quantized by chiral symmetry~\cite{Yang2021PRB,Li2020PRL}.
The HOTP induced by disorder is named higher-order topological Anderson insulator,
following the topological Anderson insulator found in first-order topological phases~\cite{Shen2009PRL}.
In an earlier study~\cite{Fulga2019PRB}, the authors studied a second-order topological superconductor with Majorana corner modes
and found that an initially trivial phase can transition into a HOTP upon adding disorder.
The disorder-driven topological phase transitions can be explained by the renormalization of Hamiltonian parameters due to disorder.
A subsequential study~\cite{ZhangXD2021PRL} considered a modified Haldane model with disorder in the hopping phases and
found that the disorder can drive a transition from a Chern insulator to
a HOTI with corner states, which was further observed in experiment.

Apart from crystals, recent years have witnessed the rapid progress on topological phases
in non-crystalline systems lacking translational symmetries, where the conventional topological band theory is not applicable.
One typical category of non-crystalline systems discovered in nature are quasicrystals, which have no translational symmetry
but still possess quasi long-range order~\cite{QCbook}.
A quasicrystalline lattice can usually be viewed as a projection of a higher-dimensional lattice.
In 2Ds, quasicrystals can respect rotational symmetry forbidden in periodic crystals,
such as five-fold or eight-fold rotations.
A number of topological phases have been studied in quasicrystals~\cite{kraus2012topological,lang2012edge,verbin2013observation,kraus2013four,tran2015topological,fuchs2016hofstadter,fulga2016aperiodic,
huang2018quantum,huang2018theory,chen2019topological,he2019quasicrystalline,huang2020aperiodic,duncan2020topological,cao2020kohn,
ghadimi2021topological,Fan2022FOP}.
Another large class of non-crystalline systems is amorphous systems, which have no any crystalline symmetry and no long-range order~\cite{zallen2008physics}.
In nature, there exist abundant materials belonging to amorphous systems, for instance, the glass-like materials.
An amorphous system can be modeled as a lattice composed of randomly distributed sites, which can be generated upon adding structural disorder to a regular lattice.
Despite the lack of any spatial symmetry, there have been an amount of topological phases reported in amorphous systems~\cite{Shenoy2017PRL,Chong2017PRB,Fan2017PRB,Irvine2018NP,Prodan2018JPA,Ojanen2018NC,Fritz2019PRB,Xu2019PRL,
Zhang2019PRB,Chern2019EPL,Fazzio2019NL,Ohtsuki2019JPSJ,Bhatt2020PRB,Ojanen2020PRR,Grushin2020PNAS,Liu2020Research,Ojanen2020PRR2,Zhang2020LSA,
Lewen20212DM,Griffin2021PRB,Varjas2021SPP,LiK2021PRL,Fazzio2021PRB,Irvine2021PRE,Grushin2022review,Huang2022PRL,Roy2022arXiv,
Grushin2023PRB,Kvorning2022PRL,Lee2022PRB,Carp2022SPP,Huang2022PRB,Palacios2022arXiv,Wulles2022PRA,
Kante2023SA,Fleury2023SA,Grushin2023arXiv,Hellmann2023NM,Grushin2023EPL,Chowdhury2023PRL,Ojanen2023arXiv,Grushin2023arXiv2,Jionghao2023arXiv}.
In addition to these two relatively real systems, people have also studied topological phases in other artificial non-periodic structures,
including fractal lattices~\cite{song2014topological,brzezinska2018topology,pai2019topological,yang2020photonic,fremling2020existence,manna2020anyons,iliasov2020hall,
sarangi2021effect,fischer2021robustness} and hyperbolic lattices~\cite{Urwyler2022PRL,Zhou2022PRB,Bzdusek2022PRB}.
Fractals exhibit self-similar structures with non-integer dimensions~\cite{falconer2004fractal}.
Remarkably, though introduced in mathematics, fractal structures have been observed in real materials recently~\cite{nunez2023topological}.
Hyperbolic lattices live in the hyperbolic plane of negative curvature, which can be partly described by hyperbolic band theory~\cite{Rayan2021scia}.
As will be introduced in the following, these non-crystalline systems can also harbor a variety of HOTPs,
including topological phases without crystalline counterparts.

Specifically, the authors in Ref.~\cite{Agarwala2020PRR} showed that a HOTI with quantized quadrupole moment and zero-energy corner modes can exist in
a 2D amorphous system, for which the higher-order topology is actually protected by chiral symmetry.
Later, the authors in Ref.~\cite{WangJH2021PRL} found SOTIs with either chiral or helical hinge states in a 3D amorphous system,
which are characterized by nontrivial topological invariants, quantized longitudinal conductance, and in-gap hinge states.
More intriguingly, it was found that the structural disorder can induce a SOTI starting
from a trivial insulator on a regular lattice. The topological phase transition occurs through a bulk energy gap closure
guaranteed by an average $C_4 T$ symmetry.
Recently, it has been realized that the average symmetry plays an important role in the classification of symmetry-protected topological phases~\cite{WangChong2023PRX}.

Apart from amorphous systems without any spatial symmetry,
the HOTPs with in-gap corner modes have been found to exist
in other non-crystalline systems such as quasicrystalline~\cite{Fulga2019PRL,Xu2020PRL,Cooper2020PRR,Xu2020PRB,Liu2021NanoLett,Zhou2021PRB,LiuFeng2022PRL,Sassetti2022Symmetry,Ziani2022PRB,Jiang2022PRAp}, hyperbolic~\cite{Xu2023PRB,ZRLiu2023PRB}, and fractal lattices~\cite{pai2019topological,manna2022higher,lijk2022higher,zheng2022observation,chenh2023higher,ma2023elastic,nunez2023topological}.
Since quasicrystalline and hyperbolic lattices can possess rotational symmetries that do not exist in regular crystals,
they can harbor topological phases without crystalline counterparts.
For instance, it was found that a HOTP hosting eight corner modes exists in a quasicrystal with eight-fold rotational symmetry,
which can be characterized by a $\mathbb{Z}_2$ invariant protected by particle-hole symmetry~\cite{Fulga2019PRL,Xu2020PRL}.
Similar phases were also found in 2D amorphous lattices with an average rotational symmetry~\cite{Tao2023Average} and hyperbolic lattices
with various rotational symmetries~\cite{Xu2023PRB,ZRLiu2023PRB}.
In Ref.~\cite{MaoYF2023arXiv}, the authors studied a 3D quasicrystalline lattice constructed by stacking 2D quasicrystals and
found a 3D SOTI with eight hinge states or helical pairs of hinge states. It was found that these phases
are protected by the winding number of the generalized quadrupole moment and a $\mathbb{Z}_2$ invariant
based on transformed lattice sites, respectively. The authors also shown that the 2D HOTI on a quasicrystalline lattice
can be characterized by the generalized quadrupole moment~\cite{MaoYF2023arXiv}.

There have been several papers reviewing the topics of HOTPs in crystalline systems.
In Refs.~\cite{Brouwer2019PRX,Brouwer2021review}, the authors introduced the general theory of higher-order bulk-boundary correspondence to classify the bulk and boundary topology of HOTPs, mainly for HOTPs protected by crystalline symmetry.
Another excellent review~\cite{ChenYF2021NRP} introduced celebrated models of HOTPs and especially focused on experimental realizations
in various classical platforms, such as metamaterials and electric circuits.
Both of them have focused on HOTPs in crystalline systems, which can be well understood using conventional topological band theory.
In comparison, our review includes HOTPs in both crystalline and non-crystalline systems and can be roughly divided into two parts.
The first part introduces several prototypical models for HOTPs in crystalline systems. We demonstrate the higher-order boundary states of these models and present corresponding topological invariants beyond symmetry indicators, which have not been reviewed in detail before.
The second part discusses HOTPs in systems with quenched disorder and various kinds of non-crystalline systems lacking translational symmetry,
which have not been covered in previous reviews.
We have comprehensively reviewed the research progress on HOTPs in disordered and non-crystalline systems including amorphous, quasicrystalline, hyperbolic, and fractal lattices.

The review is organized as follows.
In Sec.~\ref{sec2}, we review several prototypical models of HOTPs in crystals and introduce their topological properties.
In Sec.~\ref{sec3}, we review the studies on disorder effects on HOTPs and demonstrate disorder-induced HOTPs.
In Sec.~\ref{sec4}, we review the studies on HOTPs in non-crystalline systems including amorphous lattices,
quasicrystalline lattices, hyperbolic lattices, and fractal lattices.
In Sec.~\ref{sec5}, we briefly summarize the experimental progress on HOTPs in both solid state materials and metamaterials.
Finally, in Sec.~\ref{sec6}, we give a summary and perspective.

\section{Higher-order topological phases in crystalline systems}\label{sec2}

In this section, we will review several prototypical models of HOTPs,
including multipole TIs, chiral-symmetric HOTPs,
3D HOTIs with chiral or helical hinge modes,
and higher-order topological semimetals.

\subsection{Multipole topological insulators}
Here we will introduce a class of HOTPs named quantized electric multipole insulators
or multipole TIs. They are characterized by quantized multipole moment and zero-dimensional corner states,
which are generalizations of 1D TIs with quantized polarization.
We mainly focus on a prototypical model of quadrupole TIs known as the BBH model [see Fig.~\ref{fig:BBH1}(a)]
proposed in Ref.~\cite{Benalcazar2017Science},
and discuss its bulk and boundary properties and corresponding topological invariants.
The model for octupole TIs can be constructed analogously [see Fig.~\ref{fig:BBH1}(b)], which belongs to third-order topological phases~\cite{Benalcazar2017Science}.

\subsubsection{Benalcazar-Bernevig-Hughes model}\label{sec:BBH}

\begin{figure*}[t]
\centering
\includegraphics[width=1.0\linewidth]{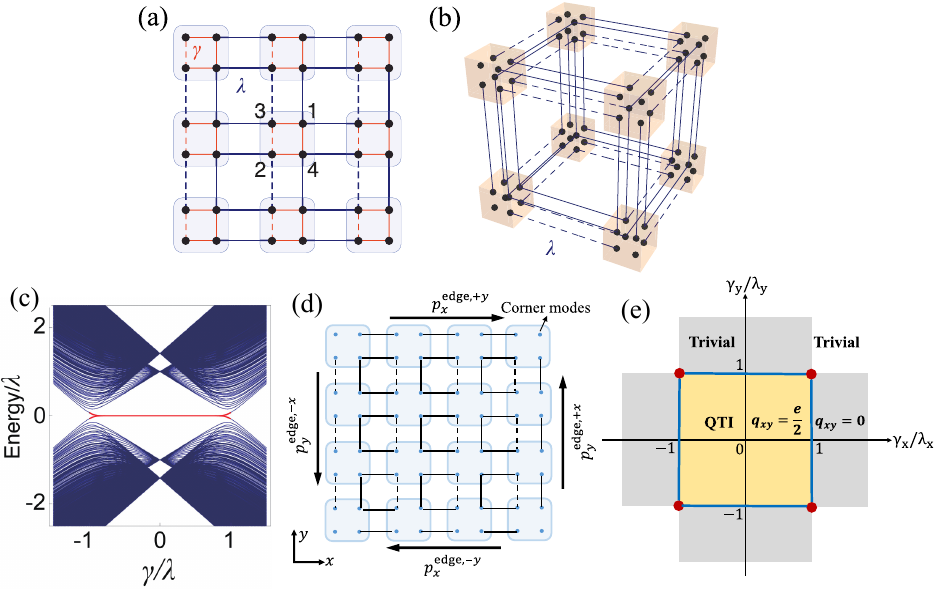}
\caption{(a) Schematic illustration of the 2D BBH model with quantized octupole moment where
	$\gamma$ and $\lambda$ represent intracell and intercell hopping, respectively.
	The dashed lines represent the hopping terms with negative signs.
	The numbers indicate the basis for $\Gamma$ matrices [see Eq.~(\ref{eq:H_BBH})].
	(b) Schematic of the tight-binding model with quantized octupole moment.	
	(c) The energy spectrum of the BBH model with open boundary conditions along $x$ and $y$ as a function
	of $\gamma/\lambda$ with $\gamma_{x,y}=\gamma$ and $\lambda_{x,y}=\lambda$.
	The red lines describe four degenerate corner states. From Ref.~\cite{Benalcazar2017Science}.
	(d)  Schematic illustration of the BBH model in a limiting case with $\gamma_x=\gamma_y=0$, showing how
	corner states and edge polarizations arise.
(e) The phase diagram of the BBH model defined in Eq.~(\ref{eq:H_BBH}),
where QTI represents the quadrupole topological insulating phase.
The phase transition between the QTI phase in yellow region and trivial phase in white and grey region occurs when either the bulk energy gap closes at the red points
or the edge energy gap closes at the blue lines.
}
\label{fig:BBH1}
\end{figure*}

The BBH model is a tight-binding model on a square lattice~\cite{Benalcazar2017Science,Benalcazar2017PRB}
\begin{align}\label{eq:H_BBH_tb}
\hat{H}_{\textrm{BBH}}
&=\sum_{\bf r}\left[\gamma_x \left(\hat{c}^\dagger_{{\bf r},1}\hat{c}_{{\bf r},3}+ \hat{c}^\dagger_{{\bf r},2}\hat{c}_{{\bf r},4} \right)\right. \nonumber\\
&+\gamma_y \left(\hat{c}^\dagger_{{\bf r},1}\hat{c}_{{\bf r},4}- \hat{c}^\dagger_{{\bf r},2}\hat{c}_{{\bf r},3} \right) \nonumber\\
&+\lambda_x \left(\hat{c}^\dagger_{{\bf r},1}\hat{c}_{{\bf r+{\bf e}_x},3}+ \hat{c}^\dagger_{{\bf r},4}\hat{c}_{{\bf r+{\bf e}_x},2} \right) \nonumber\\
&+\lambda_y \left.\left(\hat{c}^\dagger_{{\bf r},1}\hat{c}_{{\bf r+{\bf e}_y},4}- \hat{c}^\dagger_{{\bf r},3}\hat{c}_{{\bf r+{\bf e}_y},2} \right) \right] + \mathrm{H.c.},
\end{align}
where $\hat{c}^\dagger_{{\bf r},\nu}$ ($\hat{c}_{{\bf r},\nu}$) with $\nu=1,2,3,4$ are creation (annihilation) operators for
the four sites within a unit cell labelled by ${\bf r}$, as shown in Fig.~\ref{fig:BBH1}(a).
Here, ${\bf e}_x=(1,0)$ and ${\bf e}_y=(0,1)$ are lattice vectors. We take lattice constants $a_{x,y}=1$.
$\gamma_{x,y}$ and $\lambda_{x,y}$ are nearest-neighbor hopping amplitudes.
In momentum space, the corresponding Bloch Hamiltonian is
\begin{align}\label{eq:H_BBH}
H_{\textrm{BBH}} (\bm{k}) = & (\gamma_x + \lambda_x \cos k_x)\Gamma_4 + \lambda_x \sin k_x \Gamma_3 \nonumber \\
+ &(\gamma_y + \lambda_y \cos k_y)\Gamma_2 + \lambda_y \sin k_y \Gamma_1,
\end{align}
where $\Gamma_1=-\tau_2 \sigma_1$, $\Gamma_2=-\tau_2 \sigma_2$, $\Gamma_3=-\tau_2 \sigma_3$, $\Gamma_4 = \tau_1 \sigma_0$
are mutually anticommuting Hermitian matrices~\cite{Benalcazar2017Science,Benalcazar2017PRB}.
Here $\tau_{j}$, $\sigma_{j}$ are Pauli matrices representing the degrees of freedom within a unit cell.
This model describes spinless electrons on a square lattice with $\pi$-flux per plaquette and the hoppings are dimerized along $x$ and $y$ directions.
The Hamiltonian $H_{\textrm{BBH}}(\bm{k})$ has two energy bands and each band is two-fold degenerate enforced by
the TRS and inversion symmetry (see the following proof).
The model is an insulator at half filling unless the bulk energy gap closes at $|\gamma_x/\lambda_x|= |\gamma_y/\lambda_y|= 1$.

The Hamiltonian (\ref{eq:H_BBH}) satisfies two mirror symmetries ${M}_{x}$ and ${M}_{y}$ represented by $U_{M_{x}}$ and $U_{M_{y}}$, respectively,
\begin{align}
&U_{M_{x}}^{\dagger} H_{\textrm{BBH}}(k_x,k_y) U_{M_{x}} = H_{\textrm{BBH}}(-k_x,k_y), \nonumber \\
&U_{M_{y}}^{\dagger} H_{\textrm{BBH}}(k_x,k_y) U_{M_{y}} = H_{\textrm{BBH}}(k_x,-k_y),
\end{align}
with $U_{M_x} = \tau_1 \sigma_3$ and $U_{M_{y}} = \tau_1 \sigma_1$.
When $\gamma_x=\gamma_y$ and $\lambda_x=\lambda_y$,
the Hamiltonian (\ref{eq:H_BBH}) respects
a four-fold rotational symmetry ${C}_4$ with $({C}_4)^4 = -1$,
\begin{equation}
U_{C_4}^{\dagger} H_{\textrm{BBH}}(k_x,k_y) U_{C_4} =   H_{\textrm{BBH}}(-k_y,k_x),
\end{equation}
where $U_{C_4}=
\begin{pmatrix}
  0 & \sigma_0 \\
  -i\sigma_2 & 0
\end{pmatrix}
$
and $C_4 \bm{k} = (-k_y,k_x)$.
The Hamiltonian (\ref{eq:H_BBH}) also has spinless
TRS $\mathcal{T}$ represented by
the complex conjugation $\kappa$, particle-hole symmetry $\mathcal{P}$ represented by $\tau_3 \sigma_0 \kappa$, and
chiral symmetry $\mathcal{C}$ represented by $\tau_3 \sigma_0$~\cite{Benalcazar2017Science}.

We now prove that the TRS and the inversion symmetry
$U_I=\sigma_2$, which anticommute with each other, guarantee that each energy band is at least two-fold degenerate.
Let $\ket{u^n_{{\bm k}}}$
be an eigenstate of the Hamiltonian (\ref{eq:H_BBH}) with the energy $E^n_{{\bm k}}$ at ${\bm k}$. Then
$U_I \mathcal{T}\ket{u^n_{{\bm k}}}$ corresponds to an eigenstate with the same energy
at ${\bm k}$. Assume that there were no degeneracy at ${\bm k}$, which is equivalent to
saying that the two states refer to the same one up to a phase $\theta$, i.e.
$U_I \mathcal{T}\ket{u^n_{{\bm k}}}=e^{i\theta} \ket{u^n_{{\bm k}}}$.
Using $\mathcal{T} \equiv \kappa$,
we obtain $(U_I \mathcal{T})^2 \ket{u^n_{{\bm k}}} = U_I \mathcal{T} e^{i\theta} \ket{u^n_{{\bm k}}}=e^{-i\theta}U_I \mathcal{T}\ket{u^n_{{\bm k}}}=\ket{u^n_{{\bm k}}}$.
On the other hand, due to the anticommutation relation of $U_I$ and $\mathcal{T}$, $(U_I \mathcal{T})^2 \ket{u^n_{{\bm k}}}=-\ket{u^n_{{\bm k}}}$,
leading to $\ket{u^n_{{\bm k}}} = -\ket{u^n_{{\bm k}}}$, which is impossible. We therefore conclude that each energy band is at least
two-fold degenerate.

Due to the $\pi$-flux in each plaquette, the mirror symmetries $M_x$ and $M_y$ anticommute with each other, namely, $\{M_x,M_y\}=0$,
which guarantees the two-fold degeneracy of energy bands at high-symmetry points and leads to the gapped Wannier bands~\cite{Benalcazar2017Science,Benalcazar2017PRB}.
The mirror symmetries or fourfold rotational symmetry can enforce the quantization of the bulk quadrupole moment $q_{xy}$ (see definition in Sec.~\ref{sec:Qxy}).

Owing to the mirror symmetries, the BBH model has zero bulk polarization and zero Chern number so that there are no first-order edge states~\cite{Benalcazar2017Science}.
The absence of the Chern number can also be seen from the existence of the TRS.
However, when $|\gamma_x/\lambda_x|<1$ and $|\gamma_y/\lambda_y|<1$, the model supports zero-energy corner modes
localized at four corners in the energy spectrum with full open boundary conditions [see Fig.~\ref{fig:BBH1}(c)].
For clarity, we consider a limiting case with $\gamma_x=\gamma_y=0$ to illustrate these nontrivial signatures
for the quadrupole topological phase illustrated in Fig.~\ref{fig:BBH1}(d).
It is clear to see the existence of four isolated sites, contributing to four zero-energy corner modes.
In addition, we also easily see that there is a 1D Su-Schrieffer-Heeger (SSH) chain with mirror symmetry localized at each edge,
which yields a quantized polarization $\pm e/2$ tangent to the edge.
The quantized edge polarization comes from the nontrivial topology of the 1D boundary.
The edge polarization of the 1D boundary also indicates the existence of boundary states of boundary, which form the corner modes.

In fact, in the entire topological region, the phase
has a quantized bulk quadrupole moment $q_{xy}=e/2$,
quantized edge polarizations $p_x^{\textrm{edge}}=p_y^{\textrm{edge}}=e/2$ tangent to the boundaries along $x$ and $y$,
and zero-energy corner states~\cite{Benalcazar2017Science}.
The nontrivial phase is dubbed a quadrupole topological insulator (QTI).
When $|\gamma_x/\lambda_x|>1$ or $|\gamma_y/\lambda_y|>1$, the model is topologically trivial without boundary signatures.
Figure~\ref{fig:BBH1}(e) shows the phase diagram of the BBH model with respect to the parameter ratios $\gamma_x/\lambda_x$ and $\gamma_y/\lambda_y$,
which includes the topologically nontrivial QTI phase with $q_{xy}=e/2$ and the trivial phase with $q_{xy}=0$.
The quantization of the quadrupole moment is protected by the mirror symmetries or chiral symmetry (see Sec.~\ref{sec:Qxy}).
The edge polarizations are the polarizations localized along the edges when
we take a cylinder geometry with open boundary in $x$ ($y$) and periodic boundary in $y$ ($x$) (see Sec.~\ref{sec:EdgePolarization}).
The integrated polarization near one edge are quantized to $\pm e/2$ by effective mirror symmetries inherited from the bulk.
The edge polarization $p_x^{\textrm{edge}}$ and $p_y^{\textrm{edge}}$ jumps to $e/2$ from zero as we
change parameters $\gamma_x/\lambda_x$ and $\gamma_y/\lambda_y$ across the phase boundaries at
$|\gamma_x/\lambda_x|=1$ with $|\gamma_y/\lambda_y|<1$ and $|\gamma_y/\lambda_y|= 1$ with $|\gamma_x/\lambda_x|< 1$.
This happens either through the edge energy gap closing or the Wannier gap closing of the Wannier bands
(see Sec.~\ref{sec:Wannier_band}).

The BBH model can be generalized to 3Ds [see Fig.~\ref{fig:BBH1}(b)], which is a model for octupole TIs with quantized octupole moment~\cite{Benalcazar2017Science}.
The model hosts eight zero-energy corner modes.

\subsubsection{Corner modes and corner charges} \label{Sec:Cornermodes}
Here we will follow Refs.~\cite{Benalcazar2017Science,Benalcazar2017PRB} to give the analytic solution of the corner modes for the BBH model (\ref{eq:H_BBH}).
For simplicity, we will solve a continuum model for the lattice Hamiltonian (\ref{eq:H_BBH})
by expanding the Hamiltonian near $\gamma_x=\gamma_y=-1$ and $(k_x,k_y)=(0,0)$ where the bulk gap closes up to
the first order,
\begin{equation}\label{H_BBH_Dirac}
H = k_x \Gamma_3+ m_x\Gamma_4 +k_y\Gamma_1+ m_y\Gamma_2,
\end{equation}
where we take $\lambda_x=\lambda_y=1$ and assume that $m_x=\gamma_x+1$ and $m_y=\gamma_y+1$.
The parameters $m_{x,y}$ are small and are positive (negative) for the topological (trivial) phase of the BBH model.
To solve the corner states under open boundary conditions, we introduce the edges normal to the $x$ or $y$ direction,
namely, the $x$-edge or $y$-edge forming domain walls and change
$m_x$ and $m_y$ from positive to negative values across
the domain walls. Note that for positive $m_x$ and $m_y$, the system is in a topological phase while for
negative ones, it is in a trivial phase.

We first use the continuum model (\ref{H_BBH_Dirac}) to solve the states localized at the $x$-edge ($x=0$) in the absence of $y$-edges.
The wave function localized at the $x$-edge takes the form $\Psi(x,k_y)=f(x)\Phi_{x}(k_y)$ where $f(x)$ is a scalar function and $\Phi_{x}(k_y)$ is a spinor.
We require the wave function to satisfy the Schrodinger equation,
\begin{align}
&\left(-i {\partial_x f(x)} \Gamma_3+m_x(x){f(x)}\Gamma_4\right) \Phi_x(k_y) \nonumber\\
&+ \left( k_y\Gamma_1+ m_y\Gamma_2 \right) {f(x)}\Phi_x(k_y) = \epsilon {f(x)}\Phi_x(k_y),
\end{align}
where we replace $k_x$ with $-i\partial_x$ and $\epsilon$ is the eigenenergy.
This equation has a solution $f(x)=\exp (\int_{0}^{x} m_{x}(x')dx')$ and the spinor $\Phi_x$ should satisfy the equation
$(\Gamma_4-i\Gamma_3)\Phi_{x}=0$, i.e. $(\mathbb{I}-\tau^{z}\otimes \sigma^z)\Phi_{x}=0$.
We have two solutions $\Phi_{x1}=(1,0,0,0)^T$ or $\Phi_{x2}=(0,0,0,1)^T$.
We can project the Hamiltonian into the subspace of these edge states to obtain the low-energy effective Hamiltonian for the $x$-edge
\begin{equation}\label{eq:H_BBH_Dirac_edgex}
H_{\textrm{edge},\hat{x}}=-k_y \mu^y+m_y\mu^x,
\end{equation}
where $\mu^{x,y}$ are Pauli matrices in the basis $\{\Phi_{x1}, \Phi_{x2}\}$.
We can perform an analogous calculation for the $y$-edge and have the equation for the spinor
$(\mathbb{I}-\tau_0\otimes\sigma^z)\Phi_{y}=0$, which has two solutions $\Phi_{y1}=(1,0,0,0)^T$ or $\Phi_{y2}=(0,0,1,0)^T$.
Similarly, the $y$-edge Hamiltonian is obtained by projecting the Hamiltonian into the basis $\{\Phi_{y1}, \Phi_{y2}\}$,
\begin{equation}\label{eq:H_BBH_Dirac_edgey}
H_{\textrm{edge},\hat{y}}=-k_x \gamma^y+m_x\gamma^x,
\end{equation}
where $\gamma^{x,y}$ are Pauli matrices in the basis.

Both of these edge Hamiltonians take the form of massive 1D Dirac models for a 1D topological insulator
with the 1D mirror symmetry.
Now we consider a corner that is the intersection of the right $x$-edge and the upper $y$-edge.
The zero-energy corner mode can be seen as either a $y$-domain wall of the $x$-edge or a $x$-domain wall of the $y$-edge,
which have the identical solution as
\begin{equation}
\Psi^{\textrm{corner}}(x,y)= e^{\int_0^x m_x(x') dx'}e^{\int_0^y m_y(y') dy'}(1,0,0,0)^T.
\end{equation}
Therefore, we have a single corner mode at each corner of a square geometry, which is a simultaneous eigenstate of both edge Hamiltonians.

In general, the boundary states of a HOTI can be understood from the view of domain walls of Dirac mass
protected by crystalline symmetries~\cite{Langbehn2017PRL,Song2017PRL,Schindler2018SA,Brouwer2019PRX}.
Here we illustrate this picture using a Dirac Hamiltonian form of the 2D BBH model with fourfold rotational symmetry.
In the presence of the $C_4$ symmetry ($\gamma_{x,y}=\gamma$ and $\lambda_{x,y}=\lambda$),
the BBH Hamiltonian (\ref{eq:H_BBH}) can be rewritten in an alternative form of Dirac Hamiltonian~\cite{Song2017PRL},
\begin{align}\label{eq:H_BBH_C4T}
H(\bm{k})=& \left[ m+ t \left(\cos k_x +\cos k_y \right)\right]\tau_3 \sigma_0 + t_x \sin k_x \tau_1 \sigma_1 \nonumber \\
&+t_y \sin k_y \tau_1 \sigma_2 + \Delta \left(\cos k_x -\cos k_y \right) \tau_2 \sigma_0,
\end{align}
by performing the transformations
$\Gamma_4 \rightarrow (\tau_3 \sigma_0+\tau_2 \sigma_0)/\sqrt{2}$,
$\Gamma_2 \rightarrow (\tau_3 \sigma_0-\tau_2 \sigma_0)/\sqrt{2}$,
$\Gamma_3 \rightarrow \tau_1 \sigma_1$,
and $\Gamma_1 \rightarrow \tau_1 \sigma_2$; these new matrices constitute
a new set of Dirac matrices anticommmuting with each other.
Meanwhile, we define the parameters
$m=\sqrt{2}\gamma$ and $t=\Delta=\lambda/\sqrt{2}$.
Similarly to the BBH model, this Hamiltonian has a HOTP with zero-energy corner modes protected by $C_4 T$ symmetry
when $|m/t| < 2$ and $\Delta \neq 0$~\cite{Song2017PRL}.

Without the mass term, i.e. $\Delta=0$, the Hamiltonian (\ref{eq:H_BBH_C4T}) has a TRS represented by $i\sigma_y \kappa$,
belonging to the class AII (time-reversal invariant topological insulator with spin).
When $|m/t| < 2$, gapless helical edge state arise.
The $\tau_2 \sigma_0$ term breaks the TRS as well as the $C_4$ rotational symmetry represented by $e^{-i\frac{\pi}{4}\sigma_z }$,
but preserves the symmetry of their combination.
The breaking of the TRS leads to the opening of the gap of the helical modes, resulting in an effective mass for the edge
Dirac Hamiltonian.
Due to the $C_4 T$ symmetry, the Dirac mass of the edge Hamiltonian changes sign under a $C_4$ rotation.
Thus, there exists a domain wall at a corner between two edges related by the $C_4$ rotation,
which gives rise to a corner state. There are a total of four corner states for an open boundary compatible
with the $C_4$ symmetry.

The corner states lead to filling anomaly~\cite{Khalaf2021PRR} or fractional charges~\cite{Benalcazar2019PRB,Peterson2020Science} localized at the corners.
Specifically, if we require the system to be gapped and symmetric, it is inevitable to have the four degenerate corner
states to be either totally occupied or totally unoccupied, resulting in excess fractional charges localized at each corner,
which is the manifestation of filling anomaly. On the other hand, if we require the constraints of neutrality rather than
symmetry, we can add an infinitesimal symmetry-breaking term in the Hamiltonian to break the degeneracy so that
two diagonal corner states are occupied. As a result, fractional corner charges of either $e/2$ or $-e/2$ occur near the corners.

In practice, the corner charge for one corner (similarly for the other three corners) is numerically calculated by performing the integration
of the charge density over a quadrant of the system~\cite{Benalcazar2017PRB},
\begin{equation}
Q^{\mathrm{corner}~-x,-y}=\sum_{R_x=1}^{N_x/2}\sum_{R_y=1}^{N_y/2}\rho({\bf R}).
\end{equation}
Here, $\rho({\bf R})=2e-e\sum_{n=1}^{N_{\mathrm{occ}}}\sum_{\alpha=1}^4|[u^{n}]^{{\bf R},\alpha}|^2$ is
the charge density. The first and second terms arise from the atomic positive charges and the electron distribution, respectively.
The latter is calculated by the occupied state $|u^{n}\rangle$ of the Hamiltonian under open boundary conditions
($[u^{n}]^{{\bf R},\alpha}$ represents the $\alpha$th component at the site $\bf R$ of the $n$th occupied state).
For the corner charge computation, an infinitesimal term should be added to break the symmetry to lift the fourfold
degeneracy of corner modes. In other words, two corner states have positve energy and the other two have negative energy.
One thus can fill two negative corner states with electrons at half filling,
leading to fractional corner charges of $\pm e/2$ (we consider the existenc of positive charges of $2e$ in each unit cell).

The topological corner modes of the quadrupole TIs can also be probed by the in-gap modes of
the nested entanglement spectrum~\cite{Schindler2018SA,Yang2020PRR} or the entanglement spectrum between
a quarter part and its complement~\cite{Tao2023Quadrupole}. In other words, for most models, both the
energy spectrum of the Hamiltonian and the entanglement spectrum support in-gap corner modes in the topological region.
However, Tao and coworkers propose a model Hamiltonian with a staggered
$\mathbb{Z}_2$ gauge field and find that this model leads to a quadrupole insulator without corner modes in the energy spectrum
but with in-gap corner modes in the entanglement spectrum~\cite{Tao2023Quadrupole}. In fact, the entanglement spectrum has a one-to-one
correspondence with the energy spectrum of the flattened Hamiltonian rather than that of
the original Hamiltonian. When we change the flattened Hamiltonian to the original Hamiltonian by varying the
eigenenergies, the edge energy gap closes and reopens, leading to the disappearance of the corner modes~\cite{Tao2023Quadrupole}.
During this process, the bulk states remain unchanged so that the quadruple moment does not change.

\subsubsection{Quadrupole moment as a topological invariant}\label{sec:Qxy}
Here we review a real-space formula of the electric quadrupole moment and show that
its value should be quantized in the presence of some crystalline symmetries or chiral symmetry.
The formula of the quadrupole moment defined here may not correctly evaluate the physical quadrupole moment for a real system.
There have been some discussions on the subtlety of proper definition of the bulk quadrupole moment~\cite{Watanabe2019PRB,Vanderbilt2021PRB,Walet2023PRB,Oshikawa2023arXiv}.
Nevertheless, it can be used as an indicator for the second-order topology of quadrupole insulators like the BBH model.

The real-space quadrupole moment is defined by a generalization of the Resta formula for the electric polarization~\cite{Resta1998PRL}.
Proposed in Refs.~\cite{Kang2019PRB,Wheeler2019PRB},
the quadrupole moment $q_{xy}$ (in units of $e$) for a many-electron system is defined as
\begin{equation}\label{eq:Qxy_manybody}
q_{xy}=\frac{1}{2\pi}\mathrm{Im}\log \langle \Psi_G| e^{i 2\pi \hat{Q}_{xy}} |\Psi_G\rangle \mod 1,
\end{equation}
where $\hat{Q}_{xy}=\sum_{\bm{r}=(x,y)} \frac{x y}{L_x L_y} \left(\hat{n}_{\bm{r}}-\bar{n}\right)$ with $\hat{n}_{\bm{r}}$ denoting the electron number operator at site $\bm{r}=(x,y)\in(0,L_x]\times (0,L_y]$
in a system of size $L_x$ by $L_y$.
$\bar{n}$ denotes the average electron filling per unit cell counting the atomic positive charge contribution that we need to subtract.
$|\Psi_G\rangle$ is the many-body ground state of electrons in a periodic system.
Similar to the polarization from the Resta formula, the quadrupole moment $q_{xy}$ is only defined in $[0,1)$ because of the periodicity of complex phases.
Note that to have a well-defined quadrupole moment (invariance under translations), we require that the bulk polarizations vanish~\cite{Benalcazar2017Science}.
The octupole moment can be defined analogously by replacing $\hat{Q}_{xy}$
with $\hat{O}_{xyz}=\sum_{\bm{r}=(x,y,z)} \frac{x y z}{L_x L_y L_z} (\hat{n}_{\bm{r}}-\bar{n})$~\cite{Kang2019PRB,Wheeler2019PRB}.

For noninteracting fermion systems, the many-body wave function can be written as a Slater determinant of occupied single-particle eigenstates.
Then, it can be shown that the quadrupole moment is evaluated by~\cite{Roy2019PRR,Li2020PRL,Yang2021PRB}
\begin{equation}\label{eq:Qxy}
{q}_{xy}=\left[ \frac{1}{2\pi}\mathrm{Im}\log \det \left(U_{o}^\dagger \hat{D} U_{o} \right)-q_{xy}^{(0)} \right] \mod 1,
\end{equation}
where we define an $n_a\times n_c$ matrix $U_{o}=\left(|\psi_1 \rangle,|\psi_2 \rangle,\cdots,|\psi_{n_c} \rangle\right)$
representing the occupied states of an $n_a\times n_a$ single-particle Hamiltonian under periodic boundary conditions.
$\hat{D}=\text{diag}\{e^{i 2\pi x_j y_j/(L_x L_y)}\}_{j=1}^{n_a}$ is
an $n_a\times n_a$ diagonal matrix [$(x_{j},y_{j})$ is the real-space coordinate of the $j$th degree of freedom].
Here, $n_c$ is the total number of occupied states,
and $n_a$ is the total dimension of the single-particle Hamiltonian.
In the calculation, one also needs to deduct the contribution from background positive charge distribution, that is,
$q_{xy}^{(0)}=\sum_{j=1}^{n_c} x_j y_j/(L_x L_y)$ where $(x_j,y_j)$ is the position of the $j$th positive charge.

We can apply the formula (\ref{eq:Qxy}) to evaluate the quadrupole moment of typical models of quadrupole insulators.
For the BBH model (\ref{eq:H_BBH}) with chiral symmetry, we have $q_{xy}=e/2$ for the quadrupole topological phase
when $\gamma_x/\lambda, \gamma_y/\lambda$ are both in the interval $(-1, 1)$.
For the SOTI with $C_4 T$ symmetry in Eq. (\ref{eq:H_BBH_C4T}), $q_{xy}=e/2$
when $|m/t|<2$ and $|\Delta|>0$, which correctly characterizes the topologically nontrivial phase with corner modes.
Note that the quadrupole moment as a topological invariant can predict the topological phase transitions between
the topological and trivial phases of quadrupole insulators when either the bulk energy gap or edge energy gap closes.
In fact, the quadrupole moment characterizes the topology of the flattened Hamiltonian, and thus its topology
manifests in the existence of in-gap corner modes in the flattened Hamiltonian or the entanglement spectrum~\cite{Tao2023Quadrupole}.
For most models, the original Hamiltonian also supports corner states because it is connected to the flattened version
without experiencing any edge energy gap closure. However, there exists a model where quadrupole insulators
do not have corner states in the energy spectrum~\cite{Tao2023Quadrupole}.

\paragraph{Quantization of $q_{xy}$ protected by crystalline symmetries}
Under mirror symmetry, $M_x: (x,y) \rightarrow (-x,y)$ or $M_y : (x,y) \rightarrow (x,-y)$,
we see that the operator $\hat{Q}_{xy}$ in Eq.~(\ref{eq:Qxy_manybody}) transforms into $-\hat{Q}_{xy}$.
Thus, we have $q_{xy} \rightarrow -q_{xy}$. If the system is invariant under the mirror transformation,
we have $q_{xy} = -q_{xy} \mod 1$ so that $q_{xy}$ takes the quantized value of either $0$ or $1/2$.
Similarly, we see that the four-fold rotation $C_4 : (x,y) \rightarrow (-y,x)$ also flips the sign of $q_{xy}$.
Thus, the $C_4$ symmetry enforces that $q_{xy} = -q_{xy} \mod 1$, indicating that
$q_{xy}$ is quantized in the presence of the $C_4$ symmetry.

\paragraph{Quantization of $q_{xy}$ protected by chiral symmetry}
Apart from crystalline symmetries, the quadrupole moment can be quantized by internal symmetries alone.
In the following, we present the proof of quantization of $q_{xy}$ (\ref{eq:Qxy}) protected by chiral symmetry by following Refs.~\cite{Yang2021PRB,Li2020PRL}.
Let $H$ be a generic single-particle Hamiltonian in real space preserving chiral (sublattice) symmetry, that is,
there exists a unitury operator ${\Pi}$ such that ${\Pi} H {\Pi}^{-1}=-H$.
If we consider an occupied state $|\psi_n\rangle$ with energy $E_n$, then
we can obtain an unoccupied state by applying $\Pi$ to the occupied one, that is,
${\Pi}|\psi_n\rangle$ with energy $-E_n$.
Since the energy spectrum is symmetric about the zero energy, we consider the half-filling case with $n_a=2 n_c$.
Then, we have $q_{xy}^{(0)}=\frac{1}{2}\sum_{j=1}^{n_a} x_j y_j/(L_x L_y)=\frac{1}{4\pi}\mathrm{Im}\log \det \hat{D} \mod 1$.
The set
$\{{\Pi}|\psi_1\rangle,{\Pi}|\psi_2\rangle,\cdots, {\Pi}|\psi_{n_c}\rangle\}$ therefore constitutes the unoccupied states.
These unoccupied states can be represented by
$U_{u}=\left({\Pi}|\psi_1\rangle,{\Pi}|\psi_2\rangle,\cdots, {\Pi}|\psi_{n_c}\rangle\right)={\Pi}U_{o}$.
The quadrupole moment (\ref{eq:Qxy}) can be written as
\begin{equation}
q_{xy}=\frac{1}{2\pi}\mathrm{Im}\log \left[\det \left(U_{o}^\dagger \hat{D} U_{o} \right) \sqrt{\det \hat{D}^\dagger}\right].
\end{equation}
Because the unitary matrix $\hat{D}$ commutes with the chiral (sublattice) symmetry transformation $\Pi$ with $\Pi^\dagger \Pi=1$, i.e. $\left[\hat{D},\Pi\right]=0$,
\begin{align}
q_{xy}&=\frac{1}{2\pi}\mathrm{Im}\log \left[\det \left(U_{u}^\dagger \Pi \hat{D} \Pi^\dagger U_{u} \right) \sqrt{\det \hat{D}^\dagger}\right] \nonumber \\
&=\frac{1}{2\pi}\mathrm{Im}\log \left[\det \left(U_{u}^\dagger \hat{D} U_{u} \right) \sqrt{\det \hat{D}^\dagger}\right].
\end{align}

Next, we will prove that $2q_{xy} = 0 \mod 1$.
\begin{proof}
We define an $n_a \times n_a$ unitary matrix $U_{t}=\left(U_{o},U_{u}\right)$.
It can be easily seen that $\det \hat{D}^\dagger = \det \left(U_t^\dagger \hat{D}^\dagger U_t \right)$.
Then, we have the following relations
\begin{eqnarray}
4\pi q_{xy}&=&\mathrm{Im}\log \left[\det \left(U_{o}^\dagger \hat{D} U_{o} \right) \sqrt{\det \hat{D}^\dagger}\right] + \nonumber \\
&&\mathrm{Im}\log \left[\det \left(U_{u}^\dagger \hat{D} U_{u} \right) \sqrt{\det \hat{D}^\dagger}\right] \nonumber \\
&=& \mathrm{Im}\log \left[ \det \left(U_{o}^\dagger \hat{D} U_{o} \right) \det (U_{u}^\dagger \hat{D} U_{u})\right] +  \nonumber \\
&& \mathrm{Im}\log \det \left(U_t^\dagger \hat{D}^\dagger U_t \right) \nonumber \\
&=& \mathrm{Im}\log \det
\begin{pmatrix}
  U_{o}^\dagger \hat{D} U_{o} & U_{o}^\dagger \hat{D} U_{u} \\
  0 & U_{u}^\dagger \hat{D} U_{u} \\
\end{pmatrix}+ \nonumber \\
&&\mathrm{Im}\log \det
\begin{pmatrix}
  U_{o}^\dagger \hat{D}^\dagger U_{o} & U_{o}^\dagger \hat{D}^\dagger U_{u} \\
  U_{u}^\dagger \hat{D}^\dagger U_{o} & U_{u}^\dagger \hat{D}^\dagger U_{u} \\
\end{pmatrix} \nonumber \\
&=&\mathrm{Im}\log \det
\begin{pmatrix}
  \mathbb{I} & 0 \\
  U_{u}^\dagger \hat{D} U_{u} U_{u}^\dagger \hat{D}^\dagger U_{o} &
  U_{u}^\dagger \hat{D} U_{u} U_{u}^\dagger \hat{D}^\dagger U_{u} \\
\end{pmatrix} \nonumber \\
&=&\mathrm{Im}\log \det \left(U_{u}^\dagger \hat{D} U_{u} U_{u}^\dagger \hat{D}^\dagger U_{u} \right) \nonumber \\
&=&\mathrm{Im}\log \det \left(U_{u}^\dagger \hat{D} U_{u} \right) + \mathrm{Im}\log \left[ \det \left(U_{u}^\dagger \hat{D} U_{u} \right)\right]^{*} \nonumber \\
&=&0 \mod 2\pi.
\end{eqnarray}
In the derivation, the following orthonormal properties are used,
$U_{o}^\dagger U_{o}=U_{u}^\dagger U_{u}=\mathbb{I}$, $U_{o}^\dagger U_{u}=0$ and $U_{o}U_{o}^\dagger + U_{u}U_{u}^\dagger = \mathbb{I}$.
\end{proof}

Therefore, we get the conclusion that $2q_{xy}=0 \mod 1$, namely,
$q_{xy}$ is quantized to $0$ or $1/2$ up to an integer.
One can also apply the above procedure to prove that the octupole moment is quantized in 3Ds in the presence of chiral symmetry.
Similar conclusion can be generalized to systems with particle-hole symmetry~\cite{Li2020PRL}.
Since the chiral symmetry alone can protect the quantization of the quadrupole moment,
we can have quadrupole TIs in disordered systems lacking
crystalline symmetry~\cite{Yang2021PRB,Li2020PRL,Agarwala2020PRR}.

\subsubsection{Wannier bands and topological invariants}\label{sec:Wannier_band}
Here we review the topological properties of Wannier bands for quadrupole insulators protected by mirror symmetries.

\paragraph{Wilson loop and Wannier band}
In general, we consider a Bloch Hamiltonian $H({\bf k})$ which has $N_{\mathrm{orb}}$ degrees of freedom and $N_{\mathrm{occ}}$ occupied bands with Bloch functions $\ket{u_{{\bf k}}^{n}}$ for $n=1,\cdots,N_{\mathrm{occ}}$.
The Wilson loop is closely related to the position operator projected into the occupied space~\cite{Resta1998PRL}.
In the following, we will focus on the Wilson loop along $k_x$, $\mathcal{W}_{x, {\bf k}}$,
and the Wilson loop $\mathcal{W}_{y, \bf k}$ along $k_y$ can be evaluated similarly.
The Wilson loop $\mathcal{W}_{x, {\bf k}}$ starting from the base point ${\bf k}=(k_x,k_y)$ of the loop in Brillouin zone,
is defined under full periodic boundary conditions as follows
\begin{equation}\label{eq:Wilson}
\mathcal{W}_{x, \bf k} = F_{x,{\bf k}+(N_x-1){\bf \delta k_x}} \cdots F_{x,{\bf k} + {\bf \delta k_x}} F_{x,\bf k},
\end{equation}
where $[F_{x,\bf k}]^{mn} =\braket{u^{m}_{{\bf k}+{\bf \delta k_x}}}{u^n_{{\bf k}}}$, for ${\bf \delta k_x}=(2\pi/N_x,0)$.
Since the Wilson loop is unitary in the thermodynamic limit,
it can be defined as an exponential of a Hermitian matrix
\begin{equation}\label{eq:Wannier_Hamiltonian}
\mathcal{W}_{x, \bf k}\equiv e^{iH_{\mathcal{W}_x}(\bf k)}.
\end{equation}
Then, we refer to $H_{\mathcal{W}_x}(\bf k)$ as the Wannier Hamiltonian.
It has been illustrated that the Wannier Hamiltonian's spectrum has a close connection with the spectrum
and topology of the boundary perpendicular to the $x$ direction~\cite{Fidkowski2011PRL}.

The Wilson loop operator can be diagonalized as
\begin{equation}\label{eq:Wilson_loop_eigensystem}
\mathcal{W}_{x, \bf k} \ket{\nu^j_{x,\bf k}} = e^{i 2\pi \nu^j_x(k_y)} \ket{\nu^j_{x,\bf k}},
\end{equation}
where $\ket{\nu^j_{x,\bf k}}$ is the eigenvectors with components $[\nu^j_{x,\bf k}]^n$ for $n=1,\cdots,N_{\mathrm{occ}}$.
The Wannier Hamiltonian $H_{\mathcal{W}_{x}}({\bf k})$ has eigenvalues $2\pi \nu^{j}_{x}(k_y)$, $j=1,\cdots, N_{\mathrm{occ}}$,
which only depend on $k_y$ of the base point ${\bf k}$. The eigenvalues $\nu^{j}_{x}(k_y)$ defined modulo $1$ are referred to as the Wannier centers
and compose the Wannier bands.

For the BBH model with two occupied bands, the phases $\nu^{j=\pm}_x(k_y)$ of the Wilson loop $\mathcal{W}_{x, {\bf k}}$ over two occupied bands
form a pair of opposite values $\nu_x^-(k_y) = -\nu_x^+(k_y) \mod 1$ due to the mirror symmetry $M_x$,
leading to the vanishing of the bulk polarizations.
The two Wannier bands $\nu^{j=\pm}_x(k_y)$ have a Wannier gap at both $\nu_x=0$ and $\nu_x=1/2$ over the Brillouin zone.
Therefore, we can have a 1D gapped Wannier Hamiltonian $H_{\mathcal{W}_x}(k_y)$ for $k_y \in [-\pi,\pi]$,
which may host nontrivial band topology.

\paragraph{Nested Wilson loop and Wannier-sector polarization}
When the Wannier bands $\nu^j_x(k_y)$ are gapped for $k_y \in (-\pi,\pi]$,
we can define two Wannier sectors separated by the gap.
For quadrupole insulators with two anticommuting mirror symmetries $M_x$ and $M_y$,
the Wannier bands $\nu^j_x(k_y)$ are gapped for $k_y \in (-\pi,\pi]$ at both $\nu_x=0$ and $\nu_x=1/2$.
Then, we can define two Wannier sectors
$\nu^-_x \in (0,1/2)$ and  $\nu^+_x \in (1/2,1)$. In the BBH case, there is only one band in each Wannier sector.

The topological invariant for a Wannier sector $\nu_x$ can be evaluated based on the nested Wilson loops.
With the Wannier bands $\nu^{l}_x(k_y)$ and Wilson loop eigenstates $\ket{\nu^l_{x,\bf k}}$,
the nested Wilson loop is a Wilson loop of one subspace of Wannier bands along $k_y$.
Specifically, for the two Wannier sectors $\nu^{\pm}_x$ split by the Wannier gap,
we define the Wannier band basis in the Wannier sector $\nu_x^-$ as
\begin{equation}\label{eq:Wannier_basis}
\ket{w^l_{x,\bf k}} = \sum_{n=1}^{N_{\mathrm{occ}}}\ket{u^n_{\bf k}} [\nu^l_{x,\bf k}]^n
\end{equation}
for $l \in 1\ldots N_W$. $N_W$ is the number of the Wannier bands in the sector $\nu_x^-$.
This basis obeys $\braket{w^l_{x,\bf k}}{w^{l'}_{x, \bf k}} = \delta_{l l'}$.
Then, the nested Wilson loop along $y$ for the Wannier sector $\nu_x$ over the Wannier band basis is defined as
\begin{equation}\label{eq:nested_Wilson_loop}
\tilde{\mathcal{W}}^{\nu_x}_{y, k_x} = F^{\nu_x}_{y,{\bf k}+(N_y-1){\bf \delta k_y}} \cdots F^{\nu_x}_{y,{\bf k} + {\bf \delta k_y}} F^{\nu_x}_{y,\bf k},
\end{equation}
where $[F^{\nu_x}_{y,\bf k}]^{rs} = \braket{w^r_{x,{\bf k}+{\bf \delta k_y}}}{w^s_{x,\bf k}}$ with
$r,s \in 1 \ldots N_W$ over all Wannier bands in the Wannier sector $\nu_x$, and ${\bf \delta k_y}=(0, 2\pi/N_y)$.
The total polarization of the Wannier sector $\nu_x$ is
\begin{equation}
p^{\nu^\pm_x}_y= \frac{1}{N_x}\sum_{k_x} \frac{1}{2\pi}\mathrm{Im}\log \det \left [\tilde{\mathcal{W}}^\pm_{y, k_x} \right],
\end{equation}
which, in the thermodynamic limit, becomes an integral
\begin{equation}
p^{\nu_x}_y =-\frac{1}{(2\pi)^2} \int_{\textrm{BZ} } \Tr \left[\tilde{\mathcal{A}}^{\nu_x}_{y,\bf k}\right] d^2\bf k,
\end{equation}
where $\tilde{\mathcal{A}}^{\nu_x}_{y,\bf k}$ is the Berry connection of Wannier bands $\nu_x$ having components
\begin{align}
[\tilde{\mathcal{A}}^{\nu_x}_{y,\bf k}]^{l l'} = -i \bra{w^l_{x,\bf k}} \partial_{k_y} \ket{w^{l'}_{x,\bf k}},
\end{align}
where $l,l' \in 1 \ldots N_W$ run over the Wannier bands in Wannier sector $\nu_x$~\cite{Benalcazar2017Science,Benalcazar2017PRB}.

Under the mirror reflection $M_y$ and $M_x$, the Wannier-sector polarizations satisfy
$p^{\nu_x}_y \equiv -p^{\nu_x}_y \mod 1$ and $p^{\nu_y}_x \equiv -p^{\nu_y}_x \mod 1$, respectively~\cite{Benalcazar2017Science,Benalcazar2017PRB}.
Hence, $p^{\nu_x}_y$ and $p^{\nu_y}_x$ are quantized to $0$ or $1/2$.
For the topological nontrivial phase of the BBH model (\ref{eq:H_BBH}),
we have $(p^{\nu_x}_y, p^{\nu_y}_x)=(1/2,1/2)$ in consistent with the quadrupole moment $q_{xy}=1/2$.
The transition between the phase with different Wannier-sector polarizations is driven by the gap closing of the Wannier bands.

The mirror symmetries can also simplify the computation of the topological invariants.
Since the 1D Wannier Hamiltonian $H_{\mathcal{W}_{x}}({\bf k})$ is invariant under the mirror reflection $M_y$ inherited from the bulk,
the Wannier-sector polarization can be obtained from the reflection eigenvalues
of Wannier band basis (\ref{eq:Wannier_basis}) at mirror invariant momenta~\cite{Benalcazar2017Science,Benalcazar2017PRB}.

\paragraph{Symmetry constraints of Wannier Hamiltonian}
Consider a generic Bloch Hamiltonian $H({\bf k})$ with a symmorphic lattice symmetry $g$: ${\bf r}\rightarrow D_g{\bf r}$, described by
$g H({\bf k}) g^\dagger = H(D_g{\bf k})$,
where $g$ is a unitary operator for the symmetry.
We now define a Wilson line following a path $C$ in momentum space from ${\bf k}_i$ to ${\bf k}_f$ as
\begin{equation}
\mathcal{W}_{{\bf k}_f \leftarrow {\bf k}_i}=F_{{\bf k}_f-\delta {\bf k}}\cdots F_{{\bf k}_i+\delta {\bf k}}F_{{\bf k}_i},
\end{equation}
where $[F_{{\bf k}}]^{mn}=\langle u_{{\bf k}+\delta {\bf k}}^m|u_{{\bf k}}^n\rangle$ with $m,n$ labeling the occupied bands and
$\delta{\bf k}=({\bf k}_f - {\bf k}_i)/N$.
We take $N\rightarrow \infty$ in the thermodynamic limit.
The Wilson line operator transforms under the symmetry as~\cite{Benalcazar2017Science,Benalcazar2017PRB}
\begin{equation}
B_{g,{\bf k}_f} \mathcal{W}_{{\bf k}_f \leftarrow {\bf k}_i} B_{g,{\bf k}_i}^\dagger = \mathcal{W}_{D_g {\bf k}_f \leftarrow D_g {\bf k}_i}.
\end{equation}
Here a unitary sewing matrix $B_{g,{\bf k}}^{mn}=\langle u_{D_g {\bf k}}^m| g |u_{\bf k}^n\rangle$
is defined. The matrix reflects the connection between eigenstates at ${\bf k}$ with eigenstates at $D_g {\bf k}$.
In particular, a Wilson loop along the contour $\mathcal{C}$ at the base point ${\bf k}$ satisfies
\begin{equation}
B_{g,{\bf k}} \mathcal{W}_{\mathcal{C},{\bf k}} B_{g,{\bf k}}^\dagger = \mathcal{W}_{D_g \mathcal{C},D_g \bf k} .
\end{equation}

Now we focus on the topology of the Wannier band $\nu_x(k_y)$ from the Wilson loop $\mathcal{W}_{x,{\bf k}}$ in mirror-symmetric systems.
The results for $\nu_y(k_y)$ can be obtained analogously.
Under the mirror symmetry $M_x$, we have the transformation for the Wilson loop as
\begin{equation}
B_{m_x,{\bf k}_f} \mathcal{W}_{x,{\bf k}} B_{m_x,{\bf k}_i}^\dagger =\mathcal{W}_{-x,M_x{\bf k}},
\end{equation}
where $x$ and $-x$ label the direction that a Wilson loop is obtained.
We then define a Wannier Hamiltonian for the Wilson loop $\mathcal{W}_{x,(k_x=0,k_y)}$.
The Wannier Hamiltonian in Eq.~(\ref{eq:Wannier_Hamiltonian}) is a multivalued function due to the logarithm, and thus we
need to redefine it relative to a branch cut $\epsilon$~\cite{Yang2020PRR},
\begin{equation}
H_{\mathcal{W}_x}^\epsilon (k_y) \equiv -i\log_{\epsilon}(\mathcal{W}_{x,{\bf k}}),
\end{equation}
where we take $\log_{\epsilon} e^{i\phi}=i\phi$, for $\epsilon\le \phi<\epsilon+2\pi$.
With the symmetries, it is proved that the Wannier Hamiltonian should satisfy the following
constraints~\cite{Yang2020PRR},
\begin{align}
&B_{m_x,(0,k_y)}H_{\mathcal{W}_x}^\epsilon (k_y)B_{m_x,(0,k_y)}^\dagger = -H_{\mathcal{W}_x}^{-\epsilon}(k_y) +2\pi I_{N_{\textrm{occ}}} \nonumber \\
&B_{m_x,(0,k_y)}H_{\mathcal{W}_x}^\epsilon (k_y)B_{m_x,(0,k_y)}^\dagger = -H_{\mathcal{W}_x}^{-\epsilon+2\pi}(k_y) +4\pi I_{N_{\textrm{occ}}}.
\end{align}

\paragraph{Wannier-gap winding number}
The Wannier bands obey the periodicity from the relation $\nu_x \equiv \nu_x \mod 1$,
reminiscent of the effective Hamiltonian of Floquet TIs~\cite{Levin2013PRX,Yao2017PRB}.
It is thus possible that the Wannier bands under open boundary exhibit in-gap modes
at both $\nu_x=0$ and $\nu_x=1/2$ in a cylinder geometry.
As a result,
the Wannier-sector polarization are trivial so that they fail to identify the topology of the Wannier bands.
The quadrupole TIs with this type of Wannier bands are called anomalous quadrupole TIs~\cite{Fulga2018PRB}.
They resemble the anomalous topological phases in a periodically driven system, where
the edge states persist even though the traditional topological invariant of the effective Hamiltonian vanishes~\cite{Levin2013PRX}.

To correctly identify the topology of Wannier bands, Yang and coworkers introduce a topological invariant
which can fully characterize the change of the Wannier band topology due to the Wannier gap closing~\cite{Yang2020PRR}.
Specifically,
a Wilson line with respect to $\epsilon$ is introduced,
\begin{equation}
\mathcal{W}_{x,k_x\leftarrow 0}^{\epsilon}(k_y) \equiv \mathcal{W}_{x,k_x\leftarrow 0}(k_y) \exp\left[-iH_{\mathcal{W}_x}^{\epsilon}(k_y)\frac{k_x}{2\pi}\right],
\end{equation}
where $\mathcal{W}_{x,k_x\leftarrow 0}(k_y)\equiv \mathcal{W}_{(k_x,k_y)\leftarrow (0,k_y)}$.
It is shown that $\mathcal{W}_{x,2\pi\leftarrow 0}^{\epsilon}(k_y)=I_{N_{\textrm{occ}}}$,
and one arrives at the fact that
$\mathcal{W}_{x,k_x\leftarrow 0}^{\epsilon}(k_y)$ is periodic versus $k_x$.
The symmetry further imposes constraints on the Wilson line with respect to $\epsilon$ for $k_x=\pi$ and $\epsilon=0,\pi$
such that~\cite{Yang2020PRR}
\begin{align}
&B_{m_x,(\pi,k_y)}\mathcal{W}_{x,\pi \leftarrow 0}^{\epsilon=0}(k_y) B_{m_x,(0,k_y)}^{\dagger} = -\mathcal{W}_{\pi\leftarrow 0}^{\epsilon=0}(k_y),
\nonumber \\
&B_{m_x,(\pi,k_y)}\mathcal{W}_{x,\pi \leftarrow 0}^{\epsilon=\pi}(k_y) B_{m_x,(0,k_y)}^{\dagger} = \mathcal{W}_{\pi\leftarrow 0}^{\epsilon=\pi}(k_y).
\end{align}

For the quadrupole insulator respecting two anticommuting mirror symmetries with $N_{\textrm{occ}}=2$,
it can be shown that there exists a basis in which the sewing matrices along the mirror invariant lines take the form~\cite{Yang2020PRR}
\begin{equation}
B_{m_x,(\pi,k_y)}=B_{m_x,(0,k_y)}=\left(
  \begin{array}{cc}
    1 & 0 \\
    0 & -1 \\
  \end{array}
\right).
\end{equation}
As a result, the Wilson loops are $2\times 2$ matrices taking forms as
\begin{eqnarray}
\mathcal{W}_{x,\pi \leftarrow 0}^{\epsilon=0}(k_y)&=&\left(
  \begin{array}{cc}
    0 & U_+^{\epsilon=0}(k_y) \\
    U_-^{\epsilon=0}(k_y) & 0 \\
  \end{array}
\right),\\
\mathcal{W}_{x,\pi \leftarrow 0}^{\epsilon=\pi}(k_y)&=&\left(
  \begin{array}{cc}
    U_+^{\epsilon=\pi}(k_y) & 0 \\
    0 & U_-^{\epsilon=\pi}(k_y) \\
  \end{array}
\right).
\end{eqnarray}
The topological invariant for the Wannier gaps at $\epsilon=0,\pi$ can thus be defined as a winding number
\begin{equation}\label{eq:WindFormula}
W_{\nu_x}^{\epsilon}=\frac{1}{2\pi i}\int_0^{2\pi} dk_y\partial_{k_y}\log U_{+}^{\epsilon}(k_y).
\end{equation}
Since the Wilson line depends on the eigenstates, the winding number is not gauge invariant.
Thus, when one varies the system parameter to observe the topological phase transitions changing the winding number,
the global phase of wave functions are required to be continuous along mirror invariant lines in Brillouin zone.
The numerical procedure is described in detail in Ref.~\cite{Yang2020PRR}.

It has been shown that the change of the winding number can reflect the Wannier gap closing at either $\nu=0$
or $\nu=\pm 1/2$.
The winding number only responds to the closure of the bulk energy gap or the Wannier gap
rather than the edge energy gap closing.

\subsubsection{Edge polarizations}\label{sec:EdgePolarization}

\paragraph{Edge polarization from the Wilson loop}
The edge polarization $p_{x}^{\mathrm{edge~}\beta}$ ($p_{y}^{\mathrm{edge~}\alpha}$) describes the
 edge polarization per unit length along the $x$ ($y$) direction at the boundaries perpendicular to
 $y$ ($x$). The Greek letters $\beta=\pm y$ ($\alpha=\pm x$) label the $y$-normal ($x$-normal)
 edges with the sign identifying the top or bottom (left or right) boundaries.
We now review how the edge polarization is calculated based on the Wilson loop in a cylinder geometry following Refs.~\cite{Benalcazar2017Science,Benalcazar2017PRB}.
Specifically, for a 2D insulator with $N_x\times N_y$ unit cells, we evaluate the polarization $p_x$ along $x$ with respect to the position along $y$.
We take the periodic boundary condition along $x$ and the open boundary condition along $y$.
The system can be treated as a pseudo-1D Hamiltonian with $N_{\mathrm{orb}} N_y$ internal orbitals and $N_x$ unit cells along $x$,
and $k_x$ remains a good quantum number.
We can compute the Wilson loop $\mathcal{W}_x$ for the pseudo-1D system with $N_{\mathrm{occ}} N_y$ occupied bands.

With the obtained Wilson loop of the pseudo-1D system,
we can construct the hybrid Wannier functions as~\cite{Benalcazar2017PRB}
\begin{align}
\ket{\Psi^j_{R_x}} = \frac{1}{\sqrt{N_x}} \sum_{n}\sum_{k_x} \left[ \nu^j_{k_x} \right]^n e^{-i k_x R_x} \ket{n,k_x},
\label{eq:WannierFunctions_pseudo1D}
\end{align}
where $\left[ \nu^j_{k_x} \right]^n$ is the $n$th component of the $j$th eigenstate $\ket{\nu^j_{k_x}}$ for the Wilson loop with the Wannier center $\nu^j_{x}$
for $j=1,\cdots,N_{\mathrm{occ}}N_y$,
and $\ket{n,k_x}$ is the $n$th eigenstates of the pseudo-1D Hamiltonian at $k_x$.
One can calculate the density of the hybrid Wannier functions (\ref{eq:WannierFunctions_pseudo1D}) to extract
the spatially resolved polarization~\cite{Benalcazar2017PRB}. The density is given by
\begin{equation}
\rho^{j}(R_y) =\frac{1}{N_x} \sum_{n, k_x, \alpha} \left| [u^n_{k_x}]^{R_y, \alpha}[\nu^j_{k_x}]^n\right|^2,
\end{equation}
where $[u^n_{k_x}]^{R_y, \alpha}$ is the corresponding component of the $n$th eigenstate of the pseudo-1D Hamiltonian.
Then, we can calculate the polarization along $x$ at each site $R_y$ as~\cite{Benalcazar2017PRB}
\begin{align}
p_x(R_y) = \sum_j \rho^j(R_y) \nu_x^j.
\end{align}
In the calculation, we need to break the mirror symmetries infinitesimally to split the degeneracy of two Wannier centers at $\pm 1/2$
so that they gives opposite polarizations localized at opposite edges.
One then can compute
the edge polarization by summing $p_x(R_y)$ over a half of the system along $y$,
i.e.
\begin{eqnarray}\label{edgePEq}
p_x^{\mathrm{edge~}-y}&=&\sum_{R_y=1}^{N_y/2}p_x(R_y) \\ \nonumber
&=&-p_x^{\mathrm{edge~}+y}=-\sum_{R_y=N_y/2+1}^{N_y}p_x(R_y).
\end{eqnarray}
It is quantized to $0$ or $0.5$ up to an integer. One can similarly calculate the
edge polarization along $y$.

\paragraph{Edge polarization and Wannier band topology}
Because of the mirror symmetry, the nontrivial Wannier bands can have mid-gap edge modes at $0$ or $\pm 1/2$.
However,
only the edge modes at $1/2$ are responsible for nontrivial edge polarizations.
Specifically, if the Wannier bands $\nu_x$ under open boundaries perpendicular to $y$ have edge modes at $\pm 1/2$,
these modes give rise to the edge-localized polarizations along $x$.
Thus, the winding number in Eq.~(\ref{eq:WindFormula}) calculated based on bulk wave functions can characterize
the change of the corresponding edge polarization
due to bulk gap closing or Wannier band gap closing.
However, the edge polarization will also change when the edge energy gap closes and reopens.
The edge energy gap closure is usually associated with the change of the quadrupole moment.
If the correspondence is true, then one can take into account the effects of the edge energy gap closure
by defining the change of edge polarization (a $\mathbb{Z}_2$ quantity) with respect to some system parameter $\gamma$ as~\cite{Yang2020PRR}
\begin{eqnarray}
p_x^{\mathrm{edge}}\left(\gamma_1\right)-p_x^{\mathrm{edge}}\left(\gamma_0\right)
=&&e \left\{ \left [W_{\nu_x}^{\epsilon=\pi}(\gamma_1)-W_{\nu_x}^{\epsilon=\pi}(\gamma_0) \right. \right. \nonumber \\
&&\left.\left. -\Delta N_{q,x} \right]/2 \mod 1 \right\}.
\label{winding}
\end{eqnarray}
Here one uses $\Delta N_{q,x}$ to count the number of times that the change of the quadrupole moment
occurs from the $y$-normal edge energy gap closing,
as one varies the system parameter $\gamma$ from $\gamma_0$ to $\gamma_1$.
Due to the fact that the right sides only depend on the bulk property,
one may conclude that the bulk topolgy determines the edge polarization.
In a system with mirror symmetries, one can also evaluate
$\Delta N_{q,x}$ by counting the number of times of the change in a parity
(the eigenvalue of the reflection symmetry operator along $x$)
at the high-symmetric points $k_x=0$ or $k_x=\pi$ for a state localized at a $y$-normal boundary.
If the system is in a trivial phase with $p_x^{\mathrm{edge}}(\gamma_0)=0$
at $\gamma_0$, the formula in Eq.~(\ref{winding}) can be reduced to
\begin{equation}\label{eq:windingR}
p_x^{\mathrm{edge}}(\gamma_1)
=e \left \{ \left[W_{\nu_x}^{\epsilon=\pi}(\gamma_1)
-\Delta N_{q,x} \right]/2 \mod 1 \right \}
\end{equation}
by choosing a gauge such that $W_{\nu_x}^{\epsilon=\pi}(\gamma_0)=0$ in the topologically
trivial phase. It has been shown that
the edge polarization calculated based on
the formula (\ref{eq:windingR}) coincides with the Wannier spectrum and the
edge polarization computed using the hybrid Wannier functions in a cylinder geometry~\cite{Yang2020PRR}.

\begin{figure*}[t]
	\includegraphics[width=0.6\linewidth]{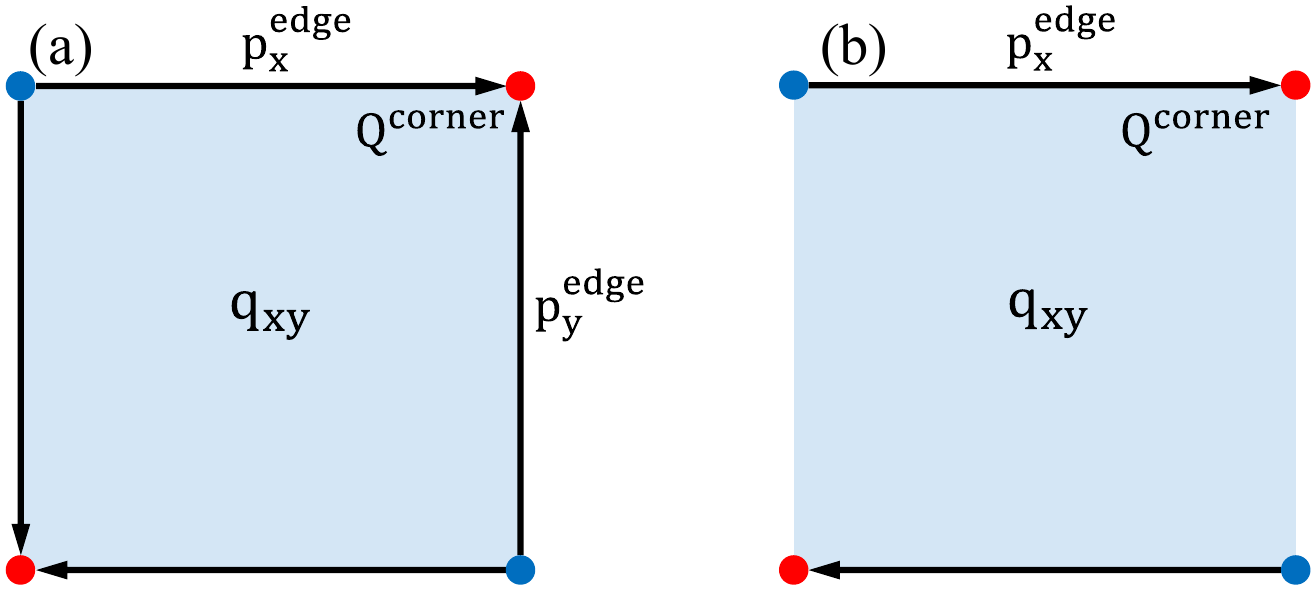}
	\caption{Schematics of edge polarizations and corner charges.
		(a) Edge dipole moments exist at all boundaries in
		type-I QTI and (b) they exist only
		at the boundaries perpendicular to $y$ in type-II QTIs. Corner
		charges $Q^{\mathrm{corner}}=\pm e/2$ (marked by different colors) appear in both phases.
	From Ref.~\cite{Yang2020PRR}.}
	\label{fig:type2QTI}
\end{figure*}

\subsubsection{Type-II quadrupole topological insulators}
The authors in Ref.~\cite{Benalcazar2017Science} derive a general relation among the quadrupole moment $q_{xy}$, corner charge $Q^{\mathrm{corner~}\pm x,\pm y}$
and edge polarizations $p_y^{\mathrm{edge~} \pm x}$ or $p_x^{\mathrm{edge~} \pm y}$
for a 2D square classical system by the multipole expansion of an electric potential in classical electromagnetic theory.
The relation reads
\begin{equation}
Q^{\mathrm{corner~}+x,+y}=p_y^{\mathrm{edge~}+x}+p_x^{\mathrm{edge~}+y}-q_{xy}.
\end{equation}
For the BBH model, the relation reduces to
\begin{equation}
|q_{xy}|=|p_{x}^{\mathrm{edge~} \pm y}|=|p_{y}^{\mathrm{edge~}\pm x}|=|Q^{\mathrm{corner~}\pm x,\pm y}|
\end{equation}
 (see Fig.~\ref{fig:type2QTI}). In a quantum system, the relation is enforced by the correspondence that the
Wannier band and the edge energy spectrum close their gaps simultaneously~\cite{Khalaf2021PRR}. The correspondence can be
illustrated by transforming the Bloch Hamiltonian of the BBH model to the following form~\cite{Yang2020PRR}
\begin{equation}\label{eq:BBH2}
	H_{\textrm{BBH}}^\prime(\bm{k}) =\sigma_0\otimes H_{\textrm{SSH}}(k_y,\gamma_y)+H_{\textrm{SSH}}(k_x,\gamma_x)\otimes \sigma_3,
\end{equation}
where $H_{\textrm{SSH}}(k,t)=(t+\cos k)\sigma_1+\sin k\sigma_2$ is the SSH Hamiltonian
and $H_{\textrm{BBH}}^\prime=U H_{\textrm{BBH}} U^\dagger$ with
\begin{equation}
U=\left(
\begin{array}{cccc}
	0 & i & 0 & 0 \\
	0 & 0 & -i & 0 \\
	0 & 0 & 0 & i \\
	i & 0 & 0 & 0\\
\end{array}
\right).
\end{equation}
In this model, the mirror symmetries are represented by
$U_{M_x}=\sigma_1 \otimes\sigma_0$ and $U_{M_y}=\sigma_3\otimes\sigma_1$.
At the high-symmetric momenta for $k_x$ (similarly for $k_y$), one can write the Hamiltonian as the following direct sum form
\begin{equation}
	[H_{\textrm{BBH}}^\prime (k_x=k^*,k_y)]_\beta= H_1\oplus H_2
\end{equation}
relative to a basis $\beta$ consisting of eigenvectors of $U_{M_x}$.
For example, one can choose $\beta=\{|\uparrow_1\uparrow_3\rangle,|\uparrow_1\downarrow_3\rangle,|\downarrow_1\uparrow_3\rangle,|\downarrow_1\downarrow_3\rangle\}$
as the basis, where $|\uparrow_\lambda\rangle$ and $|\downarrow_\lambda\rangle$ are eigenvectors of
$\sigma_\lambda$ (with $\lambda=1,2,3$) corresponding to eigenvalues $1$ and $-1$, respectively.
In this basis,
\begin{eqnarray}
H_1 &= &H_{\textrm{SSH}}(k_y,\gamma_y)+(\gamma_x+\cos k^*)\sigma_3, \\
H_2 &=& H_{\textrm{SSH}}(k_y,\gamma_y)-(\gamma_x+\cos k^*)\sigma_3.
\end{eqnarray}
Suppose that $|\gamma_y|<1$ so that the SSH model is in a topologically nontrivial phase.
We now decrease $\gamma_x$ across $1$. It is clear to see that the second term in both $H_1$ and $H_2$
vanishes at $\gamma_x=1$ and $k^*=\pi$ so that both $H_1$ and $H_2$ become the SSH model.
If we impose open boundaries along $y$, then there exist two pairs of zero-energy edge states, implying
the edge energy gap closing at the $y$-normal boundary. At the same time, there appear the
Berry phase of $\pi$ for the Hamiltonian $H_1$ and $H_2$, which is equivalent to saying that
the Wannier bands $\nu_y(k_x)$ close their gap at $\nu_y=1/2$ when $k_x=\pi$.

However, the correspondence between the Wannier band and the edge energy spectrum
 may break down for systems with either particle-hole or
chiral symmetry~\cite{Yang2020PRR} or systems without these symmetries~\cite{Wieder2018Fragile}.
Such a breakdown leads to a distinct type of quadrupole topological insulator (dubbed type-II)
with $q_{xy}=|Q^{\mathrm{corner~}\pm x,\pm y}|=|p_x^{\mathrm{edge~}\pm y}|=e/2$,
but $p_y^{\mathrm{edge~}\pm x}=0$,
violating the classical relation (see Fig.~\ref{fig:type2QTI})~\cite{Yang2020PRR}.
A similar discrepancy between the Wannier bands and edge energy spectrum has
also been observed in a generalized BBH model~\cite{Li2020PRB}.

We now follow Ref.~\cite{Yang2020PRR} to show how the breakdown occurs.
Consider a four-band Hamiltonian
\begin{equation}
\hat{H}_{II}= \sum_{\bm{k},\alpha,\beta} \hat{c}^{\dagger}_{\alpha}(\bm{k}) \left[H_{II}(\bm{k})\right]_{\alpha\beta} \hat{c}_{\beta}(\bm{k}),
\end{equation}
where $\hat{c}^{\dagger}_{\alpha}(\bm{k})$ with $\alpha=1,2,3,4$ represents the creation operator for the $\alpha$th component at momentum $\bm{k}$ in the Brillouin zone.
The Bloch Hamiltonian (the Hamiltonian $H_{II}$ in Ref.~\cite{Yang2020PRR}) is
\begin{eqnarray}
	H_{II}(\bm{k})=H_L(\bm{k})
	+\sigma_2\otimes H_{\textrm{SSH}}^\prime(k_x),
	\label{SimHam}
\end{eqnarray}
where
\begin{eqnarray}
	H_L(\bm{k})=&&\left(\gamma+\cos k_y\right)\sigma_1\otimes\sigma_0+b_2\cos 2k_y\sigma_3\otimes\sigma_2 \nonumber \\
	&&+\left( \sin k_y+b_2\sin 2k_y\right)\sigma_3\otimes\sigma_1,
\end{eqnarray}
and $H_{\textrm{SSH}}^\prime(k_x)=(\gamma+1-\gamma_c+\cos k_x)\sigma_2 + \sin k_x \sigma_3$.
Here,  $\gamma$, $\gamma_c$ and $b_2$ are system parameters with $b_2$ specifying the strength of the next-nearest-neighbor
intercell hopping.
This Hamiltonian respects the reflection symmetries $U_{M_x}=\sigma_1\otimes\sigma_3$ and
$U_{M_y}=\sigma_1\otimes\sigma_1$
and the particle-hole symmetry $\mathcal{P}=\sigma_3\otimes\sigma_0 \kappa$.

Due to the mirror symmetry, we can write the Hamiltonian at $k_x=\pi$ as
\begin{equation}
	[H_{II}]_\beta=\left(
	\begin{array}{cc}
		H_{\textrm{NNN}}(k_y)+t_0\sigma_1 & 0_{2\times2} \\
		0_{2\times2} & \sigma_3 H_{\textrm{NNN}}(-k_y)\sigma_3-t_0\sigma_1 \\
	\end{array}
	\right)
\end{equation}
under a basis $\beta=\{|\uparrow_1\uparrow_3\rangle, |\downarrow_1\downarrow_3\rangle,
|\uparrow_1\downarrow_3\rangle, |\downarrow_1\uparrow_3\rangle\}$.
Here, $t_0=\gamma-\gamma_c$, and
\begin{eqnarray}
	H_{\textrm{NNN}}(k_y)&=&\left(\gamma+\cos k_y\right)\sigma_3+\left( \sin k_y+b_2\sin 2k_y \right)\sigma_1 \nonumber \\
	&&+b_2\cos 2k_y\sigma_2.
	\label{HNN}
\end{eqnarray}
When $\gamma=\gamma_c$, $H_{\textrm{NNN}}(k_y)$ exhibits zero-energy edge modes under open boundary conditions,
indicating that $H_{II}$ closes its edge energy gap at $y$-normal boundaries.
However, the corresponding Berry phase is not equal to $\pi$, due to the existence of $b_2$.
As a result, the correspondence breaks down. The breakdown not only leads to the type-II quadrupole insulator,
but also a new insulator with only edge polarizations but without corner modes~\cite{Yang2020PRR}.

\begin{figure*}[t]
	\centering
	\includegraphics[width=\linewidth]{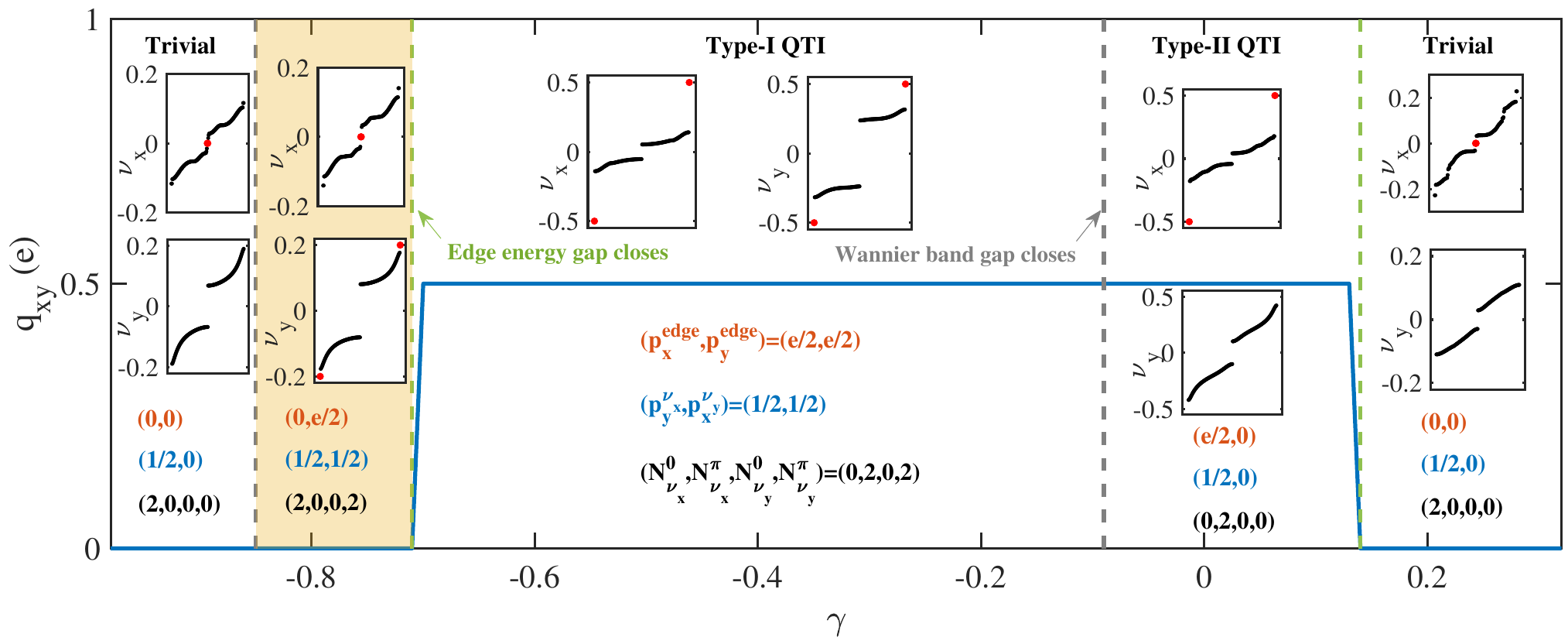}
	\caption{(a) The phase diagram of the Hamiltonian (\ref{HamSim2}) versus a system parameter $\gamma$.
		In the phase diagram, we observe the topologically trivial insulator, the type-I quadrupole insulator, the type-II quadrupole insulator
		with only edge polarizations along $x$, and a nontrivial phase with nonzero edge polarizations but without the quadrupole moment
		and zero-energy corner modes in the orange region.
		Distinct phases are identified by the quadrupole moment (blue line), edge polarizations
		$(p_x^{\mathrm{edge}},p_y^{\mathrm{edge}})$ illustrated by the Wannier spectrum $\nu_x$ ($\nu_y$) obtained in a
		cylinder geometry with periodic boundaries along $x$ ($y$) and open ones along $y$ ($x$).
		In the Wannier spectrum, the isolated edge Wannier centers are highlighted by red solid circles.
		In addition, the Wannier-sector polarization
		$(p_y^{\nu_x},p_x^{\nu_y})$ and the number of edge modes in the Wannier Hamiltonian $N_\nu \equiv (N_{\nu_x}^0, N_{\nu_x}^{\pi}, N_{\nu_y}^0, N_{\nu_y}^{\pi})$
		are displayed.
		The vertical dashed lines mark out the critical points where the closure of the edge energy gap or the Wannier band gap happens.
		Reproduced from Ref.~\cite{Yang2020PRR}.
	}
	\label{fig:type2QTI_phase}
\end{figure*}

Figure~\ref{fig:type2QTI_phase} shows the phase diagram (reproduced from Ref.~\cite{Yang2020PRR}) of the model with the following Bloch Hamiltonian
with respect to the parameter $\gamma$
\begin{equation}
	H_{III}(\bm{k})=H_L(\bm{k})-\sigma_3\otimes H_{\textrm{SSH}}^\prime(k_x).
	\label{HamSim2}
\end{equation}
When $\gamma$ decreases cross $\gamma=0.13$, while the $y$-normal edge energy gap vanishes,
the Wannier band gap remains open at $\nu_y = ±1/2$, resulting in a type-II quadrupole insulator
with nonzero quadrupole moment and nonzero edge polarizations along only one pairs of edges.
The specific configurations of edge polarizations can be seen from
the Wannier spectra $\nu_x$ and $\nu_y$ displayed in the insets:
In the type-II phase, only $\nu_x=\pm1/2$ occurs but not for $\nu_y$, while
in the type-I phase, both $\nu_x$ and $\nu_y$ exhibit isolated eigenvalues at $\nu_x=\pm 1/2$.
In the phase diagram, one can also observe a nontrivial phase with only nonzero $p_y^{\textrm{edge}}$
but without the quadrupole moment and corner modes, which arises from the Wannier gap closing alone from the trivial phase.
Such phases are protected by the topology of Wannier bands rather than energy bands.

\subsection{Chiral-symmetric higher-order topological phases}
We have seen that chiral symmetry can protect the quantized quadrupole moment of quadrupole insulators.
In the following, we will review a systematic method to construct HOTPs with
chiral symmetry and their topological characterizations.

\subsubsection{Construction of higher-order topological phases with chiral symmetry}\label{sec:chiral_symmetry}
Here we follow Ref.~\cite{Okugawa2019PRB} to present how to systematically construct
the lattice models for HOTPs protected only by chiral symmetry.
We start from the following 2D Bloch Hamiltonian in momentum space,
\begin{equation}
H_C(\bm{k})=H_x(k_x)\otimes \Pi _y + 1_x \otimes H_y(k_y),
\end{equation}
where $H_{i}(k_{i})$ with $i=x,y$ are two 1D Hamiltonian with chiral symmetry
represented by $\Pi _{i}$ such that $\{H_{i}, \Pi _{i}\}=0$.
Here, $\Pi _{i}^2=1_i$, and $1_i$ is the identity matrix with the same size as $H_{i}(k_i)$ for $i=x,y$.
The Hamiltonian $H_C(\bm{k})$ thus has chiral symmetry $\Pi = \Pi _x\otimes \Pi _y$.
Due to chiral symmetry, the energy spectrum is symmetric about zero energy.
When $H(\bm{k})$ is gapped at zero energy, it can be characterized by a $\mathbb{Z}$-valued topological invariant given by
$\nu _{2D}=w_x w_y$,
where $w_{i=x,y}$ are 1D winding numbers for $H_{i}(k_i)$ protected by chiral symmetry.
The winding numbers characterize the first-order topology of the 1D chiral symmetric Hamiltonian $H_{i=x,y}$.
The topological invariant $\nu _{2D}$ can change when the energy band gap closes at zero energy in the bulk or the edges~\cite{Okugawa2019PRB}.

The nonzero $\nu _{2D}$ indicates the existence of the zero-energy corner states
when we take the open boundary conditions along both the $x$ and $y$ directions.
One can use real-space chiral symmetry operator $\Pi _y^{\mathrm{OBC}}$ and
identity operator $1_x^{\mathrm{OBC}}$ to construct
a real-space Hamiltonian (with gapped bulk and edge energy spectra) under open boundary conditions,
\begin{equation}
{H}_C^{\mathrm{OBC}}=H_x^{\mathrm{OBC}} \otimes \Pi _y^{\mathrm{OBC}}
	+ 1_x^{\mathrm{OBC}} \otimes H_y^{\mathrm{OBC}}.
\end{equation}
The zero-energy corner modes arise when both $H_x(k_x)$ and $H_y(k_y)$ are in
a topological phase with nonzero 1D winding number.
Specifically, due to the nontrivial property of $H_i^{\mathrm{OBC}}$ ($i=x,y$), there exists
a zero-energy edge state $|\psi_i^{\mathrm{zero}}\rangle$ spatically residing at one edge.
Then, it is easy to see that $|\psi^{\mathrm{corner}}\rangle=|\psi_x^{\mathrm{zero}}\rangle \otimes |\psi_y^{\mathrm{zero}}\rangle$
is a zero-energy eigenstate of the 2D Hamiltonian ${H}^{\mathrm{OBC}}$, which is spatically
localized at a corner.
Because $H_{i=x,y}^{\mathrm{OBC}}$ have $|w_i|$ zero-energy modes,
we will generally have $|\nu _{2D}|=|w_x w_y|$ topological corner modes at each corner.

The above construction can be generalized to construct 3D Dirac semimetals with chiral symmetry
by stacking the 2D Hamiltonian along the $z$ direction.
The 3D bulk Hamiltonian can be written as
\begin{equation}
H_{\textrm{C,3D}}(\bm{k})=H_x(k_x, k_z)\otimes \Pi _y+ 1_x \otimes H_y(k_y, k_z).
\end{equation}
This Hamiltonian respects the same chiral symmetry as the 2D case.
If we regard $k_z$ as a parameter of the 2D Hamiltonian, then we can have the topological phase transition
characterized by the change of $\nu _{2D}(k_z)$,
which corresponds to gap-closing points in 3D momentum space.
Therefore, we can realize a 3D chiral-symmetric second-order topological semimetal which has gapless points and hinge states at zero energy~\cite{Okugawa2019PRB}.

\subsubsection{Higher-order topological Aubry-Andr\'{e}-Harper models with chiral symmetry}
Now we consider a 2D Aubry-Andr\'{e}-Harper (AAH) model with chiral symmetry constructed from 1D off-diagonal AAH models~\cite{Zeng2020PRB,Zilberberg2020PRR}.
The 1D off-diagonal AAH model only contains nearest-neighbor hopping with real-space Hamiltonian
\begin{equation}
\hat{H}_{1D}=\sum_{j} \left( t_j \hat{c}^{\dagger}_{j} \hat{c}_{j+1} + \mathrm{H.c.} \right),
\end{equation}
where $\hat{c}^{\dagger}_{j}$ denotes the creation operator for an electron at the site $j$, and
the hopping amplitude is modulated as
\begin{equation}
t_j=t \left[1+\lambda \cos (2\pi \alpha  j + \phi) \right].
\end{equation}
where $\lambda$ is the amplitude of the modulation whose period is determined by $\alpha$.
The modulation is commensurate when the parameters $\alpha=p/q$
with $p$ and $q$ being mutually prime positive integers.
When $q$ is an even number, the system respects a sublattice symmetry as chiral symmetry and can have zero-energy edge modes
protected by chiral symmetry~\cite{Ganeshan2013PRL}.
In momentum space, the chiral symmetry operator is represented by $\Pi=I_{\frac{q}{2}}\otimes \sigma_z$,
and the Bloch Hamiltonian $H_{1D}(k)$ is a $q \times q$ matrix given by
\begin{equation}
[H_{1D}(k)]_{m,n}=\delta_{m,n+1}t_n e^{-ik/q} + \delta_{m,n-1}t_{n-1} e^{ik/q},
\end{equation}
where $m,n=1,\cdots,q$ and the momentum $k \in [0,2\pi)$.
For the 1D off-diagonal AAH model with $q>2$, in addition to topological zero-energy edge modes,
we can obtain nonzero-energy chiral edge modes
in the energy spectrum with respect to the parameter $\phi$ corresponding to Chern bands~\cite{Ganeshan2013PRL}.

According to the previously introduced construction method,
the momentum-space Hamiltonian of a 2D lattice model with chiral symmetry
constructed from the 1D off-diagonal AAH models in the $x$ and $y$ directions can be written as~\cite{Zeng2020PRB}
\begin{equation}
H_{\textrm{AAH}}(\bm{k})=H_{\textrm{AAH},x} (k_x) \otimes \Pi_y + I_x \otimes H_{\textrm{AAH},y} (k_y),
\end{equation}
where $H_{\textrm{AAH},i}(k_i)$ ($i=x,y$) is the Hamiltonian of the 1D off-diagonal AAH model with $k_i \in [0,2\pi)$,
and $I_x$ is an identity matrix of size $q_x$.
Clearly, the 2D Hamiltonian has a chiral symmetry represented by $\Pi_{2D}=\Pi_x \otimes \Pi_y$
where $\Pi_{i=x,y}=I_{{q_i}/{2}}\otimes \sigma_z$.

In real space, the above 2D off-diagonal AAH model is defined on a square lattice~\cite{Zeng2020PRB},
\begin{align}\label{eq:H_2DAAH}
\hat{H}_{\textrm{AAH}}=\sum_{j_x,j_y} & \left[t_{j_x,j_y}^{x} \hat{c}^{\dagger}_{j_x,j_y} \hat{c}_{j_x+1,j_y} \right. \nonumber \\
  & \left. + t_{j_x,j_y}^{y} \hat{c}^{\dagger}_{j_x,j_y+1} \hat{c}_{j_x,j_y} \right] + \mathrm{H.c.},
\end{align}
where $\hat{c}^{\dagger}_{j_x,j_y}$ represents the creation operator for an electron at the site $(j_x,j_y)$,
and the hopping amplitudes along the $x$ and $y$ directions are modulated respectively as
\begin{align}\label{eq:H_2DAAH_hopping}
  t_{j_x,j_y}^{x} &= (-1)^{j_y} t_x [1+\lambda_x \cos (2\pi \alpha_x j_x + \phi_x)], \nonumber \\
  t_{j_x,j_y}^{y} &= t_y [1+\lambda_y \cos (2\pi \alpha_y j_y + \phi_y)],
\end{align}
where $\lambda_x$ and $\lambda_y$ are the amplitudes of the modulations.
The modulation is periodic when the parameters $\alpha_x=p_x/q_x$ and $\alpha_y=p_y/q_y$
with $p_x$ and $q_x$ ($p_y$ and $q_y$) being mutually prime positive integers,
whereas it becomes quasiperiodic when $\alpha_x$ or $\alpha_y$ is a irrational number.
When $\alpha_x=\alpha_y=1/2$, the Hamiltonian can be seen as a 2D lattice model
constructed from two 1D SSH models, and can be mapped to the BBH model.
As we have seen in the previous section, the 2D off-diagonal AAH model can have zero-energy corner modes
if both the 1D off-diagonal Hamiltonian $H_{\textrm{AAH},i=x,y}$ are topological.
In the following, we will follow Ref.~\cite{Zeng2020PRB} to review the results with $\phi_x=\phi_y=\phi$ for the commensurate case with $(\alpha_x,\alpha_y)=(1/2,1/4)$
and $(\alpha_x,\alpha_y)=(1/4,1/4)$ and the incommensurate case with $\alpha_x=\alpha_y=(\sqrt{5}-1)/2$.

\paragraph{$(\alpha_x,\alpha_y)=(1/2,1/4)$}

\begin{figure*}[t]
  \includegraphics[width=\linewidth]{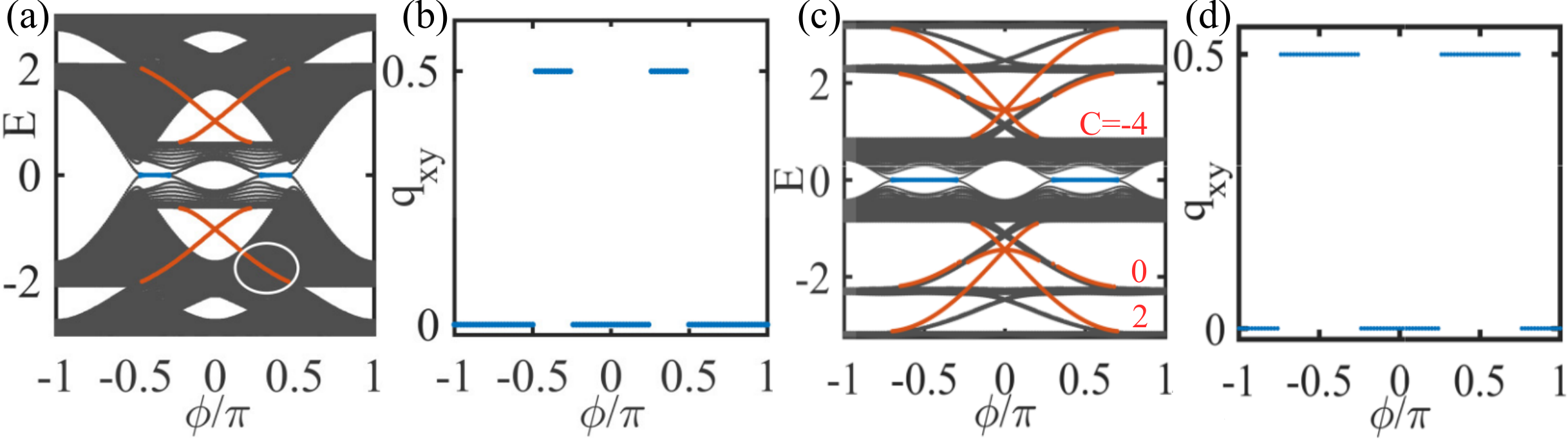}
  \caption{(a) The energy spectrum of the 2D AAH model in Eq.~(\ref{eq:H_2DAAH}) under open boundary
  	conditions along both $x$ and $y$ directions.
  	The zero-energy and nonzero-energy topological corner modes are described by the blue and red lines, respectively.
  	The nonzero-energy corner modes in a continuous bulk band are highlighted by the white circle.
  (b) The quadrupole moment with respect to $\phi$ at half filling.
  In (a) and (b), $\alpha_x=1/2$, and $\alpha_y=1/4$.
  (c) and (d) show the same quantities as (a) and (b) but for $\alpha_x=\alpha_y=1/4$.
  Here $t_x=t_y=1$, $\lambda_x=\lambda_y=0.8$.
  Reproduced from Ref.~\cite{Zeng2020PRB}.}
\label{fig:Qibo}
\end{figure*}

This is the simplest case beyond the BBH model.
Figure~\ref{fig:Qibo} illustrates the energy spectrum of the lattice model in a geometry with open boundaries along
both directions with respect to $\phi$.
We clearly see the appearance of four zero-energy states in the gap highlighted by the blue lines
when $\phi \in (-\frac{\pi}{2}, -\frac{\pi}{4})$ and $(\frac{\pi}{4}, \frac{\pi}{2})$.
These states are spatially localized at the four corners of the 2D lattice~\cite{Zeng2020PRB}.
As we have demonstrated in the previous section,
the presence of zero-energy corner states is resulted from the topology of the 1D off-diagonal AAH models.
For $H_{\textrm{AAH},x}(k_x)$ with $\alpha_x=1/2$ and  $H_{\textrm{AAH},y}(k_y)$ with $\alpha_y=1/4$,
zero-energy edge states appear when $\phi \in (-\frac{\pi}{2}, \frac{\pi}{2})$ and
$\phi \in (-\frac{\pi}{2},-\frac{\pi}{4}) \cup (\frac{\pi}{4}, \frac{\pi}{2})$, respectively.
The regions' intersection thus corresponds to the topological
parameter region observed in Fig.~\ref{fig:Qibo}(a),
corresponding to the topological invariant $\nu_{2D}=1$ with $w_x=w_y=1$.

Since the system respects chiral symmetry, one can also use the quadrupole moment to characterize
its topology given that there are four zero-energy states (one at each corner).
Note that the quadrupole moment is a $\mathbb{Z}_2$ topological invariant (similar to the Berry phase),
and thus can only distinguish between the phases with even (including zero) number of corner
modes at each corner and the ones with old number.
In Fig.~\ref{fig:Qibo}(b), we see that the quadrupole moment is equal to $1/2$ in the topological
region and vanishes in the trivial one. It changes through the $y$-normal edge gap closing at $\phi=\pm \pi/2$
and the $x$-normal edge gap closing at  $\phi=\pm \pi/4$.

If one views $\phi$ as the momentum $k_z$ along $z$, then one obtains a 3D AAH model,
which exhibits a 3D higher-order Dirac semimetallic phase. In the phase, there appear
Dirac points at both $x$-normal and $y$-normal surfaces.
As a result, zero-energy hinge arcs arise in this semimetal.
Apart from the zero-energy corner modes, the authors also find nonzero-energy corner modes in the energy spectrum as highlighted as red lines
in Fig.~\ref{fig:Qibo}(a).
These modes are chiral with respect to $\phi$, which can be seen as chiral hinge modes for the 3D system with
$\phi$ acting as $k_z$.
They arise from an effective Chern band localized on the $x$-normal surfaces, and each boundary band localized
at one boundary contributes a Chern number of $-1$~\cite{Zeng2020PRB}. They can also be characterized by the Chern number of the
Wannier bands calculated from the lowest two degenerate occupied bands. This Wannier-sector Chern number
can be interpreted as the spectral flow of the nested Wilson loop as in the case of 3D higher-order topological
insulator with chiral hinge states~\cite{Schindler2018SA}. The topological invariant is equivalent to the Chern number of
boundary bands.

These nonzero-energy corner modes can exist in the continuous bulk spectrum as highlighted by the
white circle in Fig.~\ref{fig:Qibo}(a). Such states are known as the bound states in the continuum~\cite{YangBJ2013NC,XiaoYX2017PRL,ChenZG2019PRB}.
The higher-order zero-energy bound states in the continuum have also been found in a 2D HOTI~\cite{Benalcazar2020BIC}.

\paragraph{$(\alpha_x,\alpha_y)=(1/4,1/4)$}

In this case, the zero-energy corner modes appear when $\phi \in (-\frac{3\pi}{4}, -\frac{\pi}{4})$ and
$(\frac{\pi}{4}, \frac{3\pi}{4})$, which correspond to the regime with zero-energy edge modes
in the 1D off-diagonal AAH model as shown in Fig.~\ref{fig:Qibo}(c). Similarly, as in the previous case,
the corner modes are characterized by the quadrupole moment [see Fig.~\ref{fig:Qibo}(d)], which arises from the bulk energy gap
closing, rather than the edge energy gap closing.

Similar to the previous case, the authors also find nonzero-energy chiral modes. However, different from
the previous case, there appear a group of chiral modes in the bulk gaps which are boundary bands
consisting of states localized at the 1D edges. These modes are characterized by the Chern number calculated
in the $(k_y,\phi)$ or $(k_x,\phi)$ space for each $k_x$ or $k_y$.
In addition, there exist chiral modes (highlighted by red lines) connecting different boundary bands.
These states are localized at the corners.
Compared to the previous case, the boundary states do not have a well defined Chern number
as these bands are chiral and not well separated.
Viewing $\phi$ as $k_z$ leads to a 3D model with Dirac points at zero energy in the bulk
supporting the coexistence of zero-energy hinge modes, chiral hinge modes and chiral surface modes.

\paragraph{$\alpha_x=\alpha_y=(\sqrt{5}-1)/2$}
The irrational $\alpha_x$ and $\alpha_y$ lead to a 2D quasicrystal model
where the nonzero-energy chiral modes localized at the edges and corners still exist~\cite{Zeng2020PRB}.
Similar to the commensurate AAH models, the nonzero-energy corner modes can survive across the continuous bulk spectrum.

\subsubsection{The $\mathbb{Z}$-valued multipole invariant protected by chiral symmetry}
It is well known that chiral symmetry can protect the first-order topological phases with $\mathbb{Z}$-valued
topological invariants in odd dimensions according to the tenfold classification~\cite{Chiu2016RMP}.
For example, 1D TIs with chiral symmetry can be characterized by a winding number
which gives the number of zero-energy edge modes.
In fact, from the previous sections, we see that the systematic construction of a 2D system with chiral symmetry can give
rise to a higher-order phase that is $\mathbb{Z}$ classified. However, the quadrupole moment
is a $\mathbb{Z}_2$ topological invariant.
The authors in Ref.~\cite{Benalcazar2022PRL} propose a $\mathbb{Z}$-valued multipole topological invariant to
characterize the chiral-symmetric HOTIs.
In the following, we will review the integer-valued topological invariant, which extends the $\mathbb{Z}_2$
classification for chiral-symmetric quadrupole TIs.

Let us consider a single-particle Hamiltonian $H$ respecting chiral symmetry $\Pi$ (sublattice symmetry), that is, $\Pi H \Pi = - H$.
Since $\Pi^2=1$, there are two eigenvalues of $\pm 1$ for $\Pi$ corresponding to eigenvectors $\{|\phi_n^{\pm}\rangle \}$,
respectively. The chiral symmetry operator is then represented by $\sigma_z\otimes I_{N_S}$ in the basis
$\{|\phi_n^{+}\rangle \} \cup \{|\phi_n^{+}\rangle \}$.
Under this basis, the Hamiltonian $H$ has the following form
\begin{equation}\label{eq:ChiralHamiltonian}
H=\begin{pmatrix}
0 & h \\
h^\dagger & 0
\end{pmatrix},
\end{equation}
where $h$ and $h^\dagger$ describe couplings between the states in the two subspaces
with opposite chiral charges (eigenvalues of the chiral operator).

The chiral-symmetric Hamiltonian (\ref{eq:ChiralHamiltonian}) can be easily solved by
the singular value decomposition (SVD) of the matrix $h$:
\begin{equation}\label{eq:SVD}
h=U_+ \Sigma U_-^\dagger,
\end{equation}
where $U_{S}$ for $S=\pm$ is an $N_S\times N_S$ unitary matrix written as
$U_S=(|\psi^S_1\rangle, |\psi^S_2\rangle, \cdots, |\psi^S_{N_S}\rangle)$,
and $\Sigma$ is a diagonal matrix consisting of singular values $\{\epsilon_n\}$.
Here, $\{\psi^+_n\}$ and $\{\psi^-_n\}$ are normalized vectors defined in the
$+$ and $-$ subspaces, respectively.
The eigenstates of the Hamiltonian can be constructed  as
\begin{equation}
\ket{\psi_n}=\frac{1}{\sqrt{2}}\left(\langle\psi^+_n|, \langle \psi^-_n| \right)^\dagger,
\end{equation}
which is the eigenstate of $H$ with the eigenvalue $\epsilon_n$.
Because of the chiral symmetry, for every eigenstate $\ket{\psi_n}$,
we have $\ket{\psi_n'}=\Pi \ket{\psi_n}=\frac{1}{\sqrt{2}}(\langle \psi^+_n|, -\langle \psi^-_n|)^\dagger$
as the eigenstate with the opposite eigenvalue $-\epsilon_n$.

For a 1D topological insulator with the bulk Hamiltonian $H(k_x)$, we can define a
unitary matrix $q=U_+ U_-^\dagger$ from the SVD (\ref{eq:SVD}) for each $k_x$
in the Brillouin zone. The winding number is thus evaluated by
\begin{equation}
N_x=\frac{\mathrm{i}}{2\pi} \int_0^{2\pi} d k_x \Tr \left[ q(k_x)^\dagger \partial_{k_x} q(k_x) \right].
\end{equation}
This topological invariant can also be defined in an equivalent real-space form~\cite{Benalcazar2022PRL},
\begin{equation}
N_x =\frac{1}{2\pi \mathrm{i}}\Tr \mathrm{log}\left(\bar{P}_x^+ \bar{P}_x^{-\dagger}\right),
\end{equation}
where $\bar{P}_x^S = U_S^\dagger P_x^S U_S$ and $P_x^\mathcal{S}$ is the sublattice dipole operator
associated with the Resta formula for polarization,
\begin{equation}
P_x^S=\sum_{R,\alpha \in \mathcal{S}} \ket{R,\alpha} \exp(-\mathrm{i} 2\pi R/L) \bra{R,\alpha},
\end{equation}
where the 1D system has $L$ unit cells, and $\ket{R,\alpha}$ denotes the orbital $\alpha$ of a unit cell $R$.

The 1D real-space winding number can be generalized to define the multipole chiral numbers for 2D and
3D systems with chiral symmetry. For example,
for a 2D system with chiral symmetry, one can define a quadrupole chiral invariant by
\begin{equation}\label{eq:Nxy}
N_{xy}=\frac{1}{2\pi \mathrm{i}} \Tr \mathrm{log} \left(\bar{Q}_{xy}^A \bar{Q}_{xy}^{B\dagger} \right),
\end{equation}
where $\bar{Q}_{xy}^S=U_S^\dagger Q_{xy}^S U_S$ for $S=\pm$ with the $U_S$ given by the SVD in Eq.~(\ref{eq:SVD}).
Here, $Q_{xy}^S$ for $S=\pm$ are the quadrupole operators restricted to one sublattice:
\begin{equation}\label{eq:Nxy_operator}
Q^S_{xy}=\sum_{{\bf R},\alpha \in S} \ket{{\bf R},\alpha}\exp\left(-\mathrm{i} \frac{2\pi xy}{L_x L_y}\right) \bra{{\bf R},\alpha},
\end{equation}
where $\ket{{\bf R},\alpha}$ denotes the orbital $\alpha$ of a unit cell ${\bf R}=(x,y)$ in a 2D system of size $L_x \times L_y$.
This operator is related to the definition of the $\mathbb{Z}_2$ quadrupole moment in Sec.~\ref{sec:Qxy}.
The octupole chiral invariant can be defined for 3D chiral-symmetric systems analogously~\cite{Benalcazar2022PRL}.
As a topological invariant, these multipole chiral numbers are proved to be integer-valued~\cite{Benalcazar2022PRL}.
Note that the real-space formula also works for systems without translational symmetry, similar
to the traditional quadrupole moment.

Reference~\cite{Benalcazar2022PRL} illustrates that the quadrupole chiral invariant (\ref{eq:Nxy})
can characterize the different phases of 2D second-order TIs with chiral symmetry
and indicates the number of topologically protected corner states therein.
For a model supporting quadrupole insulators,
if only the nearest-neighbor hoppings are nonzero,
we have a topological phase with the quadrupole chiral invariant $N_{xy}=1$ in consistent with the quantized quadrupole moment $q_{xy}=1/2$.
When the long-range hoppings are increased, the system will undergo a bulk gap closing and transition into another nontrivial phase with
$N_{xy} = 4$ but with $q_{xy} = 0$. With open boundaries, this new phase possesses four zero-energy states localized at each corner
within a single sublattice (see Ref.~\cite{Benalcazar2022PRL} for details).
Therefore, the $\mathbb{Z}$-valued quadrupole chiral invariant
can reveal the higher-order topology protected by chiral symmetry beyond the $\mathbb{Z}_2$-valued quadrupole moment invariant.
As shown in Ref.~\cite{Benalcazar2022PRL},
the topological phase transitions between phases of different multipole invariants accompany the gap closing at either the bulk or the boundary.

\subsection{Higher-order topological insulators with chiral hinge states} \label{Subsec:Chiral}
There exist various types of 3D HOTIs protected by crystalline symmetries~\cite{Brouwer2018PRB,Brouwer2019PRX}.
Here, we review a type of 3D HOTIs with chiral hinge modes introduced in Ref.~\cite{Schindler2018SA}.

\begin{figure*}[t]
\begin{center}
\includegraphics[width=\linewidth]{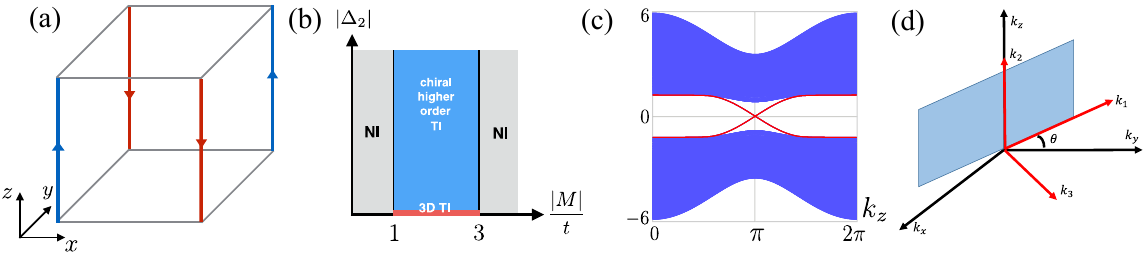}
\caption{(a) Schematic illustration of chiral hinge modes.
	(b) The phase diagram of the model~\eqref{eq:H_chiral_HOTI}, where NI represents the normal insulator.
	When $\Delta_2=0$, the chiral HOTI becomes 3D topological insulator with TRS.
	(c) The energy spectrum of the chiral HOTI model in Eq.~\eqref{eq:H_chiral_HOTI}
	with open boundaries along $x$ and $y$ and periodic boundary along $z$. The chiral hinge states are highlighted
	by the red lines.
	(d) A surface that deviates from the $yz$ plane by an angle $\theta$. $(k_1,k_2,k_3)$ denotes a
	new coordinate system with $k_3$ normal to the surface and $k_1$ and $k_2$ parallel to the surface.
(a)-(c) are reproduced from Ref.~\cite{Schindler2018SA}.}
\label{fig:HOTI1}
\end{center}
\end{figure*}

\subsubsection{Model Hamiltonian}
We consider the following 3D Bloch Hamiltonian~\cite{Schindler2018SA}
\begin{align}\label{eq:H_chiral_HOTI}
H_{\mathrm{c}}(\bm{k}) = &\Bigl(M+t\sum_i \cos k_i\Bigr) \, \tau_z \sigma_0 +\Delta_1\sum_i \sin k_i \, \tau_x \sigma_i \nonumber \\
&+\Delta_2 (\cos k_x - \cos k_y) \, \tau_y \sigma_0,
\end{align}
where $\tau_i$, $\sigma_i$ with $i=x,y,z$, are Pauli matrices describing the orbital and spin degrees of freedom, respectively.
The Hamiltonian corresponds to the tight-binding model on a cubic lattice~\cite{Schindler2018SA}
\begin{align}\label{eq:H_chiral_HOTI_tb}
\hat{H}_{\mathrm{c}} &= \frac{M}{2} \sum_{\bm{r},\alpha} (-1)^\alpha \, \hat{c}^\dagger_{\bm{r},\alpha} \hat{c}_{\bm{r},\alpha} \nonumber \\
&+ \frac{t}{2} \sum_{\bm{r},\alpha} \sum_{i=x,y,z}
 (-1)^\alpha \, \hat{c}^\dagger_{\bm{r}+\bm{e}_i,\alpha} \hat{c}_{\bm{r},\alpha}
\nonumber \\
&+ \mathrm{i} \frac{\Delta_1}{2} \sum_{\bm{r},\alpha} \sum_{i=x,y,z} \,
\hat{c}^\dagger_{\bm{r}+\bm{e}_i,\alpha+1} \, \sigma_i \, \hat{c}_{\bm{r},\alpha}
\nonumber \\
&- \frac{\Delta_2}{2 \mathrm{i}} \sum_{\bm{r},\alpha} \sum_{i=x,y,z} (-1)^{\alpha} \, n_i \, \hat{c}^\dagger_{\bm{r}+\bm{e}_i,\alpha + 1} \hat{c}_{\bm{r},\alpha}
+\mathrm{H.c.},
\end{align}
where $\alpha=0,1$ (modulo $2$) denotes the two orbitals at each site and
$c^\dagger_{\bm{r},\alpha}=(c^\dagger_{\bm{r},\alpha,\uparrow},c^\dagger_{\bm{r},\alpha,\downarrow})$ with
$c^\dagger_{\bm{r},\alpha,\sigma}$ being the creation operator for an electron with
orbital $\alpha$ and spin $\sigma$ at site $\bm{r}$.
Here, $\bm{e}_i$ for $i=x,y,z$ are lattice vectors and $\bm{n} = (1,-1,0)$.

When $\Delta_2=0$, the Hamiltonian (\ref{eq:H_chiral_HOTI}) respects TRS
represented by $i\sigma_y\kappa$ and $C_4^z$ (four-fold) rotational symmetry represented by
${C_4^z} \equiv \tau_0 e^{-\mathrm{i} \frac{\pi}{4} \sigma_z}$,
giving a 3D topological insulating phase when $1<|M|<3$.
In the presence of the $\Delta_2$ term, both TRS and rotational symmetry are broken individually,
but the Hamiltonian respects their combination $C_4^z T$ symmetry, that is,
\begin{equation}
	\begin{gathered}
		\left(C_4^z T\right) H_{\mathrm{c}}(\bm{k}) \left(C_4^z T\right)^{-1}={H}_{\mathrm{c}}(D_{\hat{C}^{z}_4 \hat{T}}\bm{k}),\\
		D_{\hat{C}_4^z \hat{T}}\bm{k} = (k_y, -k_x, -k_z).
	\end{gathered}
\end{equation}
The $C_4^z T$ symmetry is an anti-unitary symmetry with $(\hat{C}_4^z \hat{T})^4 = -1$.

Figure~\ref{fig:HOTI1}(b) shows the phase diagram of Hamiltonian (\ref{eq:H_chiral_HOTI}).
The system is in a 3D chiral higher-order topological insulating phase when $1<|M/t|<3$ and $\Delta_1,\Delta_2\neq 0$,
which hosts the gapless chiral hinge modes in the bulk gap
under open boundary conditions along $x$ and $y$ and periodic boundary conditions along $z$, as shown in Fig.~\ref{fig:HOTI1} (a) and (c).
The chiral hinge states as 1D dissipationless conduction channels will give rise to a quantized longitudinal conductance  $G=2 e^2/h$ along $z$,
which provides the transport signature of the higher-order topology~\cite{Schindler2018SA}.
The existence of chiral hinge states can be understood through the domain-wall picture as for the 2D case,
which will be detailed in the following.

\subsubsection{Hinge modes as domain walls}\label{sec:domain_wall}
The hinge states can be seen as the domain-wall states of 2D Dirac Hamiltonians resulting from the
sign change of the Dirac mass. Without the $\Delta_2$ term, the system is in a time-reversal-invariant
topological insulating phase with Dirac points on the surfaces.
When we turn on the $\Delta_2$ term, it will yield mass terms to gap out surface Dirac cones on the
$(100)$ and $(010)$ surfaces (the $(001)$ surfaces remain gapless).
Because of the $C_4^z T$ symmetry, the Dirac mass must change sign between two surfaces
related by a $C_4^z$ rotation, so that
there must be a chiral mode as a domain wall state along the hinge between the two surfaces.
In fact, before the discovery of HOTIs, it has been shown that
similar hinge modes due to domain walls can be induced by face-dependent magnetic fields
on the surfaces of TIs~\cite{Sitte2012PRL,ZhangFan2013PRL}.

We now follow Ref.~\cite{Song2017PRL} to show how the mass appears and changes its sign
based on the effective low-energy theory.
Let us consider the Dirac states on the surface parallel to $z$ which deviates from the $(y,z)$ plane by an angle $\theta$
[see Fig.~\ref{fig:HOTI1}(d)].
We write the bulk Hamiltonian (\ref{eq:H_chiral_HOTI}) in the coordinates of $\left(k_{1},k_{2},k_{3}\right)$,
where $k_{1}=-\sin\theta k_{x}+\cos\theta k_{y}$, $k_2=k_z$,
and $k_{3}=\cos\theta k_{x}+\sin\theta k_{y}$,
so that $k_1,k_2$ are parallel to the surface while $k_3$ is normal to the surface.
We then expand the Hamiltonian to
the first order of $k_{1}$, $k_{2}$ and the second order of $k_{3}$
for $M/t \gtrsim -3$ and set $t=\Delta_1=1$ for simplicity,
\begin{align}
H = &\left(m - \frac{k_{3}^{2}}{2}\right) \tau_{z}\sigma_{0} + k_{3}\tau_{x}\left(\sigma_{x}\cos\theta+\sigma_{y}\sin\theta\right)\nonumber \\
 & +k_{1}\tau_{x}\left(-\sigma_{x}\sin\theta+\sigma_{y}\cos\theta\right)+k_{2}\tau_{x}\sigma_{z} \nonumber \\
 & +\frac{\Delta\left(\theta\right)}{2}k_{3}^{2}\tau_{y}\sigma_{0},
\end{align}
where $m=M+3>0$ and $\Delta\left(\theta\right)=-\Delta_2 \cos(2\theta)$
satisfying $\Delta\left(\theta+\frac{\pi}{2}\right)=-\Delta\left(\theta\right)$
which is enforced by the symmetry
$(C_{4}^z T) \tau_{y}\sigma_{0} (C_{4}^z T)^{-1}=-\tau_{y}\sigma_{0}$.
By defining a new set of Pauli matrices $\sigma_3=\sigma_{x}\cos\theta+\sigma_{y}\sin\theta$,
$\sigma_1=-\sigma_{x}\sin\theta+\sigma_{y}\cos\theta$, $\sigma_2=\sigma_{z}$,
and the replacement $k_{3} \to -i\partial_{3}$, we obtain a continuous Hamiltonian
\begin{align}\label{eq:Heff-bulk}
H =&\left(m+\frac{\partial_{3}^{2}}{2}\right)\tau_{z}\sigma_{0}-i\partial_{3}\tau_{x}\sigma_{3}-\frac{\Delta\left(\theta\right)}{2}\partial_{3}^{2}\tau_{y}\sigma_{0}\nonumber \\
 & +k_{1}\tau_{x}\sigma_{1}+k_{2}\tau_{x}\sigma_{2}.
\end{align}

Then, we can solve the surface modes in a semi-infinite region ($x_3 \geq 0$) by neglecting the $k_{1}$, $k_{2}$, $\Delta$ terms firstly,
\begin{equation}
\left[\left(m+\frac{1}{2}\partial_{3}^{2}\right) + \tau_{y}\sigma_{3}\partial_{3}\right]\psi\left(x_{3}\right)=0
\end{equation}
with the boundary condition $\psi\left(0\right)=\psi\left(\infty\right)=0$.
It has two solutions $\psi_{n=1,2}$ as $\psi_{n}\left(x_{3}\right)= f(x_3) u_{n}$,
where
$f(x_{3})= \left( e^{-\lambda_{+}x_{3}}-e^{-\lambda_{-}x_{3}}\right)$ up to a normalization factor,
and $u_{n=1,2}$ are two spinors that satisfy $\tau_{y}\sigma_{3} u_n = u_n$  and
$\lambda_{\pm}=1\pm \sqrt{1-2m}$ with $m>0$ in the topological regime.
By projecting the remaining terms into the subspace of the two surface modes $\psi_{n=1,2}$,
one finally gets the effective surface Hamiltonian in terms of Pauli matrices in the basis of $u_{n=1,2}$
\begin{equation}
\mathcal{H}_{\mathrm{surf}} =k_{1} \tilde{s}_{2}-k_{2} \tilde{s}_{1} + \tilde{m}\left(\theta\right) \tilde{s}_{3},
\end{equation}
where the Dirac mass $\tilde{m}\left(\theta\right)$ is proportional to $\Delta(\theta)$ as
\begin{equation}
\tilde{m}\left(\theta\right) =\frac{\Delta\left(\theta\right)}{2} \int dx_{3} f^{*}(x_3) \left[-\partial_{3}^{2}f(x_3)\right].
\end{equation}
Therefore, we have $\tilde{m}\left(\theta+\frac{\pi}{2}\right)=-\tilde{m}\left(\theta\right)$
so that the mass changes the sign from a surface to another under a $C_4^z$ rotation.

\begin{figure*}[t]
	\begin{center}
		\includegraphics[width=\linewidth]{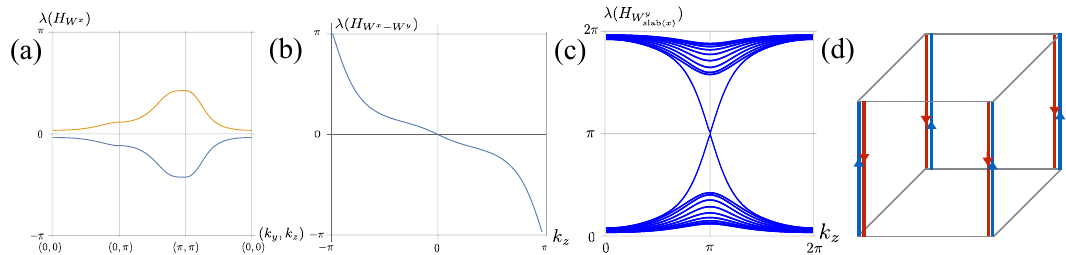}
		\caption{
			(a) The eigenvalues of the Wannier Hamiltonian $H_{\mathcal{W}_x}(k_y,k_z)$ with respect to $(k_y,k_z)$.
			(b) The eigenvalues of the nested Wilson loop along $k_y$ over the occupied subspace of the gapped Wannier
			Hamiltonian $H_{\mathcal{W}_x}(k_y,k_z)$.
			(c) The Wilson loop spectrum along $y$ in a slab geometry along $x$.
			(d) Schematic illustration of the helical hinge modes.
			Reproduced from Ref.~\cite{Schindler2018SA}.
		}
		\label{fig:HOTI_Wilson}
	\end{center}
\end{figure*}

\subsubsection{Topological characterization}
The authors in Ref.~\cite{Schindler2018SA} argue that the above HOTI protected by
the $C_4^z T$ symmetry has a $\mathbb{Z}_2$ classification for the bulk topology
in the following way.
If there exist chiral hinge modes propagating unidirectionally,
we can attach a Chern insulator on each surfaces parallel to $z$ direction; two neighboring
Chern insulators have opposite Chern numbers, so that the whole system does not break
the $C_4^z T$ symmetry. As a result, if the initial system has two chiral modes at each hinge,
these modes can be removed by attaching the Chern insulators.
However, a $\mathbb{Z}$-valued topological invariant (the winding number of the
quadrupole moment) is proposed~\cite{kang2021many,WangJH2021PRL}, making it possible to observe the phase transition
from a trivial phase to a topological phase with more than one chiral modes at each hinge if we consider long-range hopping.
It in fact remains unclear whether the attached system is topologically equivalent to a chiral HOTI with some winding number
of the quadrupole moment.
We also want to note that the gapped surfaces of 3D chiral HOTIs with one chiral mode at each hinge are distinct from Chern insulators in 2Ds
in that a single surface carries an anomalous Chern number of $\pm 1/2$ rather than an integer measured from the real-space Chern marker~\cite{pozo2019quantization}.

\paragraph{Chern-Simons invariant}

The Chern-Simons invariant is given by
\begin{equation}\label{eq:Chern-Simons}
\theta
=\frac{1}{4\pi}
\int \mathrm{d}^3\bm{k} \, \epsilon_{abc}
\Tr \left[
\mathcal{A}_a\partial_b\mathcal{A}_c + \mathrm{i}\frac{2}{3} \mathcal{A}_a\mathcal{A}_b\mathcal{A}_c
\right],
\end{equation}
where $\mathcal{A}_a$ is the Berry gauge field defined as
$[\mathcal{A}_{a}]^{n,n'}=-\mathrm{i}\langle u_{\bm k}^n|\partial_{k_a}|u_{\bm k}^{n'} \rangle$ for $a=x,y,z$,
where $\ket{u_{\bm k}^n}$ is the $n$th occupied eigenstate of the Bloch Hamiltonian.
It has been proved that owing to the $C_4^z T$ symmetry, the Chern-Simons invariant is quantized to $\theta=0,\pi \mod 2\pi$~\cite{Schindler2018SA}.

The $\mathbb{Z}_2$ topology of 3D chiral HOTIss can be characterized by a Chern-Simons invariant $\theta$,
which has been used to characterize the 3D $\mathbb{Z}_2$ TIs protected by TRS~\cite{Qi2008PRB}.
When $\theta=\pi$, the system is in the higher-order topological insulating phase with chiral hinge modes.
Moreover, in Ref.~\cite{SunKai2020PRL}, the authors propose a generalized Pfaffian topological invariant to characterize the
3D HOTIs with $C_4^z T$ symmetry.

\paragraph{Nested Wilson loop}

One can also use the nested Wilson loop to characterize the chiral higher-order topology~\cite{Schindler2018SA}.
The Wilson loop along $x$, $\mathcal{W}_{x, \bm k}$ is defined in Eq.~(\ref{eq:Wilson}) in Sec.~\ref{sec:Wannier_band}.
For the first-order TIs with TRS, the Wannier spectrum of $\mathcal{W}_{x,(k_y, k_z)}$ is gapless,
similar to the energy spectrum under open boundary conditions along $x$~\cite{Yu2011PRB,Bernevig2014PRB}.
For the chiral HOTI in Eq.~(\ref{eq:H_chiral_HOTI}),
the Wannier spectrum is fully gapped corresponding to the gapped spectrum on the surfaces perpendicular to $x$,
as shown in Fig.~\ref{fig:HOTI_Wilson}(a).

It has been found that for the nontrivial phase of the chiral HOTI, the Wannier Hamiltonian
has a Chern number $C=\pm 1$ for the half-filled Wannier bands,
which can be seen as a 2D Chern insulator. This can be verified by computing the $y$-directed
nested Wilson loop for a gapped Wannier sector of the Wannier Hamiltonian ${H}_{\mathcal{W}_x}(k_y, k_z)$.
For the nontrivial phase, the nested Wilson loop spectrum displays a spectral flow with respect to $k_z$, indicating
the existence of the Chern number for the Wannier bands, as shown in Fig.~\ref{fig:HOTI_Wilson}(b).
This nontrivial winding can also be seen in the $y$-directed Wilson loop spectrum of a slab Hamiltonian with
open boundaries along $x$ [see Fig.~\ref{fig:HOTI_Wilson}(c)], illustrating the presence of chiral edge modes
in the Wannier spectrum.

\paragraph{The winding number of the quadrupole moment}

It has been proposed that the winding number of the quadrupole moment can serve as a topological invariant
to characterize the 3D chiral topology~\cite{kang2021many,WangJH2021PRL,MaoYF2023arXiv}.
The quadrupole moment winding with respect to $k_z$ is defined as
\begin{equation}
W_Q=\int_0^{2 \pi} d k_z \frac{\partial q_{x y}(k_z)}{\partial k_z},
\end{equation}
where $q_{x y}(k_z)$ is the quadrupole moment defined in Eq.~(\ref{eq:Qxy}) for a 2D Hamiltonian at $k_z$.
For the model in Eq.~(\ref{eq:H_chiral_HOTI}), $W_Q=1$ in the nontrivial phase and $W_Q=0$ in
the trivial phase. This invariant can also be applied to many-body or disordered systems if we use a twisted
boundary condition with the flux serving as the momentum~\cite{WangJH2021PRL}. Note that this invariant does not need
a symmetry to protect its quantization. It can change through either a bulk or surface gap closing. Since
this topological invariant is $\mathbb{Z}$-valued, it is possible to observe the phase transition from
a trivial phase to a nontrivial phase with $W_Q=2$.

\subsection{Higher-order topological insulators with helical hinge states} \label{Subsec:Helical}
Helical HOTIs host helical hinge modes [see Fig.~\ref{fig:HOTI_Wilson}(d)]
protected by TRS and crystalline symmetry,
such as mirror, rotation, and inversion symmetry~\cite{Brouwer2018PRB,Khalaf2018PRB,Brouwer2019PRX}.
In the case with mirror symmetry, the system has a $\mathbb{Z}$ classification~\cite{Schindler2018SA}.
In the following, we focus on a model for helical HOTIs as a generalization of
the chiral HOTI model (\ref{eq:H_chiral_HOTI}), which respects both time-reversal
and four-fold rotational symmetry proposed in Ref.~\cite{Song2017PRL,Schindler2018SA}.
It has a $\mathbb{Z}_2$ classification of bulk topology.

\subsubsection{Model Hamiltonian}
A model for the helical HOTI with $C_4$ and $T$ symmetry takes the form as~\cite{Schindler2018SA}
\begin{align}
H_{\mathrm{h}} &= \frac{M}{2} \sum_{\bm{r},\alpha,\mu} (-1)^\alpha \, \hat{c}^\dagger_{\bm{r},\alpha,\mu} \hat{c}_{\bm{r},\alpha,\mu} \nonumber \\
&+ \frac{t}{2} \sum_{\bm{r},\alpha,\mu} \sum_{i=x,y,z}
 (-1)^\alpha \, \hat{c}^\dagger_{\bm{r}+\bm{e}_i,\alpha,\mu} \hat{c}_{\bm{r},\alpha,\mu}\nonumber \\
&+ \mathrm{i} \frac{\Delta_1}{2} \sum_{\bm{r},\alpha,\mu} \sum_{i=x,y,z} \,
\hat{c}^\dagger_{\bm{r}+\bm{e}_i,\alpha+1,\mu} \, \sigma_i \, \hat{c}_{\bm{r},\alpha,\mu}\nonumber \\
&+ \frac{\Delta_2}{2} \sum_{\bm{r},\alpha,\mu} \sum_{i=x,y,z} (-1)^{\alpha} \, {n}_i \, \hat{c}^\dagger_{\bm{r}+\bm{e}_i,\alpha + 1,\mu+1} \hat{c}_{\bm{r},\alpha,\mu}\nonumber \\
& +\mathrm{H.c.},
\end{align}
where $\alpha,\mu \in \{0,1\}$ defined modulo $2$ are the orbital indices, $\bm{n} = (1,-1,0)$, and $\hat{c}^\dagger_{\bm{r},\alpha,\mu}=(\hat{c}^\dagger_{\bm{r},\alpha,\mu,\uparrow},\hat{c}^\dagger_{\bm{r},\alpha,\mu,\downarrow})$ is a spinor
with $\hat{c}^\dagger_{\bm{r},\alpha,\mu,\sigma}$ creating an electron with spin $\sigma$ in orbital $(\alpha,\mu)$ at lattice site $\bm{r}$.
The model Hamiltonian reads in momentum space~\cite{Schindler2018SA}
\begin{align}\label{eq:H_helical_HOTI}
H_{\mathrm{h}}(\bm{k}) = &\Bigl(M+t \sum_i \cos k_i\Bigr) \, \tau_z s_0 \sigma_0 + \Delta_1 \sum_i \sin k_i \, \tau_x  s_0 \sigma_i \nonumber \\
&+\Delta_2(\cos k_x - \cos k_y) \, \tau_y s_y \sigma_0.
\end{align}
Here, the Pauli matrices $\tau_i$ and $s_i$ act on the orbital degrees of freedom for the indices $\alpha$ and $\mu$, respectively,
and $\sigma_i$ acts on the spin degree of freedom.

This Hamiltonian respects the TRS and $C_4$ rotational symmetry represented by $\mathcal{T} \equiv i  \sigma_y \kappa$
and ${C_4^z} \equiv s_x e^{-\mathrm{i} \frac{\pi}{4} \sigma_z}$,
and has the phase diagram similar to the chiral HOTI model in Eq.~(\ref{eq:H_chiral_HOTI}).
When $1<|M/t|<3$,
there exists the nontrivial HOTI phase with helical hinge modes at each of the four hinges.
The four helical pairs of hinge states will contribute a quantized longitudinal conductance $G=4 e^2/h$ along $z$~\cite{Schindler2018SA}.
Analogous to the chiral HOTI, the helical hinge modes can be understood from the perspective of domain walls~\cite{Song2017PRL}.
Without the $\Delta_2$ term, the Hamiltonian consists of two copies of a strong TI
with two Dirac cones on each surface.
This system is thus a trivial first-order TI according to the $\mathbb{Z}_2$ classification, whose Dirac cones can be gapped out by a time-reversal invariant mass, such
as the $\Delta_2$ term.
The $C_4^z$ symmetry enforces the Dirac mass to have opposite signs between two surfaces
so that there exists a helical mode at the domain wall of mass, which can be derived using the low-energy surface theory introduced in Sec.~\ref{sec:domain_wall}.

The authors in Ref.~\cite{Schindler2018SA} introduce a ${C}_4^z$-graded Wilson loop for the helical HOTI so that
the nontrivial winding of Wilson loop spectra between $(k_x,k_y)=(0,0)$ and $(\pi,\pi)$ defines a $\mathbb{Z}_2$ invariant.
In addition, they also employ the Wilson loop formalism to characterize the helical HOTI phase,
similarly to the chiral HOTI case. Specifically, one can first compute the $x$-directed Wilson loop to obtain a gapped Wannier spectrum,
and then compute the $y$-directed nested Wilson loop spectrum to obtain a pair of helical modes
as a function of $k_z$, reflecting the existence of gapless helical hinge modes.

\subsubsection{A $\mathbb{Z}_2$ topological invariant for helical higher-order topological insulators} \label{Subsec:Z2invariant}

In fact, the Hamiltonian in Eq.~(\ref{eq:H_helical_HOTI}) also obeys a $\mathbb{Z}_2$ symmetry of $s_y$ with $s_y^2=1$,
and we thus can divide the Hilbert space into two subspaces with $s_y=\pm 1$. In each subspace, the quadrupole
moment $q_{xy}^{(s)}$ can be evaluated based on Eq.~(\ref{eq:Qxy_manybody}) so that a winding number
of the quadrupole moment about $k_z$ in each subspace is defined as~\cite{WangJH2021PRL}
\begin{equation}
W_Q^{(s)}=\int_{0}^{2\pi} d\Phi_z \frac{\partial q_{xy}^{(s)}(k_z)}{\partial k_z}.
\end{equation}
It has been proved that $W_Q^{(s)}=-W_Q^{(-s)}$ due to TRS, so the spin winding number of the
quadrupole moment is defined as
\begin{equation} \label{eq:WQS}
W_{QS}=\left(W_Q^{(s=1)}-W_Q^{(s=-1)}\right)/2=W_Q^{(s=1)}.
\end{equation}
One can also include a term of $t_3\sin k_z \tau_y s_x$ to break the $\mathbb{Z}_2$ symmetry but
still preserves TRS and $C_4^z$ symmetry~\cite{WangJH2021PRL}. In this case, the authors in Ref.~\cite{WangJH2021PRL}
show that for small $t_3$, the spin winding number is still applicable.

For generic cases with TRS, a $\mathbb{Z}_2$ invariant $\nu_Q \in \{0,1\}$ is derived to characterize the helical
HOTI~\cite{WangJH2021PRL,MaoYF2023arXiv}. It is defined as
\begin{equation}
{(-1)}^{\nu_Q}=\frac{\mathrm{Pf}[A(k_z=\pi)]}{\mathrm{Pf}[A(k_z=0)]} \sqrt{\frac{\det[A(k_z=0)]}{\det[A(k_z=\pi)]}}.
\end{equation}
Here, the $k_z$ dependent matrix is defined as
\begin{equation}
[A(k_z)]_{mn}=\langle \psi_m(-k_z)| \hat{D} {\mathcal{T}} |\psi_n(k_z)\rangle,
\end{equation}
where $m,n$ run over both occupied band indices and momenta $(k_x,k_y)$ of the Hamiltonian at $k_z$,
the operator $\hat{D}$ is the same as that in the quadrupole moment formula (\ref{eq:Qxy}), and
$\mathrm{Pf}[\cdot]$ denotes the Pfaffian of an antisymmetric matrix.
For topologically nontrivial (trivial) phase, we have $\nu_Q = 1$ ($\nu_Q=0$).
Similar to the winding number of the quadrupole moment, this $\mathbb{Z}_2$ invariant can also be
generalized to systems without translational symmetry by replacing the momentum $k_z$ with a flux
twisting through a boundary along $z$~\cite{WangJH2021PRL}.

We now follow Ref.~\cite{WangJH2021PRL} to briefly review the derivation of the topological invariant based on the quadrupole moment
for a noninteracting 3D model $H_h$ with TRS represented by $\mathcal{T}=i \sigma_y \kappa$, that is,
${\mathcal{T} }{H}_h(\bm{k}){ \mathcal{T} }^{-1}={H}_h(-\bm{k})$.
Using the periodicity of the wave functions about $k_z$ and antiunitary property of the time-reversal operator,
we can prove the following relations
\begin{align}
&A(k_z)=A(k_z+2\pi), \nonumber \\
&A(k_z)=-[A(-k_z)]^{T},
\end{align}
where $[A(k_z)]^{T}$ denotes the transpose of the matrix $A(k_z)$.
The matrix $A(k_z)$ is antisymmetric at two time-reversal symmetric surfaces
$k_z=0$ and $k_z=\pi$.

We then construct a 1D Hamiltonian based on $A(k_z)$ as
\begin{equation}
{H}_{A}(k_z)=
\begin{pmatrix}
  0 & A(k_z) \\
  [A(k_z)]^\dagger & 0 \\
\end{pmatrix}.
\end{equation}
Here, we assume that the matrix $A(k_z)$ is invertible so that ${H}_{A}(k_z)$ has a gap at zero energy.
${H}_{A}$ respects the chiral symmetry, the TRS and the particle-hole symmetry,
belonging to the class DIII for free electrons, and thus its bulk topology is $\mathbb{Z}_2$ classified
according to the Altland-Zirnbauer classification~\cite{Altland1997PRB,Chiu2016RMP}.
We thus use the $\mathbb{Z}_2$ topological invariant of ${H}_{A}(k_z)$ to classify a generic 3D
SOTI with TRS.
The $\mathbb{Z}_2$ invariant $\nu_Q \in \{0,1\}$ is given by
\begin{align}\label{Z2}
(-1)^{\nu_Q}
&=\frac{\textrm{Pf}[A(\pi)]}{\textrm{Pf}[A(0)]}
\exp \left\{ -\frac{1}{2}\int_{0}^{\pi} d k_z \partial_{k_z} [\log \det A(k_z)] \right\} \nonumber \\
&=\frac{\mathrm{Pf}[A(\pi)]}{\mathrm{Pf}[A(0)]} \sqrt{\frac{\det[A(0)]}{\det[A(\pi)]}},
\end{align}
where the Pfaffian is well defined for $A(0)$ and $A(\pi)$.
Since $\det[A]=\textrm{Pf}[A]^2$, $(-1)^{\nu_Q}=\pm 1$ so that $\nu_Q$ is quantized to be $0$ or $1$.
Note that it is required that the phase of $\det[A(\Phi_z)]$ changes continuously as $k_z$ varies from
$0$ to $\pi$.
In numerical calculations, one needs to fix the gauge of the eigenstates
such that the sewing matrices for TRS take a specific form (see details in Ref.~\cite{WangJH2021PRL}).

In Ref.~\cite{WangJH2021PRL}, it was also shown that an equivalent formula
\begin{eqnarray}
(-1)^{\nu_Q}&=&\mbox{sgn}\Bigg\{\frac{\mbox{Pf}[A(\pi)]}{\mbox{Pf}[A(0)]} \exp \{-i\pi[q_{xy}(k_z=\pi) \nonumber \\
&& -q_{xy}(k_z=0)] \} \Bigg\}
\end{eqnarray}
can be used for practical calculations.
Here, $q_{xy}(k_z)$ is the quadrupole moment evaluated at $k_z$.
When the pseudospin $s_y$ is conserved, the $\mathbb{Z}_2$ topological invariant $\nu_Q$ and
spin winding number $W_{QS}$ are related by $(-1)^{\nu_Q}=\exp(i\pi W_{QS})$~\cite{WangJH2021PRL}.

\subsection{Higher-order topological semimetals} \label{Sec:HOSM}

\begin{figure*}[t]
\centering
\includegraphics[width=0.8\linewidth]{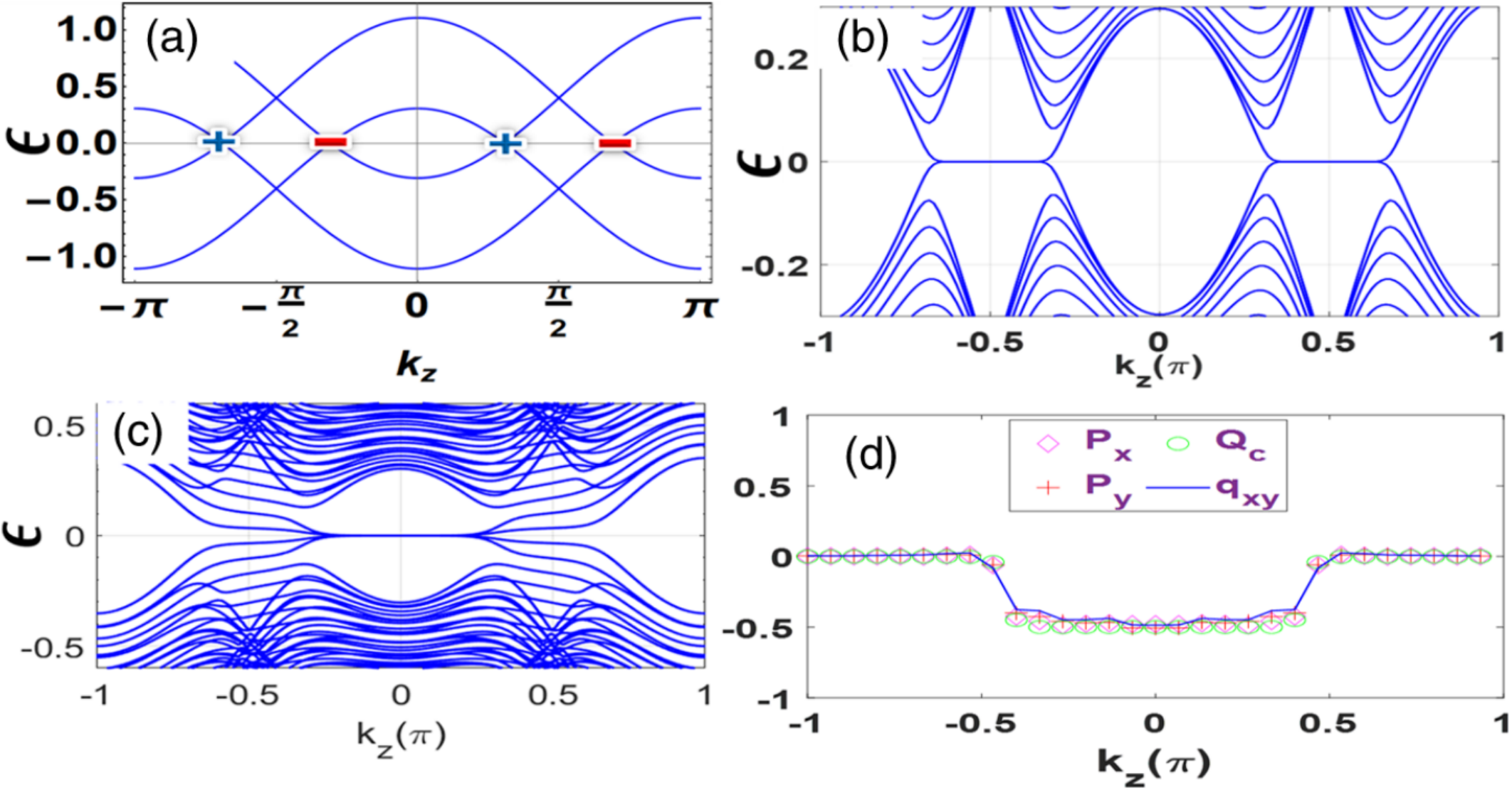}
\caption{(a)-(c) The energy spectrum of the Hamiltonian $H_{\textrm{HODSM}}$ perturbed by a term of $m_1 \sigma_2$.
	(a) The bulk band structure at $k_x=k_y=0$.
	A Dirac point splits into two Weyl point whose charges are marked out as positive and negative signs.
	The energy spectrum (b) at $k_y=0$ ($k_x=0$) with open boundaries along $x$ ($y$)
    and that (c) with open boundaries along both $x$ and $y$.
(d) The quadrupole moment $q_{xy}(k_z)$, edge polarizations $P_x(k_z)$, $P_y(k_z)$ and corner charge $Q_c(k_z)$
with respect to $k_z$.
Here, $\gamma=-1$ and $m_1=0.4$.
Reproduced from Ref.~\cite{Hughes2020PRL}.}
\label{fig:HOWSM}
\end{figure*}

After the discovery of HOTIs, a number of works have explored
higher-order topological semimetals with gapless bulk or surface nodes~\cite{Lin2018PRB,WangZJ2019PRL,Roy2019PRB,Roy2019PRR,Okugawa2019PRB,
Zilberberg2020PRR,Zeng2020PRB,Hughes2020PRL,JiangJH2020PRL,Wieder2020NC,Brouwer2022PRB}.
In contrast to surface Fermi arcs in first-order topological semimetals,
hinge Fermi arcs appear in higher-order topological semimetals.
In the following, we will briefly review how higher-order Dirac and Weyl semimetals
arise.

We first consider a model for higher-order Dirac semimetals proposed in Ref.~\cite{Lin2018PRB},
\begin{equation}
\hat{H}_{\mathrm{HODSM}}= \sum_{\bm{k}} \hat{c}^{\dagger}(\bm{k}) H_{\mathrm{HODSM}}(\bm{k}) \hat{c}(\bm{k}),
\end{equation}
where $\hat{c}^{\dagger}(\bm{k})$ is a four-component creation operator at momentum $\bm{k}$ in the Brillouin zone
and the corresponding Bloch Hamiltonian is
\begin{align}\label{eq:HODSM}
H_{\mathrm{HODSM}}(\bm{k})=& \left[\gamma_x(k_z) + \cos k_x \right]\Gamma_4+\sin k_x\Gamma_3 \nonumber \\
+& \left[\gamma_y(k_z) + \cos k_y\right]\Gamma_2+\sin k_y\Gamma_1, \\
=& \left[\gamma_x(k_z) + \cos k_x \right]\tau_1 \sigma_0-\sin k_x\tau_2 \sigma_3 \nonumber \\
-& \left[\gamma_y(k_z) + \cos k_y\right]\tau_2 \sigma_2-\sin k_y\tau_2 \sigma_1,
\end{align}
where $\gamma_\nu(k_z)=\gamma_\nu+0.5\cos k_z$ with $\nu=x,y$.
The Gamma matrices are the same as those defined in the BBH model in Eq.~(\ref{eq:H_BBH}).
The Hamiltonian constitutes a stack of the 2D BBH models with the intracell hoppings modified by the hopping
along $z$.
Similar to the BBH model, this model respects
the spinless TRS
$\mathcal{T}=\kappa$, particle-hole symmetry $\mathcal{P}=\tau_3 \kappa$,
chiral symmetry $\mathcal{C}=\tau_3 $,
reflection symmetries $U_{M_x}=\tau_1\sigma_3$,
$U_{M_y}=\tau_1\sigma_1$ and $U_{M_z}=I_4$, and inversion symmetry
$U_{I}\equiv \sigma_2$. When $\gamma_x=\gamma_y=\gamma$, it also has the $C_4^{z}$ symmetry.

The BBH model exhibits the nontrivial higher-order topology when $|\gamma_x|<1$ and $|\gamma_y|<1$~\cite{Benalcazar2017Science,Benalcazar2017PRB}.
In the presence of the $C_4^{z}$ symmetry, the system undergoes a topological phase transition at $|\gamma|=1$
through the bulk energy gap closure, implying the existence of a Dirac point in the bulk at $k_z^D$ where $|\gamma(k_z^D)|=1$.
For example, in the case with $-1.5<\gamma <-0.5$, there appear two Dirac points at $k_z^D=\pm \arccos(-2-2\gamma)$.
The zero-energy hinge modes arise when $|k_z|<|k_z^D|$, leading to hinge Fermi arcs.
Moreover, in the absence of the rotational symmetry, the Dirac points emerge at $x$-normal or $y$-normal surfaces.
These semimetals are called higher-order Dirac semimetals~\cite{Lin2018PRB}.

As we have proved in Sec.~\ref{sec:BBH}, the TRS and inversion symmetry ensures the double degeneracy of each band.
Since the model in Eq.~(\ref{eq:HODSM}) respects both of these symmetries, any touching points must be four-fold
degenerate. Therefore, to create Weyl points in the system, one has to add terms breaking one of these symmetries.
In addition, the system also respects chiral symmetry so that the Chern number at each $k_z$ cannot exist.
Therefore, we can add terms to break the TRS so that the chiral symmetry is broken but the particle-hole symmetry
is still preserved. In this case, the quadrupole moment is still quantized and the Chern number may appear.
For example, let us add a term $m_1 \sigma_2$ in the Hamiltonian in Eq.~(\ref{eq:HODSM}) that
breaks the TRS and chiral symmetry but preserves the particle-hole symmetry.
Indeed, this term can split a Dirac point into two Weyl points [see Fig.~\ref{fig:HOWSM}(a)]~\cite{Hughes2020PRL}. Between the two Weyl points,
the Chern number $C=1$ emerges in each $k_z$ slice so that there appear surface Fermi arcs connecting the projections of
two Weyl points as revealed by Fig.~\ref{fig:HOWSM}(b). The Weyl point thus
still corresponds to the anomalous quantum Hall insulating phase transition point.
In fact, two of these Weyl points also correspond to the critical point of a higher-order topological phase transition
between a quadrupole insulator with nonzero quadrupole moment and a higher-order trivial phase [see Fig.~\ref{fig:HOWSM}(d)].
Thus, a hinge Fermi arc also exists connecting these Weyl points as shown in Fig.~\ref{fig:HOWSM}(c).
By applying different perturbation terms breaking other symmetries, one can
get different configurations of Weyl nodes and Fermi arcs~\cite{Hughes2020PRL}.

\section{Effects of quenched disorder on higher-order topological phases}\label{sec3}

Disorder is ubiquitous in nature and plays an important role in quantum transports and topological classifications~\cite{Altland1997PRB,Ludwig2008PRB}.
For instance, disorder can localize particles, resulting in an Anderson insulating phase~\cite{Anderson1958,Mirlin2008RMP}.
In the context of symmetry-protected topological phases, these states are usually robust against
weak disorder preserving the corresponding symmetry~\cite{Sheng2005PRL,WuCJ2006PRL,XuCK2006PRB,Shen2009PRL,Beenakker2009PRL,Qiao2009PRB,JiangHua2009PRB,JiangHua2014PRL,ChuiZhen2015PRL,Shindou2016PRL} (or an average symmetry~\cite{Fu2012PRL,Akhmerov2014PRB}).
Yet, for a sufficiently strong disorder, a system will transition into a trivial Anderson insulator.
Remarkably, it has also been found that disorder can drive a topological phase transition
from a trivial phase to a first-order topological phase~\cite{Shen2009PRL}.
These TIs emerging from trivial insulators due to disorder are named
topological Anderson insulators (TAIs)~\cite{Shen2009PRL,Beenakker2009PRL}.

In the context of HOTPs, it has been found that these phases are stable
against weak disorder~\cite{Benalcazar2017Science,Hatsugai2019PRB,Fulga2019PRB,Li2020PRB,JiangH2019CPB,Yang2021PRB,Li2020PRL,WangXR2020PRR,Szabo2020PRR,
WangXR2021PRB,ChenCZ2021PRB,
JiangH2021PRB,Lu2023PRB,Castro2023arXiv,Qiao2023arXiv}. Interestingly, the concept of TAIs is also extended
to HOTIs with chiral symmetry~\cite{Yang2021PRB,Li2020PRL}
and higher-order topological superconductors~\cite{Fulga2019PRB}, showing that disorder can induce
a higher-order topological phase transition from a trivial one to a nontrivial one.
Meanwhile, a modified Haldane model has been shown to exhibit a disorder-induced
HOTP from an anomalous quantum Hall insulator~\cite{ZhangXD2021PRL},
which is experimentally observed in electric circuits.
In Ref.~\cite{li2022transition}, the authors showed that random flux can induce a metal-band insulator transition from a metallic phase to an extrinsic HOTI with zero-energy corner modes
in a 2D SSH model.
Recently, the disorder-induced third-order topological insulator with the quantized octupole moment in 3Ds
is also found~\cite{Castro2023arXiv}.
A recent work studied the disorder-induced phase transitions in chiral SOTIs protected by the $C_4^z T$ symmetry,
and found that the chiral HOTI is robust with quantized conductance at weak disorder and then
driven into a diffusive metal and finally an Anderson insulator~\cite{Qiao2023arXiv}.

In this section, we will consider quenched disorder which is manifested as some static random-valued parameters in the system, and
review the effects of quenched disorder on quadrupole insulators in subsection~\ref{Sec:HOTAI},
on 3D SOTIs in subsection~\ref{sec:Dis3DSOTI}, and on HOTSMs in subsection~\ref{sec:DisHOTSM}.

\subsection{Higher-order topological Anderson insulators} \label{Sec:HOTAI}

In this subsection, we will review disorder-induced quadrupole TIs,
localization properties, self-consistent Born approximation, and effective boundary Hamiltonian from the Green's function.

\subsubsection{Disorder-induced quadrupole topological insulators}

\begin{figure*}[t]
	\includegraphics[width=0.8\linewidth]{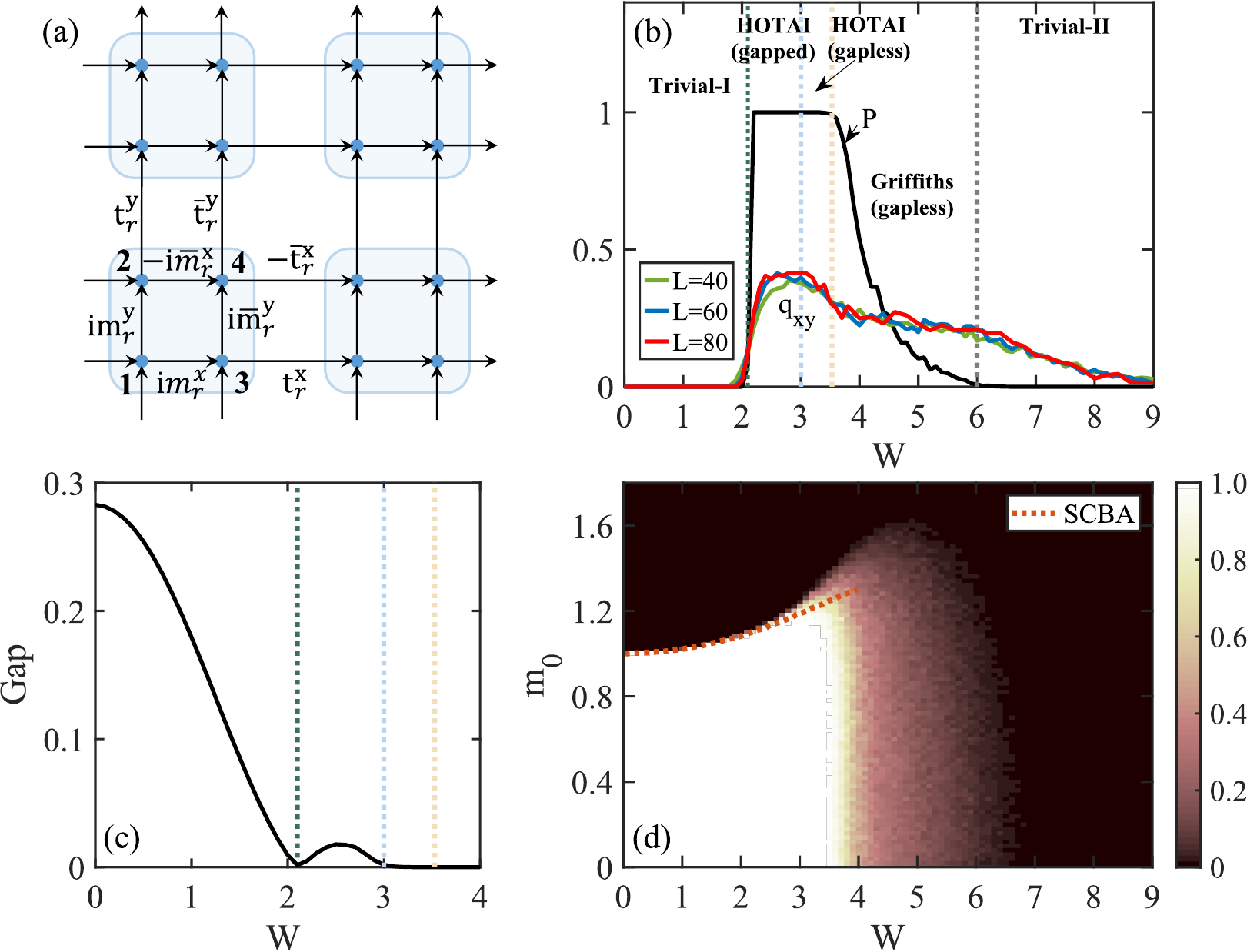}
	\caption{
		(a) Schematic illustration of the disordered tight-binding model in Eq.~(\ref{eq:Yang_model}).
		(b) The phase diagram with respect to the disorder strength $W$ mapped out by the topological invariant $P$ and
		the quadrupole moment $q_{xy}$.
		There exist distinct phases including gapped/gapless higher-order TAI, Griffiths phase, and trivial I/II phases, separated by vertical dashed lines.
		(c) The bulk energy gap versus $W$.
		In (b)-(c), $m_x=m_y=1.1$.
		(d) The topological invariant $P$ with respect to $W$ and $m_0$ with $m_x=m_y =m_0$.
		The topological phase boundary calculated by the SCBA method is highlighted by the red dotted line.
		Reproduced from Ref.~\cite{Yang2021PRB}.}	
		\label{fig_YangHOTI1}
\end{figure*}

\paragraph{Model Hamiltonian}
The authors in Ref.~\cite{Yang2021PRB,Li2020PRL} prove that chiral symmetry can protect the quantization
of the quadrupole moment (also see Sec.~\ref{sec:Qxy} for proof)
and find that disorder can drive a topological phase transition with bulk energy gap closing or edge energy gap
closing from a trivial insulator to a HOTI, namely,
a higher-order topological Anderson insulator.
To introduce the disorder effects, we consider the following tight-binding Hamiltonian
proposed in Ref.~\cite{Yang2021PRB}
\begin{equation}
	\hat{H}=\sum_{\bf r} \left[ \hat{c}^\dagger_{\bf r}h_0\hat{c}_{\bf r}
	+\left( \hat{c}^\dagger_{\bf r}h_x\hat{c}_{{\bf r}+{\bf e}_x}
	+\hat{c}^\dagger_{\bf r}h_y\hat{c}_{{\bf r}+{\bf e}_y}+\mathrm{H.c.} \right) \right].
	\label{eq:Yang_model}
\end{equation}
Here $\hat{c}^\dagger_{\bf r}=\left(
\begin{array}{cccc}
	\hat{c}^\dagger_{{\bf r} 1} & \hat{c}^\dagger_{{\bf r} 2} & \hat{c}^\dagger_{{\bf r} 3} &
	\hat{c}^\dagger_{{\bf r} 4} \\
\end{array}
\right)
$ where $\hat{c}^\dagger_{{\bf r} \nu}$ ($\hat{c}_{{\bf r} \nu}$) are creation (annihilation) operator of a particle
at the $\nu$th degree of freedom in a unit cell labelled by the position vector ${\bf r}=(x,y)$ ($x$ and $y$ are integers).
${\bf e}_x=(1,0)$ and ${\bf e}_y=(0,1)$ are unit vectors along $x$ and $y$, respectively.
\begin{equation}
	h_0=\left(
	\begin{array}{cccc}
		0 & -im_{{\bf r}}^y & -im_{{\bf r}}^x & 0 \\
		im_{{\bf r}}^y & 0 & 0 & i\bar{m}_{{\bf r}}^x \\
		im_{{\bf r}}^x & 0 & 0 & -i\bar{m}_{{\bf r}}^y \\
		0 & -i\bar{m}_{{\bf r}}^x & i\bar{m}_{{\bf r}}^y & 0 \\
	\end{array}
	\right)
\end{equation}
describes the intra-cell hopping, and
\begin{equation}
	\begin{aligned}
		&h_x=\left(
		\begin{array}{cccc}
			0 & 0 & 0 & 0 \\
			0 & 0 & 0 & 0 \\
			t_{{\bf r}}^x & 0 & 0 & 0 \\
			0 & -\bar{t}_{{\bf r}}^x & 0 & 0 \\
		\end{array}
		\right)~\text{and}~
		h_y=\left(
		\begin{array}{cccc}
			0 & 0 & 0 & 0 \\
			t_{{\bf r}}^y & 0 & 0 & 0 \\
			0 & 0 & 0 & 0 \\
			0 & 0 & \bar{t}_{{\bf r}}^y & 0 \\
		\end{array}
		\right)
	\end{aligned}
\end{equation}
describe the inter-cell hopping along $x$ and $y$, respectively [see Fig.~\ref{fig_YangHOTI1}(a)
for the hopping parameters].
The inter-cell hopping magnitude are taken to be
$t_{\bf r}^x=\bar{t}_{\bf r}^x=t_{\bf r}^y=\bar{t}_{\bf r}^y=1$,
and the other parameters $m_{{\bf r}}^x$, $\bar{m}_{{\bf r}}^x$, $m_{{\bf r}}^y$, $\bar{m}_{{\bf r}}^y$ are real numbers.
The system preserves chiral symmetry (sublattice symmetry) given that only the nearest-neighbor hopping
is involved in the model.
Since the Hamiltonian contains the hopping that takes complex values,
the spinless TRS represented by $\kappa$ does not exist.
However, the Hamiltonian can be mapped to the BBH model through a local unitary transformation~\cite{Yang2021PRB} and
thus share similar topological and localization properties with a disordered BBH model studied in Ref.~\cite{Li2020PRL}.

The disorder is introduced
in the intra-cell hopping, that is,
\begin{eqnarray}
m_{\bf r}^\nu &=& m_\nu+ W V_{\bf r}^\nu,  \\
\bar{m}_{\bf r}^\nu &=& {m}_\nu+ W \bar{V}_{\bf r}^\nu
\end{eqnarray}
with $\nu=x,y$, where $m_x=m_y=m_0$ and ${V}_{\bf r}^\nu$ and $\bar{V}_{\bf r}^\nu$ obey
independent uniform random distribution within the region of $[-0.5,0.5]$.
Here, $W$ represents the disorder strength.
Due to the random property, the system is characterized by properties averaged over large numbers of random configurations.

In the clean case with $m_{\bf r}^x=\bar{m}_{\bf r}^x=m_x$
and $m_{\bf r}^y=\bar{m}_{\bf r}^y=m_y$, the system can be described by a Bloch Hamiltonian in momentum space
\begin{equation}
	H_0({\bf k})=H_x(k_x,m_x)\otimes \sigma_3+\sigma_0 \otimes H_y(k_y,m_y),
	\label{eq:Yang_model_k}
\end{equation}
where $H_\nu(k_\nu,m_\nu)=\cos k_\nu \sigma_1+(m_\nu+\sin k_\nu)\sigma_2$ with $\nu=x,y$.
This Hamiltonian takes the similar form to the Hamiltonian in Eq.~(\ref{eq:BBH2}),
and thus can support zero-energy corner modes under open boundary conditions.
When $m_x=m_y$, it has the $C_4$ symmetry so that the quadrupole insulator is intrinsic.
Similar to the BBH model, the system is topologically nontrivial (trivial) when $|m_{0}|<1$ ($|m_0|>1$).

\paragraph{Topological characterization}\label{sec:boundary_Hamiltonian}
The higher-order topology of a quadrupole topological insulator can be characterized by its quantized quadrupole moment.
In the presence of the chiral symmetry, the system still supports quantized quadrupole moment as a topological invariant
in the presence of disorder (see Sec.~\ref{sec:Qxy}).

In addition, the authors evaluated the quantized polarization $p_x$ ($p_y$) of
effective boundary Hamiltonians (see the following introduction) for the $y$-normal ($x$-normal) boundary at half filling~\cite{Yang2021PRB}.
For disordered systems without translational symmetries, the polarization can be calculated according to
the Resta formula~\cite{Resta1998PRL,Prodan2014PRL}.
In the clean case of the model in Eq. (\ref{eq:Yang_model}), the derived effective boundary Hamiltonian at the $y$-normal ($x$-normal)
edges is equal to $H_x(k_x,m_x)$ [$H_y(k_y,m_y)$] multiplyed by a nonzero factor as shown in Ref.~\cite{Yang2021PRB}.
Evidently, the HOTP arises when these effective boundary Hamiltonians become topological with nontrivial polarizations $p_x=p_y=1/2$ which can be
evaluated by the Resta formula in real space.
Thus, the higher-order topology is characterized by the topological invariant
\begin{equation}
	P=4|p_x p_y|.
\end{equation}
When $P=1$, the system is in a HOTP, and when $P=0$, it is in a trivial phase.
Compared with the calculated quadruple moment in a system of $80\times 80$,
$P$ can be numerically evaluated in a system up to $500$.

We now introduce the effective boundary Hamiltonian proposed in Ref.~\cite{Oppen2017PRB}
and developed in Ref.~\cite{Yang2021PRB} for disordered systems.

In general, we can write the real-space Hamiltonian in a quasi-1D tight-binding form with only nearest-neighor couplings along one direction
when the unit cell is chosen to be large enough,
\begin{equation}\label{eq:tridiagonalHam}
	H=\left(\begin{array}{ccccc}
		h_{1} & V_{1}^{\dagger}\\
		V_{1} & h_{2} & V_{2}^{\dagger}\\
		& V_{2} & \ddots & \ddots\\
		&  & \ddots & h_{N-1} & V_{N-1}^{\dagger}\\
		&  &  & V_{N-1} & h_{N}
	\end{array}\right).
\end{equation}
Here, $h_n$ is the Hamiltonian for the $n$th unit cell and $V_n$ describes the hopping between the $n$th and $(n+1)$th unit cells for $n=1,\cdots,N$.
In disordered systems, the parameters in $h_n$ and $V_{n}$ can take random values.
The full Green's function of the system is defined as $\mathbf{G}(\omega)=(\omega \mathbb{I}-H)^{-1}$.
We will focus on the boundary Green's functions defined as the boundary blocks of the full Green's function matrix,
$G_{1}= \mathbf{G}_{11}$ and $G_{N}=\mathbf{G}_{NN}$, which can characterize topologically nontrivial gapped boundaries of HOTIs.

The boundary Green's function can be computed recursively through the Dyson equation~\cite{Oppen2017PRB}
\begin{equation}\label{eq:Dyson}
	\left(g_{N}^{-1} - V_{N-1}G_{N-1}V_{N-1}^{\dagger} \right) G_{N} = \mathbb{I}.
\end{equation}
Here, $G_N$ ($G_{N-1}$) is the boundary Green's function of a system with $N$ ($N-1$) unit cells,
and $g_{N}^{-1}(E) = \omega\mathbb{I}- h_{N}$ refers to the bare Green's function of the $N$th unit cell.
For a translational invariant system,
it is expected that when $N$ is sufficiently large, a fixed point $G$ will be approached for the boundary Green's function so that it satisfies the following
closed equation~\cite{Oppen2017PRB},
\begin{equation}\label{eq:fpGF}
	\left(g^{-1} - VGV^{\dagger} \right) G = \mathbb{I}.
\end{equation}
In numerical computations, the boundary Green's function can be evaluated efficiently
using the transfer matrix method.
The boundary Green's function encodes information about boundary states of TIs~\cite{Oppen2017PRB}.
We can define an effective boundary Hamiltonian from the boundary Green's function,
\begin{equation}
	H_{\mathrm{eff}} = -\left[G(E=0)\right]^{-1}.
\end{equation}
For a quadrupole topological insulator, the effective boundary Hamiltonians for both the $x$-edges and the $y$-edges have nontrivial topology.
Note that if the bulk Hamiltonian respects the chiral symmetry, the effective boundary Hamiltonian also preserves the chiral symmetry
so that we can use the 1D winding number or quantized polarization to characterize the edge topology~\cite{Yang2021PRB}.
This approach has also been applied to 3D third-order TIs with disorder~\cite{Castro2023arXiv}.

However, in disordered systems unlike the translational invariant systems in Ref.~\cite{Oppen2017PRB},
there is no fixed-point boundary Green's function in Eq.~(\ref{eq:fpGF}).
Instead, there are fluctuations due to randomness for the effective boundary Hamiltonian $H_{\mathrm{eff}}$ over iterations (\ref{eq:Dyson}).
The authors in Ref.~\cite{Yang2021PRB} use the statistical average over a large number of iteration steps to characterize the topology in disordered systems.
In each iteration,
the intra-cell hopping parts in $h_n$ and $V_{n}$ are randomly generated.
Specifically,
the polarization average along $x$ of the effective boundary Hamiltonian
at the $y$-normal boundary (similarly for $x$-normal one) is calculated,
\begin{equation}
p_x=\frac{1}{N_i}\sum_{n=1}^{N_i} |p_{x,n}|,
\end{equation}
where
\begin{equation}
p_{x,n}=\frac{1}{2\pi}\text{Im}\text{log}\langle \Psi_n|e^{2\pi i\hat{x}/L_x}|\Psi_n\rangle
\end{equation}
~\cite{Resta1998PRL} (note that the atomic positive charge contribution should be deducted),
$\hat{x}=\sum_{x}x\hat{n}_x$ with $\hat{n}_x$ denoting the particle number operator at the site $x$,
and $L_x$ is the system's length along $x$.
The polarization is evaluated for the many-body ground state $|\Psi_n\rangle$ of the boundary Hamiltonian
$H_{n}=-G_{2n}(E=0)^{-1}$ at half filling. Here, $G_{2n}$ is the $2n$th boundary
Green's function calculated via the Dyson equation. Note that one only considers the Green's functions
at even steps since two distinct layers exist in the clean limit.
As $H_{n}$ also preserves chiral symmetry, $p_{x,n}$ is still quantized to either $0$ or $0.5$
for each iteration.

\paragraph{Disorder-induced HOTIs}
To see the existence of higher-order TAI phase,
the authors consider a topologically trivial phase with $m_x=m_y>1$ in the clean limit.
The phase diagram with respect to the disorder strength is mapped out in Fig.~\ref{fig_YangHOTI1}(b),
showing remarkably the disorder-induced phase transition from a trivial insulator to a quadrupole topological insulator.
As shown in Fig.~\ref{fig_YangHOTI1}(b), the topological invariant $P$ experiences a sudden change from $0$ to $1$ near $W\approx 2.1$,
implying that a topological phase transition happens there.
The induced quadrupole insulator also manifests in the presence of corner states at zero energy.
When the disorder strength is further increased,
the system becomes gapless while still has quantized topological invariant $P$, which is named as the gapless HOTAI.
Then, the system will enter a regime with nonquantized values of $P$ due to the coexistence of
trivial and nontrivial random samples.
Finally, when the disorder is sufficiently strong, the topological invariant $P$ vanishes and the system
becomes a trivial Anderson insulator.

The results of the quadrupole moment and $P$ [see Fig.~\ref{fig_YangHOTI1}(b)]
show qualitative agreement with each other.
Yet, there exists apparent discrepancy, which arises from finite-size effects.
Though the quadrupole moment is quantized by chiral symmetry for a single random configuration,
the averaged quadrupole moment may not be quantized to $0$ or $1/2$ due to fluctuations.
A finite-size scaling performed in Ref.~\cite{Yang2021PRB} indicates that $q_{xy}$ will approach $1/2$ in the thermodynamic limit.

\paragraph{Self-consistent Born approximation (SCBA)}\label{sec:SCBA}

\begin{figure*}[t]
\includegraphics[width=0.9\linewidth]{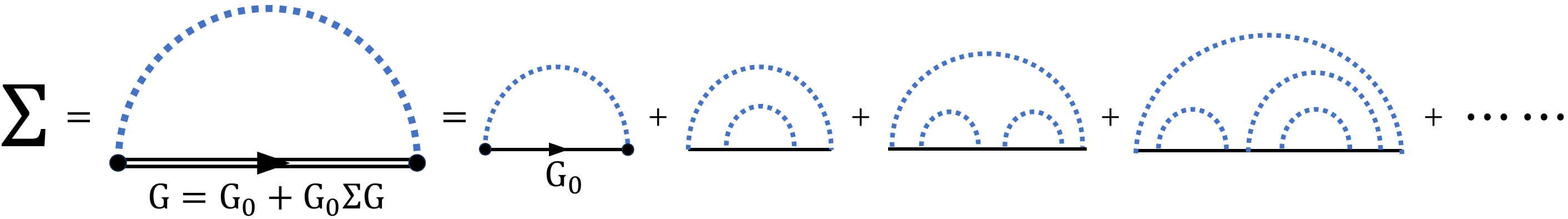}
\caption{
Schematic illustration of the rainbow diagrams for the self-energy $\Sigma$ in the SCBA.
The single solid lines represent the bare Green's function $G_0(E)\equiv (E-H_0)^{-1}$.
The double solid line represents the full Green' function $G(E)=\left(E-H_{0}-\Sigma \right)^{-1}$.
Each dotted line comes from two paired scattering lines from disorder due to ensemble averaging and contributes a factor about disorder strength.
Each vertex carries a scattering matrix over the internal degrees of freedom.}	
\label{fig_rainbow}
\end{figure*}

The disorder-induced topological phase transition can be understood based on the renormalization
of a Hamiltonian due to disorder~\cite{Beenakker2009PRL,Franz2010PRL}.
When disorder is weak, the phase boundaries modified by disorder can be evaluated by the SCBA method.
The method has been applied to study disorder-induced quantum spin Hall insulator from a metallic trivial phase~\cite{Beenakker2009PRL},
disorder-induced phase transition between 3D quantum anomalous Hall state and a Weyl semimetal~\cite{ChuiZhen2015PRL,Shindou2016PRL},
and disordered 3D topological insulators~\cite{shindou2009effects,Franz2010PRL,sbierski2014PRB}.
It has also been employed to explain the higher-order TAIs and identify their phase boundaries~\cite{Yang2021PRB,Li2020PRL}.
Figure~\ref{fig_YangHOTI1}(d) shows the phase boundaries of the disordered model in Eq.~(\ref{eq:Yang_model})
obtained by the SCBA method,
which agree well with the numerical phase boundary determined by the topological invariant.

We now introduce the effective medium theory with SCBA~\cite{Dattabook,Beenakker2009PRL,Franz2010PRL}.
For a Hamiltonian $H=H_0+V({\bf r})$ consisting of a clean system $H_0$ with translational symmetry
and a disorder perturbation $V({\bf r})$, the disorder effect at the energy $E$ can be described by the self-energy $\Sigma$
defined by
\begin{equation}\label{sigmadef}
\left(E-H_{0}-\Sigma \right)^{-1}=\langle(E-H)^{-1}\rangle,
\end{equation}
where $\langle \cdot \rangle$ denotes the disorder ensemble average.
Upon the disorder average, the system can still be described by a translational invariant Hamiltonian,
where the self-energy renormalizes parameters or create new terms.
To evaluate the self-energy, we use the SCBA which only takes into account contributions from rainbow diagrams with no crossed lines as shown in Fig.~\ref{fig_rainbow}.
The approximation neglecting the diagrams with crossed impurity lines usually works well in the regime of weak disorder.
In general, the self-energy $\Sigma$ as the summation of rainbow diagrams can be expressed in a recursive equation like
\begin{equation}\label{SE2}
\Sigma=W^2 U G(E) U,
\end{equation}
where $G=\left(E-H_{0}-\Sigma \right)^{-1}$ represents the full Green's function renormalized by disorder.
$G(E)$ is related to the bare Green's function $G_0(E)=(E-H_{0})^{-1}$ through Dyson series
yielding $G=G_0+G_0 \Sigma G$ [see double solid lines in Fig.~\ref{fig_rainbow}].
In the recursive equation, $W$ represents a factor proportional to the disorder strength
which is contributed from the ensemble average for a disorder term in the random Hamiltonian (see dotted lines in Fig.~\ref{fig_rainbow}).
$U$ represents the scattering matrix over the internal degrees of freedom due to the disorder term (see solid vertex in Fig.~\ref{fig_rainbow}).
The self-consistent equation (\ref{SE2}) for a single type of disorder term can be easily
generalized to cases for multiple disorder terms in the Hamiltonian.

We remark that the effective Hamiltonian obtained from the SCBA is a result of the
ensemble average of the Green’s function. It thus
does not correspond to the Hamiltonian for any single disorder configuration.
Nevertheless, in the deep region of a topological or trivial phase at weak disorder,
it is expected that the majority of random configurations have the similar topological
properties except for a small portion of samples due to random fluctuations. Therefore,
in these cases, the disorder averaged effective Hamiltonian can correctly describe the
bulk topology of most random configurations as well as the presence or absence of their boundary states.

Specifically, we consider the disordered model in Eq.~(\ref{eq:Yang_model}),
where the random intra-cell hopping term at each unit cell ${\bf r}$ reads
\begin{equation}
	\begin{aligned}
		V({\bm r})&=W\left[
		\begin{array}{cccc}
			0 & -iV^y({\bm r}) & -iV^x({\bm r}) & 0 \\
			iV^y({\bm r}) & 0 & 0 & i\bar{V}^x({\bm r}) \\
			iV^x({\bm r}) & 0 & 0 & -i\bar{V}^y({\bm r}) \\
			0 & -i\bar{V}^x({\bm r}) & i\bar{V}^y({\bm r}) & 0 \\
		\end{array}
		\right] \\
		&=W\left[\sum_{i = 1}^{4} V_{i}'({\bm r}) U_{i}\right].
	\end{aligned}
\end{equation}
Here, $V_1'=(V^x+\bar{V}^x)/2$, $V_2'=(V^x-\bar{V}^x)/2$, $V_3'=(V^y+\bar{V}^y)/2$, $V_4'=(V^y-\bar{V}^y)/2$,
and $U_1=\sigma_y\otimes\sigma_z$, $U_2=\sigma_y\otimes\sigma_0$, $U_3=\sigma_0\otimes\sigma_y$, $U_4=\sigma_z\otimes\sigma_y$.
For the independent random distributions considered, we require
\begin{eqnarray}
	\langle V_i'({\bm r}) \rangle&=&0   \\
	\langle V_i'({\bm r}_1) V_j'({\bm r}_2) \rangle &=&\frac{1}{24}
	\delta_{ij}\delta_{{\bm r}_1{\bm r}_2}
\end{eqnarray}
for $i,j=1,2,3,4$,
where $\langle \cdots \rangle$ represents the average over disorder configurations.

One thus can write the effective Hamiltonian at $E=0$ as
\begin{equation}
H_{\textrm{eff}}({\bm k})=H_0({\bm k})+\Sigma(E=0).
\end{equation}
Here, with disorder, one can evaluate the self-energy $\Sigma$ based on
the following self-consistent equation~\cite{Yang2021PRB}
\begin{equation}\label{SE}
	\Sigma(E)=\frac{W^2}{96\pi^2}\int_{\textrm{BZ} } d^2{\bm k} \sum_{n=1}^{4} U_n G(E) U_n,
\end{equation}
where the renormalized Green's function $G(E)=[(E+i0^+)I-H_0({\bm k})-\Sigma(E)]^{-1}$.
The integral is over the first Brillouin zone. This equation is equivalent to the general equation (\ref{SE2}) represented in momentum space.

The self-energy is independent of momentum and only renormalize the onsite terms.
At energy $E=0$, numerical results suggest the existence of only three terms in the self energy so that it can be written as~\cite{Yang2021PRB}
\begin{equation}
	\Sigma=i\Sigma_0 I + \Sigma_x \sigma_y\otimes\sigma_z + \Sigma_y \sigma_0\otimes\sigma_y,
\end{equation}
where $\Sigma_0,\Sigma_x,\Sigma_y$ take real values.
These new terms arising from disorder thus renormalize the mass terms
$m_x$ and $m_y$ as
\begin{align}
	m_x^\prime &= m_x + \Sigma_x, \\
	m_y^\prime &= m_y + \Sigma_y.
\end{align}
$\Sigma_x$ and $\Sigma_y$ thus cause topological phase transitions, since
the phase boundaries are determined by $m_x^\prime=1$ and $m_y^\prime=1$.

When disorder is weak, one can approximate the self-energy $\Sigma$ by taking $\Sigma=0$
on the right-hand side of Eq.~(\ref{SE}),
\begin{equation}
	\Sigma_\nu = -\frac{W^2}{48\pi^2}\iint_{\textrm{BZ} } d{\bm k} \frac{m_\nu+\sin(k_\nu)}{F({\bm k})},
\end{equation}
where
\begin{equation}
	F({\bm k}) = 2+\sum_{\nu=x,y}[m_\nu^2+2m_\nu\sin(k_\nu)]
\end{equation}
with $\nu=x,y$.
For $m_x>1$ and $m_y>1$,
the integrands are positive so that both $\Sigma_x$ and $\Sigma_y$ are negative.
The consequence is that for a sufficiently large disorder $W$,
the renormalized parameters will enter a topological nontrivial phase region
with $m_x^\prime<1$ and $m_y^\prime<1$ [see Fig.~\ref{fig_YangHOTI1}(d)].

\subsubsection{Band and localization properties}\label{sec:localization}
\paragraph{Introduction}

We now review some concepts and methods used in the study of localization transitions in disordered systems.

Disorder plays an important role in the behavior of materials, impacting not only
their topology, as previously discussed, but also their localization properties~\cite{Mirlin2008RMP}.
Sufficiently strong disorder can result in the absence of diffusion
in a 3D lattice with random onsite energies and
short-range coupling, as demonstrated by Anderson's pioneering work~\cite{Anderson1958}.
This phenomenon, known as Anderson localization, may arise from the
mismatch between the random potential and the hopping energy between electrons
at neighboring lattice sites~\cite{Anderson1958}.

A significant contribution to the understanding of localization is the
one-parameter scaling theory introduced by Abrahams et al.~\cite{Abrahams1979PRL}.
The key concept of this theory is the scaling function defined as
\begin{equation}
	\beta(\ln g) = \mathrm{d} \ln g/ \mathrm{d} \ln L,
\end{equation}
where $g$ is the dimensionless conductance and $L$ denotes the system size.
In this context, $g \gg 1$ and $g \ll$ 1 correspond to
extended and localized states, respectively.
By examining the asymptotic behavior of $\beta$ for large and small $g$,
one can establish an approximate expression of $\beta$ that
effectively covers all values of $g$.
In 3D systems, it is anticipated that a critical point $g_c$ exists,
separating regimes where $\beta$ is positive and negative for $g > g_c$ and $g < g_c$, respectively.
When the system size $L$ is increased, the dimensionless conductance $g$ tends to flow
towards $g = 0$ ($g = \infty$) if it initially resides in the regime $g < g_c$ ($g > g_c$).
As $g$ is inversely correlated to the strength of disorder, the result implies
the existence of a localization transition at finite disorder strength.
In 1D and 2Ds, however, it is conjectured that $\beta$ is always negative,
indicating that Anderson localization occurs for arbitrarily weak disorder.

Despite the broad applicability of the scaling theory, exceptional cases exist
in one and two-dimensional systems where states can be delocalized
in the presence of disorder.
Many of these cases are closely related to the topology of the system.
Notable examples include the boundaries of two or three-dimensional topological
insulators~\cite{Ludwig2008PRB,schnyder2009AIP,Gurarie2015JPA}.
In such cases, metallic edge or surface states persist, protected by bulk topology,
and remain immune to disorder as long as the bulk gap does not close.
Another example arises in 2D systems precisely tuned to the critical
point between two integer quantum Hall plateaus.
At this transition point, the states become critical and display universal conductance and multifractal behavior, even in the presence of disorder~\cite{Mirlin2008RMP}.

In fact, the nontrivial bulk topology of a topological insulator is intimately connected to the delocalization of
the boundary states against disorder respecting symmetry.
For example, the gapless surface states of a 3D topological insulator or the chiral edge states of a Chern insulator
are robust to scattering by random impurities.
The analysis of Anderson localization at the boundary provides
an alternative approach to obtain the Altland-Zirnbauer classification for topological insulators and superconductors~\cite{Ludwig2008PRB,schnyder2009AIP}.
In brief, to study the Anderson localization problem,
we consider noninteracting fermions in a random potential which can be described
by an action of the nonlinear sigma model (NL$\sigma$M) using replica trick in field theory.
Then, the boundary states can evade Anderson localization when a topological term is allowed in the action of NL$\sigma$M at the boundary,
which is determined by the homotopy group of the target space in NL$\sigma$M
and directly related to the classification of bulk topology~\cite{Ludwig2008PRB,schnyder2009AIP}.

Such intriguing localization properties can also manifest in disordered HOTIs.
However, before delving into these examples, it is necessary to introduce
several quantities used to characterize the localization properties of a system.
These include the localization length, level-spacing ratio (LSR),
inverse participation ratio (IPR), and fractal dimension.

The localization length serves as a measure of the extent
of localization for the eigenstates.
In the case of Anderson localization, the eigenstates are exponential localized
as
\begin{equation}
\psi(r) \sim e^{-r/\xi},
\end{equation}
where $\xi$ represents the localization length.
The localization length can be computed by partitioning the Hamiltonian into layers
and employing the formula~\cite{MacKinnon1983ZPB}
\begin{equation}
2/\xi = - \lim_{N\rightarrow \infty} N^{-1}\ln \mathrm{Tr} (G_{1N} G_{1N}^\dagger).
\end{equation}
Here, $G_{1N}$ is a submatrix of the Green's function $\mathbf{G}(\omega) = (\omega \mathbb{I} - H)^{-1}$ for a Hamiltonian
$H$ with $N$ layers [Eq.~(\ref{eq:tridiagonalHam})], which can be calculated by the iteration
$G_{1N} = G_{1, N-1} V_{N-1}^\dagger G_{N}$ and
$G_{N} = (\omega \mathbb{I} - h_{N} - V_{N-1} G_{N-1} V_{N-1}^\dagger)^{-1}$.
For two (or three) dimensions, the normalized localization length $\Lambda = \xi/L$
is commonly used, with $L$ (or $L^2$) denoting the size of the cross-section.
For extended (localized) states, $\Lambda$ increases (decreases) with $L$, while critical states exhibit a
constant value of $\Lambda$ with respect to $L$.

One can also use the LSR to identify the localization properties defined as~\cite{Huse2007PRB}
\begin{equation}
	r(E) = \frac{1}{N_E - 2} \sum_{i=1}^{N_E - 2}
	\mathrm{min}(\delta_i, \delta_{i+1})/\mathrm{max}(\delta_i, \delta_{i+1}),
\end{equation}
where $\delta_i = E_{i+1} - E_{i}$ represents the level spacing between adjacent energy levels
and $N_E$ denotes the number of selected energy levels around energy $E$.
For a random Hamiltonian belonging to the Gaussian orthogonal ensemble,
the LSR for extended states is given by $r \approx 0.53$~\cite{Huse2007PRB, Atas2013PRL}.
In the case of strong disorder where the states become localized, the LSR is
approximately $r \approx 0.386$, which arises from the Poisson statistics
of the level spacing~\cite{Huse2007PRB, Atas2013PRL}.

Another quantity to examine the localization property is the IPR defined by
\begin{equation}
	I(E) = \frac{1}{N_E} \sum_{i=1}^{N_E} \sum_{\bm r} \Big( \sum_\nu |\Psi_{E_i, {\bm r}\nu}|^2 \Big)^2,
\end{equation}
where $\nu$ represents the internal degrees of freedom.
The IPR can be used to determine whether the states around energy $E$ are localized or extended.
For an extended state, the IPR follows the scaling $I \propto 1/L^d$
where $d$ is the dimension of the system.
For example, a state that is evenly distributed in the full lattice has an IPR of $I = 1/L^d$.
While for a localized state, the IPR tends to be a constant as the system size increases
since the distribution of the state remains unchanged.
In the limiting case where the state is localized at a single site, the IPR is $I = 1$.
Based on the IPR, one can further define the fractal dimension $D_2$ such that
$I \propto 1/L^{D_2}$, with $D_2 = 0$ and $D_2 = d$ corresponding to localized
and extended states, respectively.
When $0 < D_2 < d$, the states are neither exponentially localized nor
fully extended, but exhibit multifractal behaviors~\cite{Castellani1986JPA}.

\begin{figure*}[t]
	\includegraphics[width=0.8\linewidth]{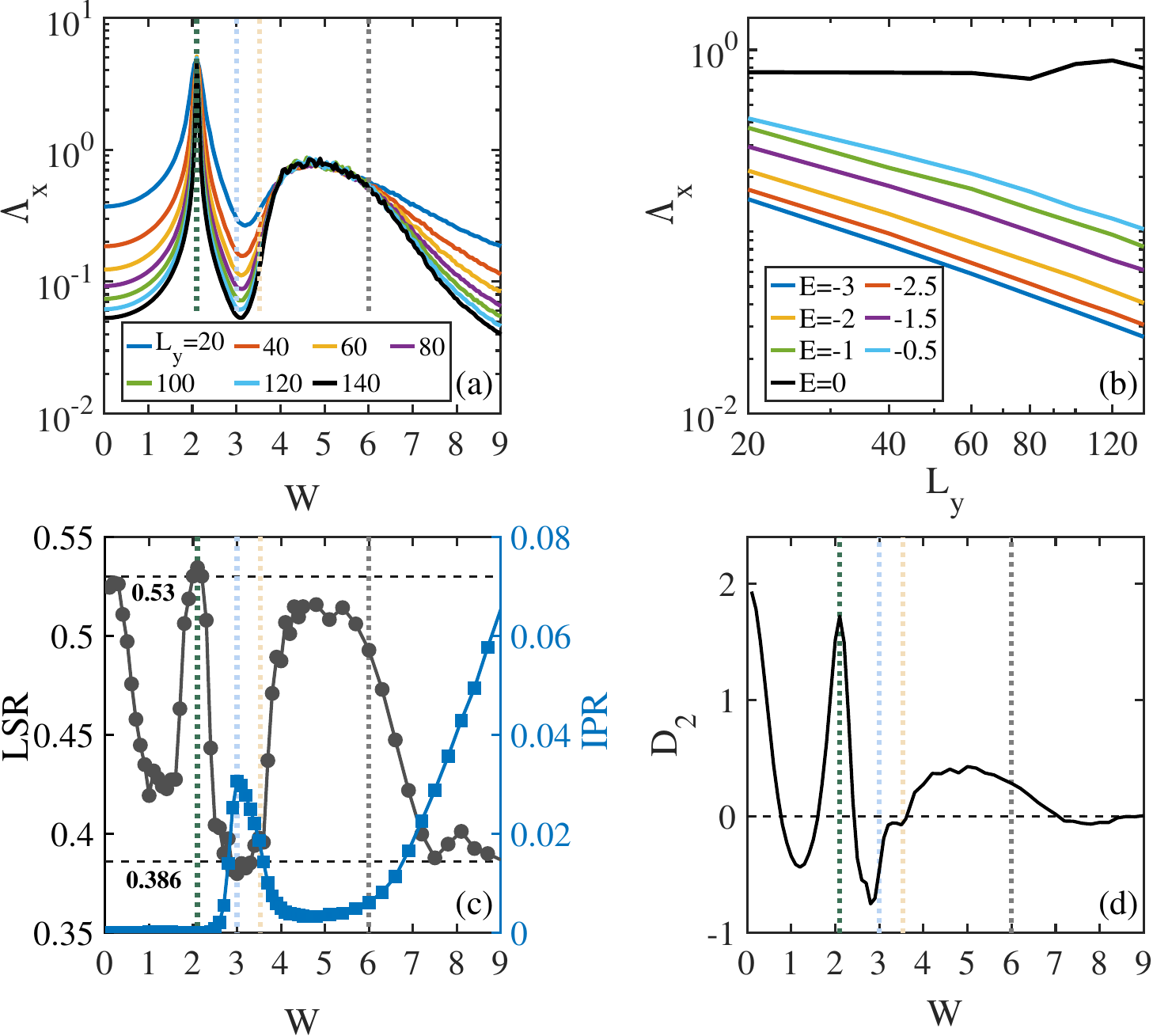}
	\caption{
		(a) The normalized localization length $\Lambda_x=\lambda_x/L_y$ at $E=0$ as a function of the disorder strength
		$W$ for several distinct $L_y$.
		(b) The scaling of $\Lambda_x$ at different energies when $W=4.6$, illustrating the multifractal behavior at
		zero energy and localized behavior at other energies.		
		(c) The LSR and IPR with respect to $W$ for the eigenstates around zero energy
		in a system with size $L=500$.
		(d) The fractal dimension $D_2$ as a function of $W$ for the eigenstates
		near zero energy.
		In (a) and (d), the vertical dashed lines mark out the boundaries of different phases shown in Fig.~\ref{fig_YangHOTI1}(b).
		Here, $m_x=m_y=1.1$. Reproduced from Ref.~\cite{Yang2021PRB}. }
	\label{fig_YangHOTI2}	
\end{figure*}

\paragraph{Localization properties at the topological transition point}
When $m_x=m_y$ in the model in Eq.~(\ref{eq:Yang_model}), the disorder-induced
higher-order topological phase transition occurs through the bulk energy gap closure,
as shown in Fig.~\ref{fig_YangHOTI1}(c). In the ensemble of disordered systems,
there in fact emerges an average $C_4$ symmetry, that is,
a Hamiltonian $\hat{H}$ for a certain disorder configuration occurs with the same probability as
its symmetry conjugate partner $\hat{C}_4\hat{H} \hat{C}_4^{-1}$.
The average symmetry enforce the bulk energy gap closure (see Sec.~\ref{sec:Amorphous} for detailed discussion).
When $m_x \neq m_y$, the symmetry is broken so that the topological phase transition
occurs via the edge energy gap closing~\cite{Li2020PRL}.

At the critical point, the normalized localization length $\Lambda_x=\lambda_x/L_y$ (similarly for $\lambda_y/L_x$) at zero energy
becomes scale free, indicating the divergence localization length in the thermodynamic limit, as shown in Fig.~\ref{fig_YangHOTI2}(a).
This is associated with the delocalized states occurring at the phase transition.
On the other hand, at both sides of the critical point, the normalized localization length
decreases as the transverse size increases,
indicating the localization of states around zero energy.


The localization properties can also be confirmed by looking at the energy level statistics
from the LSR and the real-space distribution from the IPR (see Sec.~\ref{sec:localization}).
The LSR of zero-energy states at the criticality is close to the value $0.53$, corresponding to extended states [see Fig.~\ref{fig_YangHOTI2}(c)].
The fractal dimension $D_2$ from the scaling of the IPR is close to the bulk dimension $2$, indicating the states are extended in the bulk [see Fig.~\ref{fig_YangHOTI2}(d)].
Therefore, the topological phase transition accompanies a localization-delocalization-localization (LDL) transition with the bulk gap closing.

\begin{figure*}
	\includegraphics[width=0.8\linewidth]{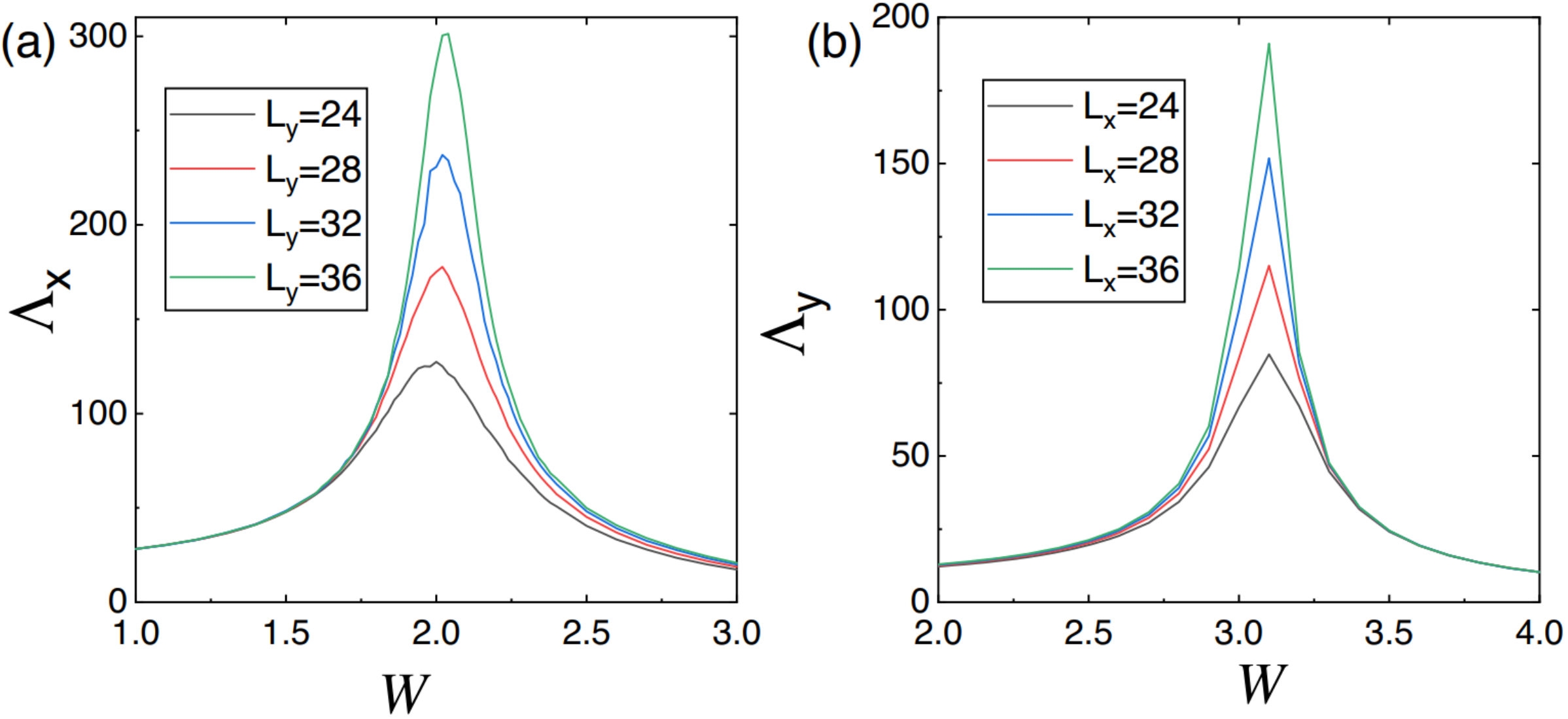}
	\caption{The localization-delocalization-localization transition at the boundaries: finite-size scaling of the unnormalized localization length at zero energy
		along the $x$ and $y$ direction.
		(a) $\Lambda_{x}$ and (b) $\Lambda_{y}$ as a function of the disorder strength $W$ with open boundary condition along transverse direction.
		From Ref.~\cite{Li2020PRL}.}
	\label{fig:Li_LDL_transition}
\end{figure*}

For the model studied in Ref.~\cite{Li2020PRL}, the disorder-driven phase transition between topologically trivial and nontrivial phase
is accompanied by a LDL transition with gap closing at open boundaries while the bulk remains insulating,
corresponding to the case with $m_x \neq m_y$ in the model in Eq.~(\ref{eq:Yang_model}).
Li et al. performed the finite-size scaling of the localization length at zero energy for a quasi-one-dimensional ribbon
with open boundaries in the transverse dimension.
As shown in Fig.~\ref{fig:Li_LDL_transition}, around the transition point, the localization length increases
monotonically with respect to the ribbon width, which signifies the divergence in the thermodynamic limit.
The divergence of the localization length under open boundary conditions indicates the occurrence
of zero-energy delocalized states at the critical point due to gap closing along the corresponding
boundaries parallel to the longitudinal direction.
In contrast, with a periodic boundary condition along transverse direction,
the normalized localization length decreases as the width increases, which indicates the absence of delocalized bulk states at the phase transition.

\paragraph{The gapless regime}

For sufficiently strong disorder, the system becomes gapless, leading to gapless higher-order
TAI, trivial-II phase and the Griffiths regime [see Fig.~\ref{fig_YangHOTI1}].
In the former two phases, the states near zero energy are localized, as revealed by the decrease of the localization length
with the transverse size in Fig.~\ref{fig_YangHOTI2}(a).
The localized nature of the states around zero energy can also be evidenced by their
LSR approaching $0.386$, and the relatively larger IPR than the regime with weak disorder [see Fig.~\ref{fig_YangHOTI2}(c)].
More precisely, the localization behavior should be characterized by the fractal dimension $D_2$
obtained from the scaling of the IPR versus the system size, which provides more information than the magnitude of IPR itself. As shown in Fig.~\ref{fig_YangHOTI2}(d),
the fractal dimension $D_2$ is around zero reflecting the states near zero energy is localized in the gapless higher-order topological regime.
Further finite-size analysis indicates that $D_2$ approaches zero in the thermodynamic limit~\cite{Yang2021PRB}.
The authors further found that the states of higher energy are all localized via the LSR and localization length calculations,
which indicates that the higher-order topology can be carried by the localized bulk states.
For other models, the existence of the mobility edge in the gapless
HOTP has also been found~\cite{ChenCZ2021PRB}

The Griffiths regime (or critical regime) has more interesting property that
the zero-energy normalized localization length displays scale invariant behavior as the transverse size increases,
suggesting a multifractal phase therein [see Fig.~\ref{fig_YangHOTI2}(a) and (b)].
This multifractal regime appears as a critical region residing between two localized phases,
which differs from the conventional case with a critical point between delocalized and localized phase.
In the Griffiths regime, other states at nonzero energy remain localized [see Fig.~\ref{fig_YangHOTI2}(b)].
The multifractal behavior at zero energy can be verified by the LSR close to $0.53$ [see Fig.~\ref{fig_YangHOTI2}(c)]
and the fractal dimension $D_2$ between $0$ and $2$ [see Fig.~\ref{fig_YangHOTI2}(d)].
Similar localization properties have also been observed in the case with $m_{x}=m_{y}<1$.

\begin{figure*}[t]
	\centering
	\includegraphics[width=0.8\linewidth]{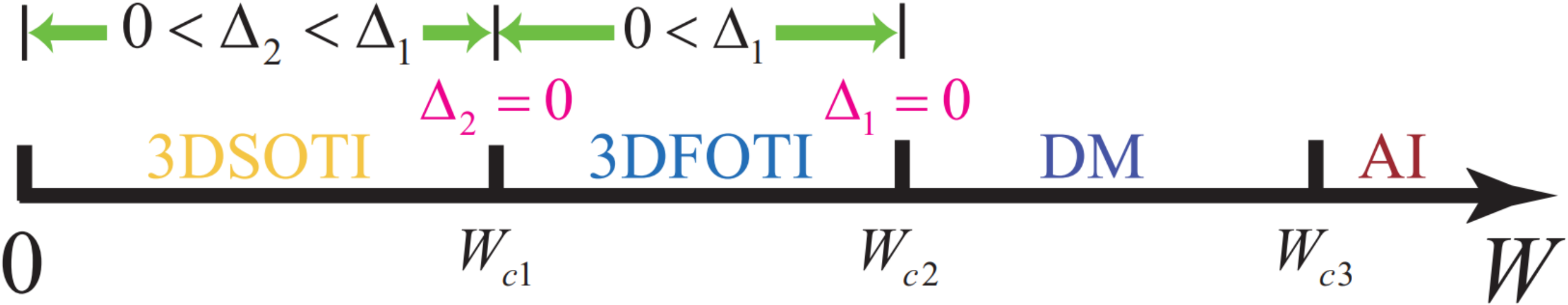}
	\caption{Schematic illustration of the phase transitions in a disordered 3D SOTI.
		There are four different phases with increasing disorder: (1) 3D SOTIs with hinge states;
		(2) 3D first-order TIs (FOTIs) with surface states;
		(3) diffusive metals (DMs) with extended states in the bulk;
		(4) Anderson insulators (AIs) with all states localized.
		Here, $\Delta_1$ and $\Delta_2$ denote the gaps of the bulk and surface
		states, respectively. From Ref.~\cite{WangXR2020PRR}.
	}
	\label{fig_dis3DSOTI_phase}
\end{figure*}

\subsection{Disordered three-dimensional second-order topological insulators}\label{sec:Dis3DSOTI}

In 3Ds, SOTIs usually support chiral hinge modes or helical hinge modes~\cite{Langbehn2017PRL,Song2017PRL,Schindler2018SA}. Note that it has also been found that
non-chiral hinge modes can appear in a higher-order Klein bottle topological insulator
protected by momentum-space glide reflection symmetry~\cite{XuChen2023arXiv}. Here, we will present the impact of onsite disorder
on chiral hinge modes in a 3D system with reflection symmetry~\cite{WangXR2020PRR} and helical hinge modes in a 3D weak SOTI system~\cite{WangXR2021PRB}.
We will discuss 3D SOTIs in amorphous lattices and structural disorder induced higher-order topological phase
transition in Sec.~\ref{sec:Amorphous}.

For a 3D SOTI protected by reflection symmetry~\cite{Langbehn2017PRL},
the authors in Ref.~\cite{WangXR2020PRR} have studied the effects of onsite disorder (white noise) on the topological
insulator. Note that such a white noise breaks the reflection symmetry for each sample, but the symmetry should be preserved
on average (see the discussion in Sec.~\ref{sec:Amorphous}). It is found that
disorder will drive a series of transitions from a SOTI to a 3D first-order topological insulator, then to a diffusive metal,
and finally to an Anderson insulator as shown in Fig.~\ref{fig_dis3DSOTI_phase}.
Specifically, at weak disorder, the SOTI has the quantized conductance of $G=e^2/h$ contributed by robust chiral hinge states,
although disorder breaks the reflection symmetry.

Meanwhile, due to the linear dispersion of chiral hinge states, the SOTI phase can be characterized by a constant density of states at weak disorder.
As the disorder increases, the surface gap closes at a critical point. As a result, the system transitions into the first-order
topological insulating phase characterized by the dominate local density of states on surfaces.
At a higher critical disorder, the bulk energy gap closes and the system turns into a diffusive metallic phase with nonzero density of states
at the Fermi level.
Finally, when the disorder is very strong, the system undergoes a 3D metal-insulator transition to an Anderson insulator.
The phase boundaries can be obtained by the SCBA method as well~\cite{WangXR2020PRR}.

For a 3D weak SOTI with TRS in the clean limit, a
first-order topological invariant of the valley Chern number is used to characterize the helical hinge states~\cite{WangXR2021PRB}.
With disorder respecting TRS, the valley Chern number is not well defined due to the absence of translational symmetry,
and the conductance is no longer quantized due to inter-valley scattering~\cite{WangXR2021PRB}.
Yet, the authors use the distribution of a state on hinges to characterize the hinge states and find
that its value remains almost unchanged for weak disorder, suggesting that the hinge states
are robust against weak disorder.
As disorder increases, the system transitions from the weak SOTI to a weak first-order topological insulator,
then to a diffusive metal, and finally turns into an Anderson insulator at strong disorder.

\begin{figure}[t]
	\centering
	\includegraphics[width=1.0\linewidth]{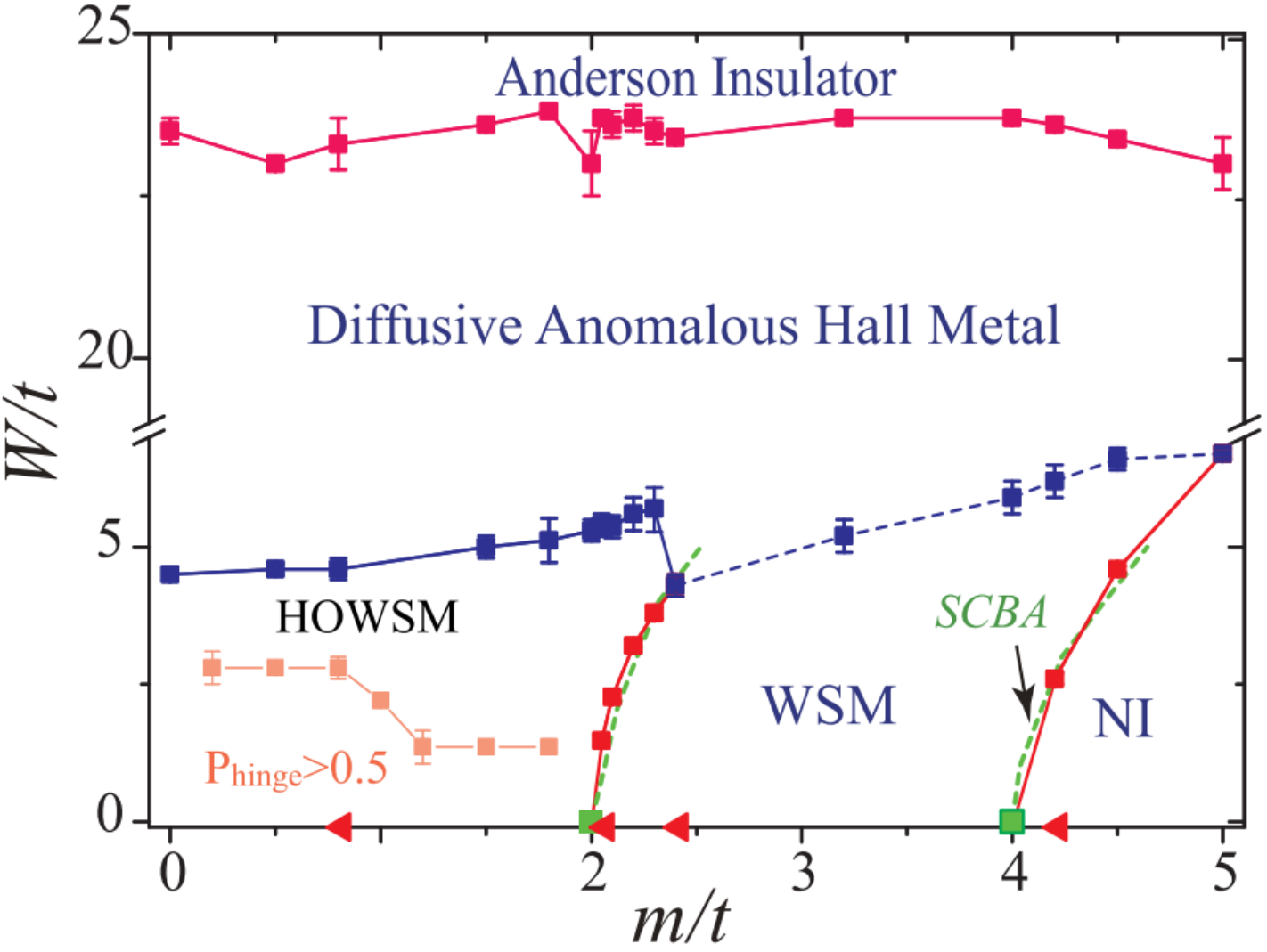}
	\caption{The phase diagram with respect to the onsite disorder strength $W$ and the mass $m$.
		The square points are obtained by the localization length scaling. The green dashed lines describe the
		phase boundaries determined by the SCBA method.
	From Ref.~\cite{JiangH2021PRB}.}
	\label{fig:disHOWSM_phase}
\end{figure}

\subsection{Disordered higher-order topological semimetals}\label{sec:DisHOTSM}

In Sec.~\ref{Sec:HOSM}, we introduce how higher-order Dirac and Weyl semimetals arise from the stacking
of the BBH model. Here, we discuss how disorder affects the phase diagram.
Consider the model described by the following Bloch Hamiltonian in the clean case~\cite{Ezawa2018PRB,Roy2019PRR}
\begin{equation}
	\begin{aligned}
		{H}&= [m+t_x\cos k_x+t_y\cos k_y+t_z \cos k_z]\tau_z\sigma_0\\
		&+B\tau_0\sigma_z+\lambda \sin k_x\tau_x\sigma_x+\lambda \sin k_y\tau_x\sigma_y\\
		&+\Delta (\cos k_x-\cos k_y)\tau_y\sigma_0,
	\end{aligned}
\end{equation}
where $m$, $t_x$, $t_y$, $t_z$, $B$, $\lambda$ and $\Delta$ are system parameters. When $B=0$,
the Hamiltonian respects chiral symmetry represented by $\tau_x \sigma_z$ so that a higher-order Dirac
semimetal can emerge. The $B\tau_0\sigma_z$ term breaks the chiral symmetry while preserves the
particle-hole symmetry represented by $\tau_y\sigma_y\kappa$, splitting a Dirac points into a pair of
Weyl points. The Hamiltonian can also realize a higher-order Weyl semimetal with only a pair of Weyl points.

The authors in Ref.~\cite{JiangH2021PRB} study the effects of onsite disorder on the higher-order
Weyl semimetals and map out its phase diagram with respect to the disorder strength $W$
and the parameter $m$ as shown in Fig.~\ref{fig:disHOWSM_phase}. Five phases are found, including
higher-order Weyl semimetals, Weyl semimetals, diffusive anomalous Hall metals, normal insulators
and Anderson insulators, by identifying their normalized localization length calculated by the transfer matrix method.
Since the disorder breaks the particle-hole symmetry, the quadrupole moment is fragile to weak disorder.
They also show that the hinge charge is unstable against disorder.
However, numerical results show that the Hall conductance remains unchanged in the semimetal phase,
indicating the stability of Weyl points against disorder.
The stability of hinge modes is also demonstrated using the neural network method.

From the phase diagram, we see that disorder can induce a phase transition from a Weyl semimetal
to a higher-order Weyl semimetal or from a normal insulator to a Weyl semimetal.
The SCBA method is also utilized to determine the phase boundaries at weak disorder.

\section{Higher-order topological phases in non-crystalline systems}\label{sec4}

We are now in a position to explore HOTPs in non-crystalline lattices,
including amorphous lattices, quasicrystalline lattices, hyperbolic lattices, and fractal lattices.

\subsection{Amorphous lattices}{\label{sec:Amorphous}}

\begin{figure*}[t]
\includegraphics[width=0.6\linewidth]{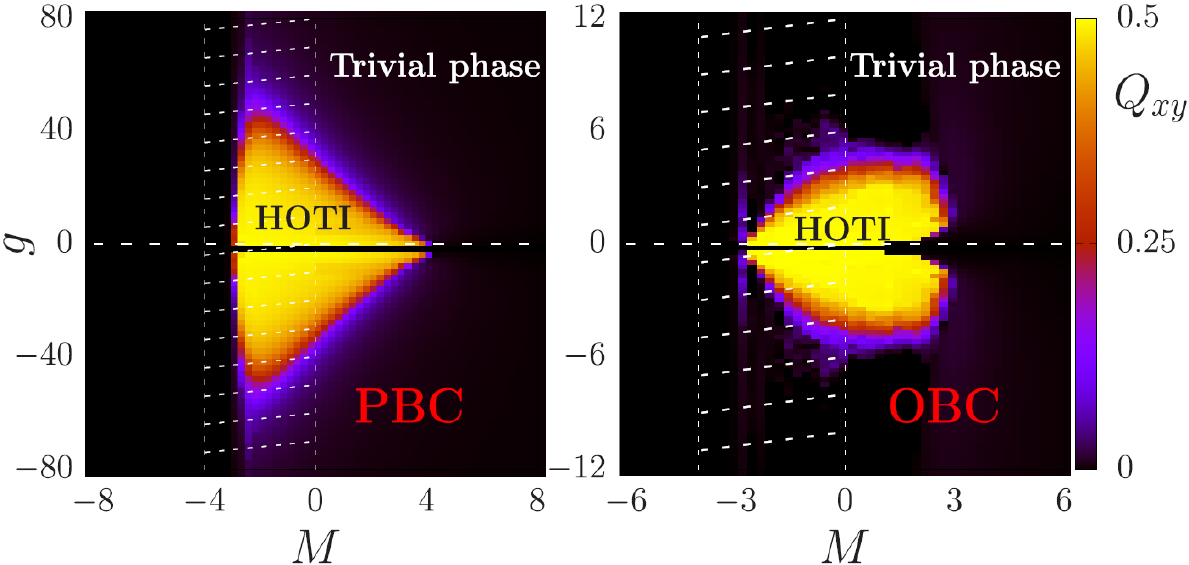}
\caption{
The phase diagram of the tight-binding model in Eq.~(\ref{Ham2}) on amorphous lattices
for a fixed scrambling radius $R_s=6$
with respect to the parameters $M$ and $g$, which is obtained by evaluating the quadrupole moment
under periodic boundary conditions (left) and open boundary conditions (right).
The HOTI phase in the crystalline system is marked by the white dashed region.
From Ref.~\cite{Agarwala2020PRR}.}
\label{figAmorphQTI}
\end{figure*}

\begin{figure*}[t]
	\includegraphics[width=0.8\linewidth]{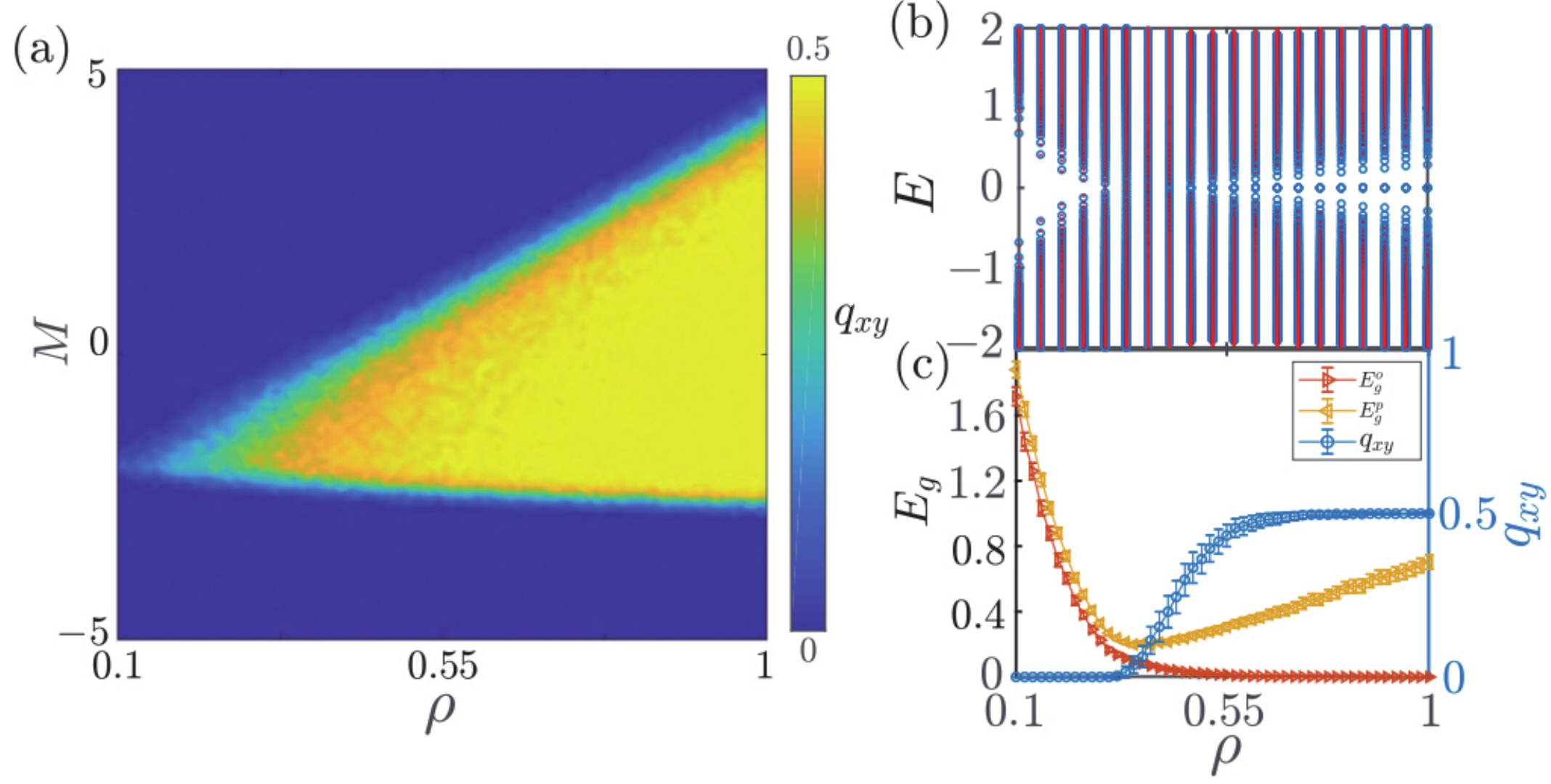}
	\caption{(a) The phase diagram of the tight-binding model in Eq.~(\ref{Ham2}) on amorphous lattices
		with respect to the density $\rho$ and the mass $M$, which is mapped out by the quadrupole moment $q_{xy}$.
		(b) The energy spectrum under open boundary conditions (blue circles) and periodic boundary conditions
		(red circles) at $M=0$.
		(c) The energy gap (red line for open boundaries and yellow line for periodic boundaries) and the quadrupole moment
		$q_{xy}$ (blue line) with respect to the density $\rho$ at $M=0$. From Ref.~\cite{Zhou2022PRBa}.}	
	\label{Fig1Zhou}
\end{figure*}

\subsubsection{Two dimensions}
Higher-order topological insulators have been first studied in amorphous lattices
based on the following tight-binding model~\cite{Agarwala2020PRR}
\begin{equation}\label{Ham2}
	\hat{H}=\sum_{\bm r} \left[\hat{c}_{\bm r}^\dagger T_0\hat{c}_{\bm r}+\sum_{\bm d}
	\hat{c}_{\bm r+\bm d}^\dagger
	T({\bm d})\hat{c}_{\bm r}	\right ],
\end{equation}
where $\hat{c}_{\bf r}^\dagger=(\hat{c}_{{\bf r},1}^\dagger,\hat{c}_{{\bf r},2}^\dagger,
\hat{c}_{{\bf r},3}^\dagger,\hat{c}_{{\bf r},4}^\dagger)$
with $\hat{c}_{{\bf r},\alpha}^\dagger$ being a creation operator of an electron of the $\alpha$th component
at the site of position ${\bf r}$,
the onsite term $T_0=\tau_3 \sigma_0(M+2t_2)$ and the hopping matrix
\begin{eqnarray}
T({\bm d})=&&\frac{t(|\bm{d}|)}{2} \left[-it_1(\tau_1 \sigma_3\cos{\theta}
+\tau_2 \sigma_0\sin{\theta})-t_2\tau_3 \sigma_0  \right. \nonumber \\
&&\left.+g\cos{(2\theta)}\tau_1 \sigma_1 \right]
\end{eqnarray}
 with
 $|\bm{d}|$ and $\theta$ denoting the distance between two sites and the polar angle of the separation vector, respectively.
Here, $t(|\bm{d}|)=\Theta(R-|\bm{d}|)\exp(-|\bm{d}|/d_0)$ with $\Theta(R-|\bm{r}|)$ being the
Heaviside step function to cut off hoppings longer than $R$.
The lattice constant $d_0$ is taken to be $d_0=1$.
For a square lattice including only nearest-neighbour hoppings with $t(|\bm{d}_{NN}|)=1$, the Bloch Hamiltonian reduces to
\begin{align}\label{Ham1}
	H(\bm{k})&=-t_1(\sin{k_x}\tau_1 \sigma_3+\sin{k_y}\tau_2 \sigma_0)\notag\\
	&+[M+t_2(2-\cos{k_x}-\cos{k_y})]\tau_3\sigma_0\notag\\
	&+g(\cos{k_x}-\cos{k_y})\tau_1 \sigma_1,
\end{align}
which is equivalent to the Dirac Hamiltonian in Eq.~(\ref{eq:H_BBH_C4T}).

To study the topological phases in amorphous lattices, the authors in Ref.~\cite{Agarwala2020PRR} consider completely randomly
distributed sites within a scrambling radius in a square.
Figure~\ref{figAmorphQTI} shows the phase diagrams obtained from the quadrupole moment under different boundary conditions in the $(M,g)$ plane.
It is found that the Hamiltonian in Eq.~(\ref{Ham2}) on amorphous
lattices can still host four corner states and its quadrupole moment is quantized at $Q_{xy}=0.5$,
indicating the existence of HOTI on amorphous lattices.

Based on the model (\ref{Ham2}), density-driven higher-order topological phase transitions are observed in Ref.~\cite{Zhou2022PRBa}.
Specifically, the authors map out the phase diagram in the $(\rho,M)$ plane ($\rho$ is the density of sites and
$M$ is the mass) based on the quadrupole moment $q_{xy}$
[see Fig.~\ref{Fig1Zhou}(a)].
The figure clearly illustrates that a system in a trivial phase at low density (e.g. $\rho=0.24$) will
become a nontrivial one at high density (e.g. $\rho=1$).
Figure~\ref{Fig1Zhou}(b) and (c) also illustrate that as the density increases,
the bulk energy gap closes and reopens at the critical density $\rho\approx 0.41$ and the
quadrupole moment changes from 0 to 1, resulting in the corner modes.

Recently, the authors in Ref.~\cite{Tao2023Average} find that amorphous systems can support
HOTPs without crystalline counterparts protected by an average $C_pT$ symmetry.
For instance, the HOTIs can host eight or twelve corner modes, which cannot exist in crystalline systems.
Specifically, they consider the Hamiltonian with $T_0=m_z\tau_3\sigma_0$
and
\begin{eqnarray} \label{Eq:2DAmorHopping}
T(\bm d)=&&f(|\bm d|) \left[t_0\tau_3\sigma_0+it_1\left(\cos \theta \tau_1\sigma_1+\sin \theta
\tau_1\sigma_2\right) \right. \nonumber \\
&&\left.+g\cos (p\theta/2)\tau_2\sigma_0\right],
\end{eqnarray}	
where $p$ takes the value of a multiple of $4$ and $f(|\bm d|)=\Theta(R-|\bm d|)\exp[-\lambda(|\bm d|/d_0-1)]/2$.
The lattice sites are positioned completely randomly with a hardcore radius $r_h=0.2$ in a regular
$p$-gon.
Although each configuration of lattice sites lacks $p$-fold rotation
($C_p$) symmetry due to the randomness of site positions, the ensemble of sample configurations
respects average $C_p$ symmetry.
As a result, the ensemble of Hamiltonians respects average $C_pM$ and $C_pT$ symmetry, i.e.
$(C_pM)H(\mathcal{R})(C_pM)^{-1}=(C_pT)H(\mathcal{R})(C_pT)^{-1}=H(D_{C_p}\mathcal{R})$,
where $M=\tau_z\sigma_z$, $T=i\sigma_y\kappa$
and $H(\mathcal{R})$ is the Hamiltonian on lattice configuration $\mathcal{R}$ (see Sec.~\ref{sec:3DASOTI}
for detailed discussions on average symmetry).

\begin{figure*}[t]
\includegraphics[width=0.7\linewidth]{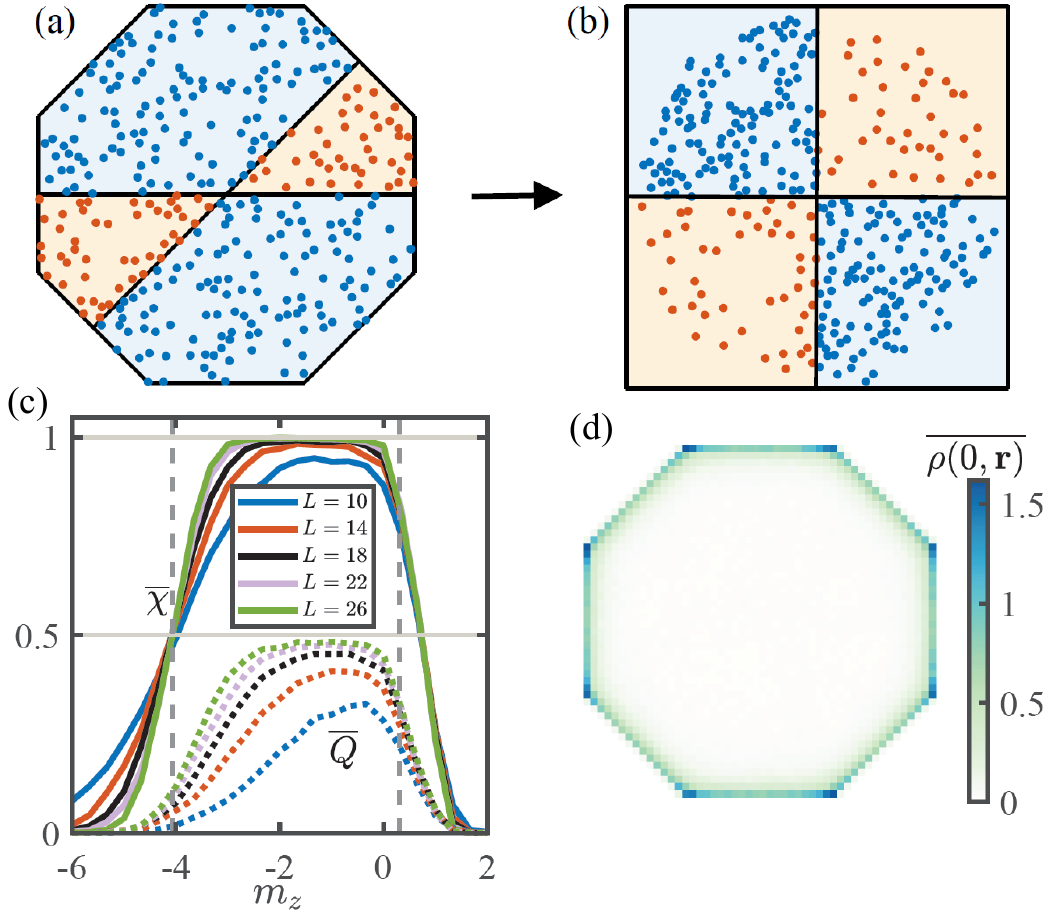}
\caption{
(a),(b) Schematics of changing site positions in a regular octagon with $p=8$ for correctly computing the quadrupole moment.
The site positions in the orange (blue) region in (a) are transformed to the first or third (second or fourth) quadrant in (b).
The quadrupole moment is calculated using the deformed lattices and the original eigenstates.
(c) The configuration averaged $\mathbb{Z}_2$ topological invariant $\overline{\chi}$ (solid
lines) and quadrupole moment $\overline{Q}$ (dotted lines) calculated from the reconstructed lattices of different sizes with respect to $m_z$ for
the amorphous HOTI model with the average $C_8 M$ symmetry.
(d) The configuration averaged local DOS at zero energy in the amorphous HOTI phase on a regular octagon.
Reproduced from Ref.~\cite{Tao2023Average}.
}
\label{FigAveAmorph}
\end{figure*}

To characterize the topological properties, the authors build a $\mathbb{Z}_2$ topological invariant
based on the $C_p M$ symmetry by imposing twisted boundary conditions~\cite{Tao2023Average}. In addition,
since the model
has the chiral symmetry represented by $\tau_1\sigma_3$, one may ask whether the topological phase
can be characterized by the traditional quadrupole moment defined in Eq.~(\ref{eq:Qxy}).
However, when directly evaluating the quadrupole moment, it consistently yields zero.
This may be attributed to the presence of an even number of corner states in each quadrant within a nontrivial phase.
To address this issue, the authors in Ref.~\cite{Tao2023Average} propose a solution whereby the positions of sites within a $1/p$
sector and its opposite counterpart are relocated to the first and third quadrants, respectively. Likewise,
the sites in other sectors are relocated to the second and fourth quadrants, respectively [see Fig.~\ref{FigAveAmorph}(a,b)
for an example of an octagon geometry with $p=8$].

The authors evaluate the averaged $\mathbb{Z}_2$ topological invariant $\overline{\chi}$ and the
quadrupole moment $\overline{Q}$ based on the transformed site positions, which correctly characterize the higher-order topological amorphous insulators
in an octagon with $p=8$,
as shown in Fig.~\ref{FigAveAmorph}(c).
The HOTI phase is further supported by the existence of corner modes indicated by
the profile of local density of states (DOS) [see Fig.~\ref{FigAveAmorph}(d)].
Moreover, the average symmetry protected HOTIs are found to exist in amorphous lattices of a dodecagon for $p=12$ as well~\cite{Tao2023Average}.
In addition, it is shown that the higher-order topology in 2D quasicrystalline lattices can also be characterized by the generalized quadrupole moment~\cite{MaoYF2023arXiv}.
In 3Ds, the winding number of the generalized quadrupole moment and a $\mathbb{Z}_2$ invariant
based on transformed site positions are defined to characterize the chiral and helical modes without
crystalline counterparts~\cite{MaoYF2023arXiv} (see Sec.~\ref{sec:Quasicrystal} for a detailed discussion).

\begin{figure*}[t]
	\includegraphics[width=0.8\linewidth]{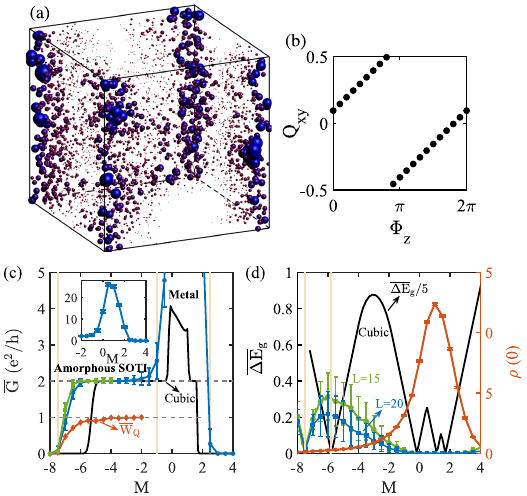}
	\caption{(a) The density distribution of four states near zero energy in the SOTI phase, exhibiting the hinge modes.
		(b) The quadrupole moment $Q_{xy}$ with respect to the flux $\Phi_z$, reflecting the nontrivial
		winding number of the quadrupole moment.
		(c) Sample averaged two-terminal conductance $\overline{G}$ [blue (system size $L=30$) and green lines ($L=40$)] and winding number of the
		quadrupole moment $\overline{W}_Q$ [red line ($L=16$)] of the 3D amorphous
		model as a function of the mass $M$. The black line describes the conductance of the Hamiltonian on a cubic lattice.
		(d) Sample averaged bulk energy gap (green and blue lines) of the 3D amorphous model with respect to $M$ compared with the
		results (black line) on a cubic lattice The red line depicts the DOS at zero energy. Reproduced from Ref.~\cite{WangJH2021PRL}.
	}
	\label{FigWang_SOTI1}
\end{figure*}

\begin{figure*}[t]
	\includegraphics[width=0.8\linewidth]{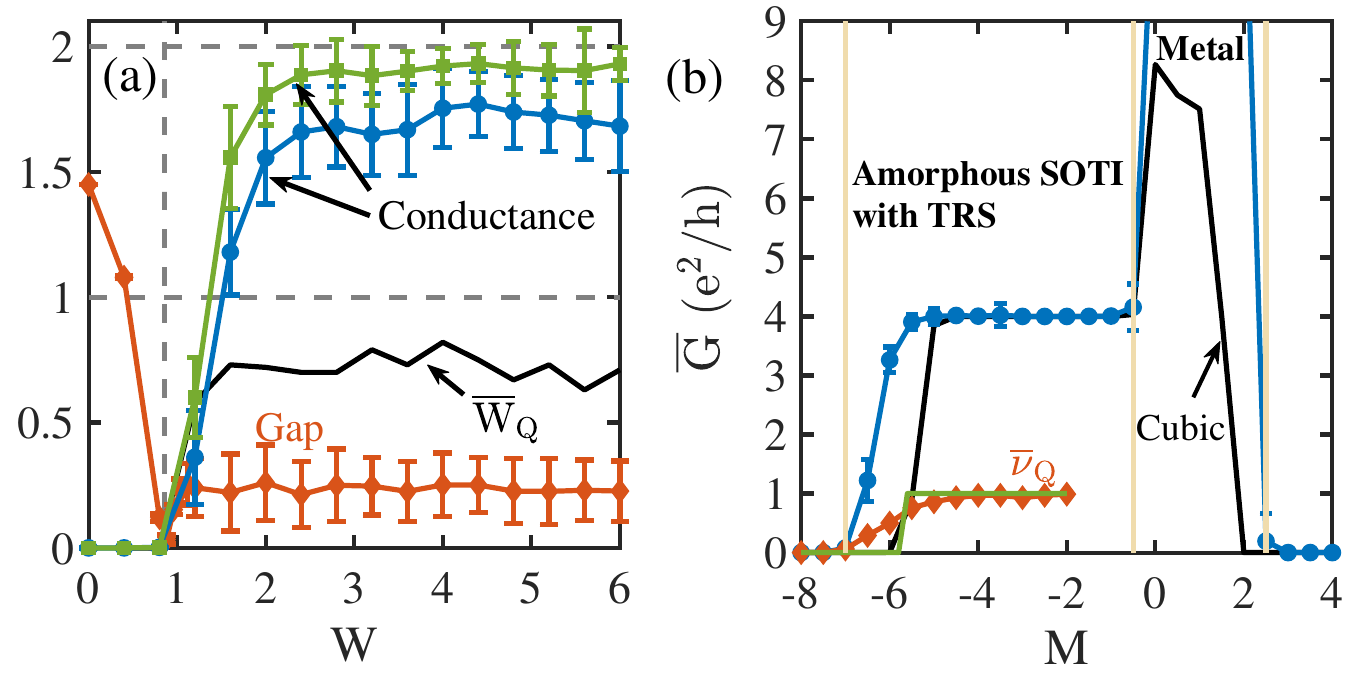}
	\caption{(a) Structural disorder induced SOTI, illustrated by the sample averaged two-terminal conductance $\overline{G}$ (blue and green lines),
		winding number of the quadrupole moment $\overline{W}_Q$ (black line) and bulk energy gap (red line)
		of the 3D amorphous model without TRS with respect to the disorder strength $W$.
		(b) Sample averaged two-terminal conductance $\overline{G}$ (blue line) and $\mathbb{Z}_2$
		topological invariant (red line) of the 3D amorphous model with TRS as a function of $M$.
		The black and green lines describe the conductance and $\mathbb{Z}_2$ topological invariant of the model
		on a cubic lattice, respectively. Reproduced from Ref.~\cite{WangJH2021PRL}.
	}
	\label{FigWang_SOTI2}
\end{figure*}

\subsubsection{Three dimensions}\label{sec:3DASOTI}
In Sec.~\ref{Subsec:Chiral} and \ref{Subsec:Helical}, we present a chiral and helical model
in 3D crystalline systems to show the existence of chiral and helical hinge modes.
There, we see that the crystalline symmetry such as the $C_4^z T$ symmetry is required
to protect the quantization of the Chern-Simons invariant, implying that the topological
invariant is not well defined if we consider amorphous lattices, which break the symmetry.
Fortunately, the winding number of the quadrupole moment is not dependent on any
symmetry, making it possible to realize chiral hinge modes in amorphous lattices.
Indeed, the authors in Ref.~\cite{WangJH2021PRL} predict the existence of SOTIs without TRS
[see the hinge states in Fig.~\ref{FigWang_SOTI1}(a)] and
with TRS in amorphous lattices and structural disorder induced second-order topological
phase transition.

\paragraph{Amorphous second-order topological insulator}
The tight-binding Hamiltonian $\hat{H}_c$ studied in Ref.~\cite{WangJH2021PRL} includes the
onsite mass term $T_0=M\tau_z\sigma_0$ and the hopping matrix
\begin{equation}
T_c(\bm d)=f(|\bm d|)
\left[ t_0\tau_z\sigma_0+it_1\tau_x(\bm{\hat{d}\cdot\sigma})+t_2(\hat{d}_x^2-\hat{d}_y^2)\tau_y
\sigma_0 \right]
\end{equation}
from site $\bm r$ to $\bm{r+d}$ with $\bm{\hat{d}}=\bm d/|\bm d|=(\hat{d}_x,
\hat{d}_y,\hat{d}_z)$ and $f(|\bm d|)=\Theta (R-|\bm d|)\exp[-\lambda(|\bm d|/d_0-1)]/2$.
For regular cubic lattices with only nearest-neighbor hoppings, the Hamiltonian reduces to the
paradigmatic 3D SOTI model in Eq.~(\ref{eq:H_chiral_HOTI}).
The SOTI phase is protected by the winding number of the quadrupole moment about the
momentum $k_z$, which requires no spatial symmetry, and thus it is possible for SOTIs to persist in
completely random lattice sites.
In amorphous systems, there is no well-defined momentum $k_z$ due to the lack of translational
symmetry.
Therefore, the authors insert a flux $\Phi_z$ through a twisted the boundary condition
along $z$ in place of $k_z$ and
calculate the winding number of the quadrupole moment with respect to $\Phi_z$~\cite{WangJH2021PRL}, defined as
\begin{equation}\label{winding1}
	W_Q=\int_{0}^{2\pi} d\Phi_z \frac{\partial Q_{xy}(\Phi_z)}{\partial \Phi_z},
\end{equation}
where $Q_{xy}(\Phi_z)$ is the quadrupole moment in the $(x,y)$ plane, which is a function of the inserted flux $\Phi_z$.
The flux can also be added by introducing a phase factor $\exp(i\Phi_zd_z/L_z)$ to the hopping term.
In a SOTI phase, a nontrivial winding number of $Q_{xy}(\Phi_z)$ arises when one tunes $\Phi_z$ from 0
to $2\pi$, as illustrated in Fig. \ref{FigWang_SOTI1}(b) for a disorder configuration.

The study finds the existence of a second-order topological region with almost quantized winding number
of one and
quantized two-terminal conductance of $G=2e^2/h$ along the $z$ direction arising from
chiral hinge states [see Fig. \ref{FigWang_SOTI1}(c)].

In the presence of TRS, the authors construct a tight-binding Hamiltonian~\cite{WangJH2021PRL}
\begin{equation}\label{HamTRS}
	\hat{H}_h=\sum_{\bm r} \left[\hat{c}_{\bm r}^\dagger T_0\hat{c}_{\bm r}+\sum_{\bm d}
	\hat{c}_{\bm r+\bm d}^\dagger
	T_h({\bm d})\hat{c}_{\bm r}	\right ],
\end{equation}
where $\hat{c}_{{\bf r}}^\dagger=(\hat{c}_{{\bf r},1}^\dagger,\cdots,\hat{c}_{{\bf r},8}^\dagger)$,
the onsite mass term $T_0=M\tau_zs_0\sigma_0$
and the hopping matrix
\begin{eqnarray}
T_h(\bm d)=&&f(|\bm d|)
\left[t_0\tau_zs_0\sigma_0+it_1\tau_xs_0(\bm{\hat{d}\cdot\sigma}) \right. \nonumber \\
&&\left.+t_2(\hat{d}_x^2-\hat{d}_y^2)
\tau_ys_y \sigma_0+it_3\hat{d}_z\tau_ys_x\sigma_0 \right]
\end{eqnarray}
with $\{ s_\nu\}$ denoting another set of Pauli matrices. The TRS represented by $\hat{T}$ is respected,
 i.e.
$\hat{T}\hat{H}_h\hat{T}^{-1}=\hat{H}_h$, where $\hat{T}i\hat{T}^{-1}=-i$ and
$\hat{T}\hat{c}_{\bm r}\hat{T}^{-1}=U_T\hat{c}_{\bm r}$ with $U_T=i\tau_0s_0\sigma_y$.
On a cubic lattice, the Hamiltonian reduces to the model in Eq.~(\ref{eq:H_helical_HOTI})
with an extra term proportional to $t_3\sin k_z \tau_y s_x$.

When $t_3=0$, the pseudospin $s_y$ is a conservative quantity and the topological properties can
be identified by the winding number of the spin quadrupole moment about a flux $\Phi_z$,
\begin{equation}
W_Q^{(s)}=\int_{0}^{2\pi} d\Phi_z \frac{\partial Q_{xy}^{(s)}(\Phi_z)}{\partial \Phi_z},
\end{equation}
where $Q_{xy}^{(s)}(\Phi_z)$ is the spin quadrupole moment as defined in
Sec.~\ref{Subsec:Z2invariant} but with the momentum $k_z$ replaced by the flux $\Phi_z$.
As a result, the HOTP can be characterized by the spin
winding number defined in Eq.~(\ref{eq:WQS}). It is also shown that
for small $t_3$, the spin winding number is still applicable even though  $s_y$ is not a conservative quantity.
Besides, a $\mathbb{Z}_2$ invariant is defined to characterize the topological phase
[see Eq.~(\ref{Z2}) in Sec.~\ref{Subsec:Z2invariant} for its definition. Here, the momentum
$k_z$ should be replaced by a flux $\Phi_z$].

Numerical calculations reveal a range of $M$ with sample averaged $\mathbb{Z}_2$ topological
invariant $\overline{\nu}_Q$ and spin winding number $\overline{W}_{QS}$ quantized at 1 up
to finite-size effects,
as shown in Fig. \ref{FigWang_SOTI2}(b),
indicating the existence of an amorphous SOTI phase with TRS.
The topological helical hinge states manifest in the two-terminal conductance quantized at
$4e^2/h$.

\paragraph{Structural disorder induced second-order topological insulator}

Structural disorder refers to position randomness of lattice sites, serving as a bridge between a crystalline
system and an amorphous system. The authors in Ref.~\cite{WangJH2021PRL} also demonstrate
that structural disorder can drive a topological phase transition from a trivial phase to a SOTI.
The induced transition can be seen from the jump of the conductance and the winding number
of the quadrupole moment from zero and the bulk energy gap closure
as shown in Fig. \ref{FigWang_SOTI2}(a). The driven SOTI also exhibits the hinge states
as revealed by the local DOS.

\paragraph{Average symmetry protected bulk energy gap closure}

In Fig.~\ref{FigWang_SOTI1}(d) and Fig.~\ref{FigWang_SOTI2}(a), we see that the higher-order topological
phase transition occurs both through the closure of the bulk energy gap, although the $\hat{C}_4^z\hat{T}$
symmetry is broken in each sample. Here, we will follow Ref.~\cite{WangJH2021PRL} to show that such a
bulk energy gap closure happens due to the $\hat{C}_4^z\hat{T}$ symmetry on average. One can also
apply similar argument to the 2D case in Sec.~\ref{Sec:HOTAI}.

To demonstrate that the Hamiltonian $\hat{H}_c$ respects the average $\hat{C}_4^z\hat{T}$ symmetry,
one can calculate the symmetry counterpart of the Hamiltonian,
\begin{widetext}
	\begin{eqnarray}
		\hat{C}_4^z \hat{T}\hat{H}_c(\hat{C}_4^z \hat{T})^{-1}
		&=&\sum_{{\bf r}\in S} \left[ M\hat{c}_{D_{\hat{C}_4}{\bf r}}^\dagger \tau_z\sigma_0 \hat{c}_{D_{\hat{C}_4}{\bf r}}
		+\sum_{\substack{{\bf d}={\bf r}_1-{\bf r} \\ {\bf r}_1 \in S, {\bf r}_1\neq {\bf r} }} \hat{c}_{D_{\hat{C}_4}({\bf r}+{\bf d})}^\dagger T_c(D_{\hat{C}_4}\hat{\bf d}) \hat{c}_{D_{\hat{C}_4}{\bf r}} \right] \\
		&=&\sum_{{\bf r}^\prime \in S^\prime } \left[ M\hat{c}_{{\bf r}^\prime}^\dagger \tau_z\sigma_0 \hat{c}_{{\bf r}^\prime}
		+\sum_{\substack{{\bf d}^\prime={\bf r}_1^\prime-{\bf r}^\prime \\ {\bf r}_1^\prime \in S^\prime, {\bf r}_1^\prime \neq {\bf r}^\prime }} \hat{c}_{{\bf r}^\prime+{\bf d}^\prime}^\dagger T_c(\hat{\bf d}^\prime)
        \hat{c}_{{\bf r}^\prime} \right].
	\end{eqnarray}
\end{widetext}
In the derivation, the results are used: $\hat{C}_4^z \hat{T} \hat{c}_{\bf r}(\hat{C}_4^z \hat{T})^{-1}=i\tau_0\sigma_y e^{i\frac{\pi}{4}\sigma_z}\hat{c}_{D_{\hat{C}_4}{\bf r}}$ and
$\hat{C}_4^z \hat{T}i(\hat{C}_4^z \hat{T})^{-1}=-i$.
We use $S$ to represent a set of all site positions in a disorder sample and then rotate all site
positions in $S$ in a counterclockwise direction about $z$ by 90 degrees to obtain a new set
$S^\prime=D_{\hat{C}_4} S\equiv \{ D_{\hat{C}_4}{\bf r}=(-y,x,z): {\bf r}\in S\}$.
If we consider cubic lattices, then $S^\prime = S$ so that $\hat{C}_4^z \hat{T}\hat{H}_c(\hat{C}_4^z \hat{T})^{-1}= \hat{H}_c$,
showing that the $\hat{C}_4^z \hat{T}$ symmetry is obeyed.
However, for a typical individual disorder configuration, $S\neq S^\prime$, and we usually have
$\hat{C}_4^z \hat{T}\hat{H}_c(\hat{C}_4^z \hat{T})^{-1} \neq \hat{H}_c$, indicating that the symmetry
is absent.

\begin{figure}[t]
	\includegraphics[width=\linewidth]{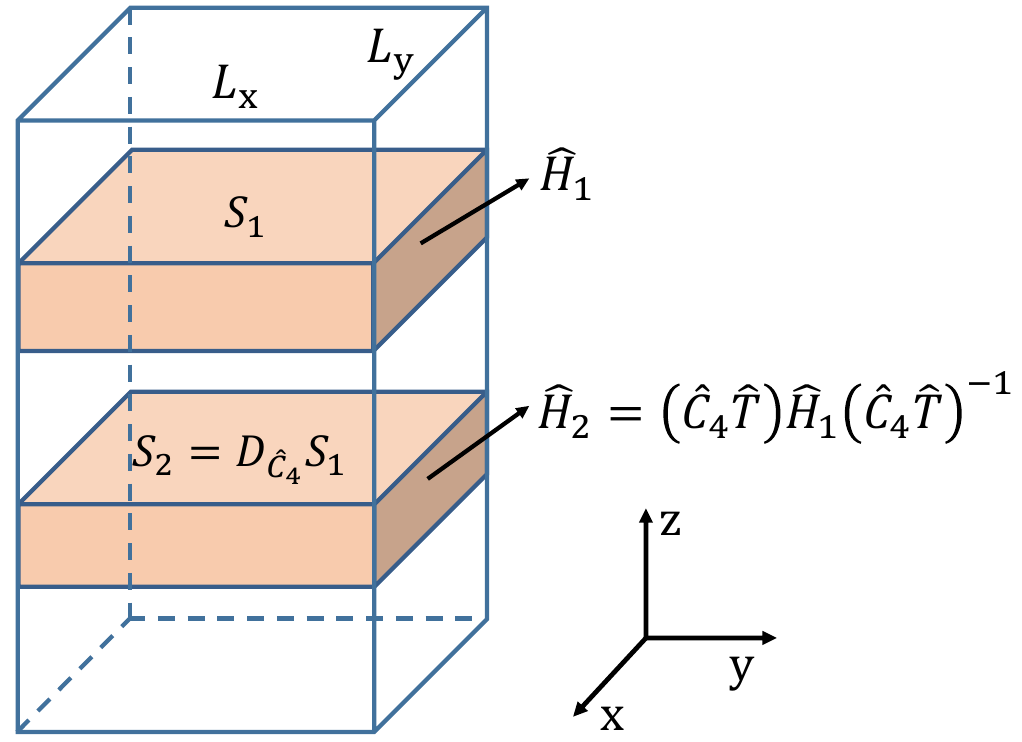}
	\caption{
		Schematic illustration of a system including two subsystems $\hat{H}_1$ and $\hat{H}_2$ on the site
		configurations of $S_1$
		and $S_2$, respectively. From Ref.~\cite{WangJH2021PRL}.
	}
	\label{figAvgSym}
\end{figure}

We now consider a statistical ensemble consisting of all systems. The ensemble is defined to respect
an average symmetry if a Hamiltonian and its symmetry conjugate partner exist in the ensemble
with the same probability~\cite{Fu2012PRL,Akhmerov2014PRB}. For instance, a configuration
$S$ and its rotational part $S^\prime=D_{\hat{C}_4} S$ clearly appear with the same probability
since we consider the structural disorder without any correlation. As a consequence,
the probability of occurrence of $\hat{H}_c$ is equal to that of its $\hat{C}_4^z \hat{T}$ conjugate partner $\hat{C}_4\hat{T}\hat{H}_c(\hat{C}_4\hat{T})^{-1}$
so that the ensemble respects the $\hat{C}_4\hat{T}$ symmetry on average.

We now argue that this average symmetry enforces the bulk energy gap closure at the phase transition point
for a single system~\cite{WangJH2021PRL}. To demonstrate it, we consider a subsystem $\hat{H}_1$ whose $z$ coordinates
are restricted in the interval $[z_0,z_0+\Delta z]$ inside a large system (see Fig.~\ref{figAvgSym}).
As a single system with a spatial configuration $S_1$, the subsystem does not respect the $\hat{C}_4^z \hat{T}$
symmetry so that the topological phase transition happens through a surface energy gap closing, e.g.
the energy gap on the $x$-normal surface. For a very large system (e.g. infinitely long along $z$),
it is reasonable to expect that there exists the spatial configuration $S_2=D_{\hat{C}_4} S_1+\alpha {\bf e}_z$,
which is obtained by rotating $S_1$ about $z$ by 90 degrees and shifting the coordinate along $z$ by $\alpha$.
This
leads to another subsystem $\hat{H}_2$ with the spatial configuration $S_2$
(see Fig.~\ref{figAvgSym}). It is clear to see that $\hat{H}_2$ is the $\hat{C}_4^z \hat{T}$ conjugate
partner of $\hat{H}_1$. Therefore, when the $x$-normal surface energy gap of $\hat{H}_1$ vanishes,
the $y$-normal surface energy gap of $\hat{H}_2$ must close.
It indicates that for the entire system including both $\hat{H}_1$ and $\hat{H}_2$, the closure
of the energy gap must simultaneously occur on both surfaces, suggesting that
the bulk energy gap should close.

\begin{figure}[t]
	\includegraphics[width=0.9\linewidth]{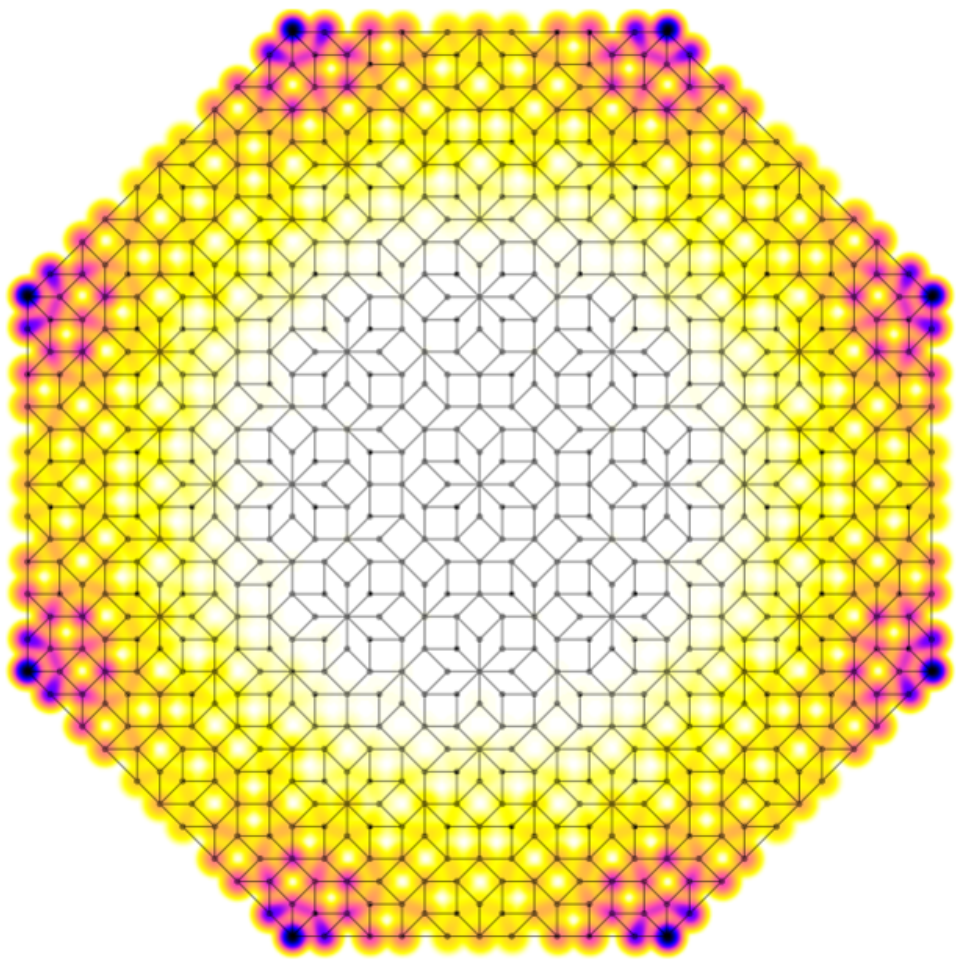}
	\caption{The spatial distributions of the corner modes of the higher-order topological superconductor
		on an AB tiling quasicrystal sample.
		From Ref.~\cite{Fulga2019PRL}.
	}
	\label{Fig1Fulga}
\end{figure}

\begin{figure*}[t]
	\includegraphics[width=\linewidth]{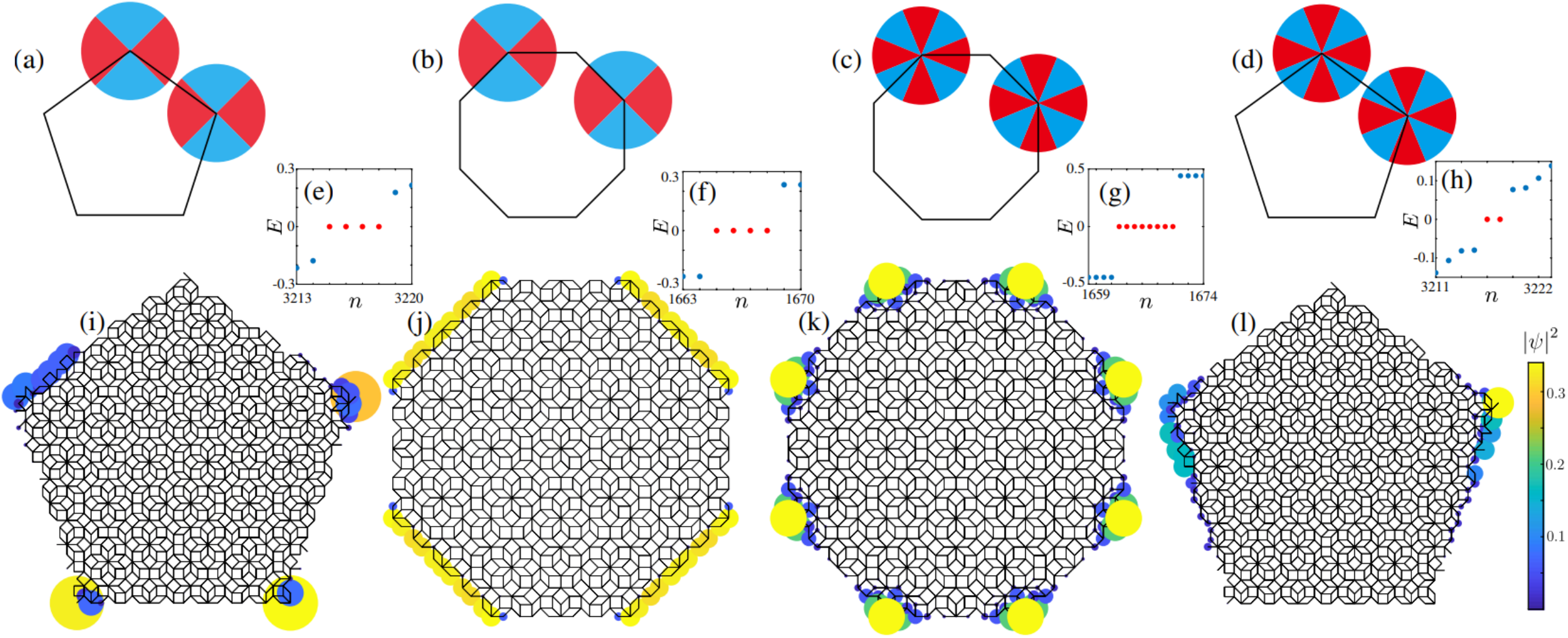}
	\caption{(a)-(d) Illustrations of the angle dependence of the sign of the effective masses near
		a corner of pentagonal and octagonal AB tiling quasicrystal samples.
		The red and blue color represents the effective mass with opposite signs.
		(e)-(h) The energy spectrum under open boundary conditions.
		(i)-(l) The corresponding real-space distributions of zero-energy states.
		For the first and second columns, $\eta =2$ and for the third and fourth columns, $\eta=4$.
		From Ref.~\cite{Xu2020PRL}.
	}
	\label{Fig1Xu}
\end{figure*}

\subsection{Quasicrystalline lattices}\label{sec:Quasicrystal}
Another class of non-crystalline materials is quasicrystals which lack translational symmetry while possess
quasi long-range order~\cite{QCbook}. Quasicrystals can also respect five-fold, eight-fold, ten-fold or twelve-fold rotational
symmetry, which are not allowed in crystalline systems. These symmetries enable the existence of new
HOTPs in both 2Ds~\cite{Fulga2019PRL,Xu2020PRL,Cooper2020PRR,Xu2020PRB,Liu2021NanoLett,
		Zhou2021PRB,LiuFeng2022PRL,Sassetti2022Symmetry,Ziani2022PRB,Jiang2022PRAp} and 3Ds~\cite{MaoYF2023arXiv,XuD2023arXiv}, which cannot appear in
crystalline systems.

\begin{figure*}[t]
	\includegraphics[width=\linewidth]{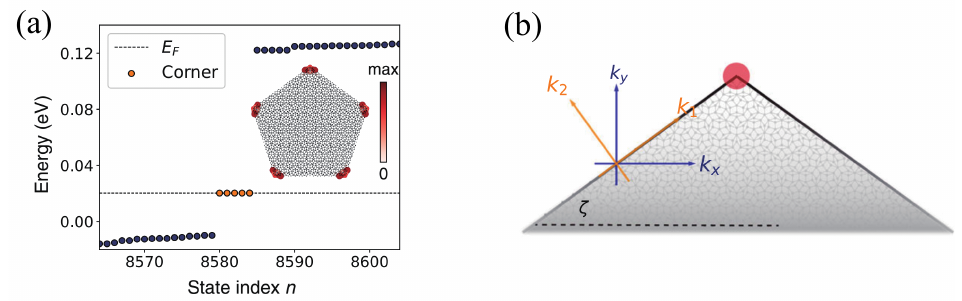}
	\caption{(a) The energy spectrum of a pentagonal Penrose tiling quasicrystal sample
		subject to an in-plane Zeeman field, exhibiting five corner modes at the energy $E_F$.
		The inset displays the spatial distribution of the corner states.
		(b) Schematic illustration of an edge whose polar angle measured relative to the positive
		$x$ axis is $\zeta$.
		Reproduced from Ref.~\cite{LiuFeng2022PRL}.
	}
	\label{Fig_FiveCorner}
\end{figure*}

Higher-order topological superconductors with Majorana
zero modes localized at eight corners of an octagonal sample can occur on a 2D Ammann-Beenker
(AB) tiling quasicrystalline lattice with eight-fold rotational symmetry (see Fig. \ref{Fig1Fulga})~\cite{Fulga2019PRL}.
The authors consider the following mean-field real-space many-body Hamiltonian on a AB tiling quasicrystalline lattice,
\begin{equation}
\hat{H}=\sum_j\hat{\Psi}_j^\dagger H_j\hat{\Psi}_j
+\sum_{\langle j,l\rangle}\hat{\Psi}_j^\dagger H_{jl} \hat{\Psi}_l ,
\end{equation}
where
$\hat{\Psi}_j^\dagger=
(\hat{c}^\dagger_{j,\uparrow},\hat{c}_{j,\uparrow},\hat{c}^\dagger_{j,\downarrow},
\hat{c}_{j,\downarrow})$ with $\hat{c}^\dagger_{j,\nu}$ ($\hat{c}_{j,\nu}$) creating (annihilating) an
electron with spin $\nu$ at site $j$, the mass term $H_j=\mu\sigma_z\tau_z$, and the hopping matrix
\begin{eqnarray}
H_{jl}=&&(t/2)\sigma_z\tau_z-(i\Delta/2) \left(\cos \theta_{jl} \sigma_z\tau_x+\sin\theta_{jl} \sigma_z\tau_y \right) \nonumber \\
&&+(V/2)\sigma_y\tau_0\cos 4\theta_{jl},
\end{eqnarray}
where $\Delta$ is the $p$-wave pairing strength, $\theta_{jl}$ is the polar angle between site
$j$ and $l$, and $\langle \cdots \rangle$ denotes a pair of sites connected by an edge on the lattice.
This Hamiltonian
can be written as $\hat{H}=\hat{\Psi}^\dagger H_{\textrm{BdG}} \hat{\Psi}$ where
$H_{\textrm{BdG}} $ is the Bogoliubov-de-Gennes (BdG) Hamiltonian and
$\hat{\Psi}^\dagger=(\hat{\Psi}_1^\dagger,\hat{\Psi}_2^\dagger,\cdots)$.
The system respects the particle-hole symmetry, i.e.
$\mathcal{P} H_{\textrm{BdG}} \mathcal{P}^{-1}=-H_{\textrm{BdG}}$
where $\mathcal{P}=\tau_x\sigma_0 I \kappa$,
and the chiral symmetry, i.e. $\Pi H_{\textrm{BdG}} \Pi^{-1} =-H_{\textrm{BdG}} $
with $\Pi=\tau_0\sigma_x I$ ($I$ is an identity matrix acting
on the degrees of freedom of lattice sites).

When $V=0$, $H_{\textrm{BdG}}$ is a model consisting of two Chern bands with
opposite Chern number, thereby contributing helical Majorana edge modes. In fact,
the helical edge modes are protected by a time-reversal like symmetry represented by $\sigma_y \tau_x I \kappa$.
The $V$ term breaks the symmetry and gaps out the edge states by introducing an
effective mass on each surface; such masses change sign across each neighboring edges,
giving rise to Majorana corner modes, as discussed in Sec.~\ref{Sec:Cornermodes}.
While the $V$ term also breaks the $C_8$ symmetry, the $C_8M$ symmetry is preserved with
$M=\sigma_z\tau_0 I$~\cite{Fulga2019PRL}.

To define a topological invariant, the authors in Ref.~\cite{Fulga2019PRL} construct an effective Hamiltonian
$H_{\textrm{eff} }=G_{\textrm{eff} }^{-1}$, where
\begin{equation}
	G_{\textrm{eff} }(k)_{mn}=\langle k,n|G|k,m\rangle
\end{equation}
with
$G=\lim_{\eta\rightarrow0}(H_{\textrm{BdG}}+i\eta)^{-1}$ being the zero-energy Green's function and
$|k,n\rangle$ being the plane-wave state with momentum $k$.
Here, $n$ labels the local components composed of spin and particle-hole degrees of freedom.
The plane-wave state is defined on a four-dimensional (4D) cubic lattice which generates the 2D quasicrystal
through the cut-and-project method~\cite{Fulga2019PRL}.
Since the system respects the $C_8M$ symmetry, one can restrict $H_{\textrm{eff} }$ at
the $C_8$-invariant momenta $\Gamma=(0,0,0,0)$ and $\Pi=(\pi,\pi,\pi,\pi)$ to the eigenspace
of $C_8M$ with eigenvalues $\omega_{\pm n}=\exp(\pm i\pi n/8)$ ($n=1,3,5,7$).
At each subspace, the restricted Hamiltonian still respects the particle-hole symmetry and
thus is classified as the zero-dimensional class D. One thus can calculate the sign (denoted by $\nu_{n,k^*}$)
of the Pfaffian of an antisymmetric Hamiltonian transformed from the original Hamiltonian.
A $\mathbb{Z}_2$ invariant is thus defined as
\begin{equation}
	\nu=\nu_1\nu_3,
\end{equation}
where $\nu_n=\nu_{n,0}/\nu_{n,\Pi}$. It is found that in the nontrivial phase with corner modes,
$\nu=-1$ whereas in the trivial one, $\nu=1$.

HOTIs can also arise in quasicrystals~\cite{Xu2020PRL}.
The authors in Ref.~\cite{Xu2020PRL} study a model Hamiltonian
with the onsite term $T_0=(M+2t_2)\sigma_0\tau_3$ and the hopping matrix
\begin{eqnarray}
T({\bm d})=&&(f(d)/2) \left\{[-it_1(\sigma_3\tau_1\cos \theta+\sigma_0\tau_2\sin \theta)-t_2\sigma_0\tau_3] \right. \nonumber \\
&&\left. +g \cos (\eta\theta) \sigma_1\tau_1 \right \}.
\end{eqnarray}
 on an AB tiling quasicrystal where $f(d)=e^{1-d/\xi}$ describes the decay of the hopping with respect to the distance $d$
 between two sites. Similarly, when $g=0$, the Hamiltonian describes a quantum spin Hall insulator
 protected by TRS represented by $i\sigma_y \kappa$ and conserved spin $\sigma_z$~\cite{Xu2020PRL},
 hosting helical edge modes. The $g$ term breaks the TRS so as to open an energy gap at boundaries.
 With this term, the system still respects the particle-hole symmetry represented by $\sigma_3\tau_1\kappa$ and chiral symmetry
 represented by $\sigma_2\tau_1$.

Approximately, the effective mass at each edge depends on $\cos (\eta\theta_{\textrm{edge}})$ where
$\theta_{\textrm{edge} }$ is the polar angle of a boundary.
For $\eta=2$, the mass's sign is positive for
$\theta_{\textrm{edge} }\in [-\pi/4,\pi/4]\cup  [3\pi/4,5\pi/4]$ (labeled by red color) while it becomes
negative for  $\theta_{\textrm{edge} }\in [-3\pi/4,-\pi/4]\cup  [\pi/4,3\pi/4]$ (labeled by blue color),
as shown in Fig. \ref{Fig1Xu}(a) and (b).
For a quasicrystal lattice with pentagon boundaries, shown in Fig. \ref{Fig1Xu}(a), (e) and (i),
there appear four zero-energy states localized at corners where the mass exhibits a sign change.
For an octagon, the masses are zero at four edges while the sign of the mass at its two neighboring
edges is opposite, leading to edge states extended along the four edges [see Fig. \ref{Fig1Xu}(b) and (j)]. In fact, these edge states
can be understood as corner states as one can continuously deform the boundary into a vertex
so that a square is formed and corner states are localized at the corners.
Similarly, when $\eta=4$,
the system exhibits eight-fold symmetric zero-energy corner states for octagon boundaries [see Fig. \ref{Fig1Xu}(k)] and
for pentagon boundaries, only two corner states appear [see Fig. \ref{Fig1Xu}(l)].
The HOTI can be characterized by the $\mathbb{Z}_2$ topological
invariant $\nu$~\cite{Fulga2019PRL}.

The authors in Ref.~\cite{Xu2020PRL} also construct a Hamiltonian on the AB tiling quasicrystalline lattice
with the onsite term $T_0=\gamma(\Gamma_2+\Gamma_4)$ and the hopping matrix
\begin{equation}
T({\bm d})= \frac{\lambda f(d)}{2}
	[ |\cos \theta|\Gamma_4 - i\cos \theta \Gamma_3
    + |\sin \theta| \Gamma_2- i\sin \theta \Gamma_1],
\end{equation}
where $\Gamma_4=\tau_1s_0$ and $\Gamma_\nu=-\tau_2s_\nu$ with
$\nu=1,2,3$.
The system also has chiral symmetry represented by $\tau_3$ and thus the quadrupole moment $q_{xy}$
is a quantized quantity~\cite{Yang2021PRB,Li2020PRL}.
Indeed, it is found that in the nontrivial phase $q_{xy}=1/2$, and four zero-energy corner states
emerge in a geometry with square boundaries~\cite{Xu2020PRL}. However, for the case with eight or more corner states,
this quadrupole moment vanishes, and one needs to calculate the generalized quadurpole moment based on
transformed lattice sites~\cite{Tao2023Average,MaoYF2023arXiv}.

We have shown that an even number of corner modes can exist in a quasicrystalline lattice. One may wonder
whether five corner modes can appear in a Penrose tiling quasicrystalline lattice as the lattice has the
five-fold rotational symmetry. Since the sign change of the mass can only give rise to even number of
corner modes, one may consider other mechanisms. Indeed, the authors in Refs.~\cite{Liu2021NanoLett,LiuFeng2022PRL}
find that a Penrose tiling quasicrystal can support five corner modes associated with a fractional
charge $Q=e/5$ as shown in Fig.~\ref{Fig_FiveCorner}(a). To generate the corner modes, they apply an in-plane Zeeman field
to a TI with helical edge states on a quasicrystal to gap out the edge states, leading to an effective
Hamiltonian at a boundary
\begin{equation}
	H_{\textrm{edge}}=-ak_1 s_z +g\left[\cos(\phi_j)s_x+\sin(\phi_j)s_y\right],
\end{equation}
where $\phi_j=\zeta_j+\theta-\pi/2$ with $\zeta_j$ being the orientation
angle of the $j$th edge [see Fig.~\ref{Fig_FiveCorner}(b)], $k_1$ is the momentum along the edge,
$\theta$ is the polar angle of the Zeeman field.
It has been shown that even though a mass kink $\Delta \zeta$ across neighboring boundaries
does not lead to the sign change of the mass, it can still result in a corner state associated with a fractional charge
of $Q=e|\Delta \zeta/2\pi |$~\cite{Moore2019PRB,Wilczek1981PRL,Semenoff1983PRL}.
It is found that for a pentagonal Penrose tiling quasicrystal, five corner modes appear associated
with a fractional charge $Q= e/5$ due to the angle change of adjacent edges $\Delta \zeta =2\pi/5$~\cite{LiuFeng2022PRL}.

While the HOTIs in 2D quasicrystals indicate the existence of new TIs without crystalline counterparts,
it remains unclear whether novel chiral and helical HOTIs without crystalline counterparts can occur in 3Ds.
Very recently, the authors in Ref.~\cite{MaoYF2023arXiv} construct a 3D model by stacking 2D models on the AB tiling quasicrystalline
lattices and find the existence of eight gapless chiral hinge modes, leading to the two-terminal conductance of
$4e^2/h$. The TI is characterized by the winding number of the generalized quadrupole moment based on
transformed lattice sites. In the presence of TRS, it is shown that eight helical pairs of hinge modes arise,
resulting in the conductance of $8e^2/h$. The phase is
characterized by a $\mathbb{Z}_2$ topological invariant as well as the spin winding number of
the quadrupole moment based on transformed lattice sites.
In addition, higher-order Weyl-like and Dirac-like semimetals in quasicrystals supporting
both surface Fermi arcs and hinge Fermi arcs have been theoretically predicted~\cite{XuD2023arXiv,MaoYF2023arXiv}.

\subsection{Hyperbolic lattices}

\begin{figure}[t]
	\includegraphics[width=\linewidth]{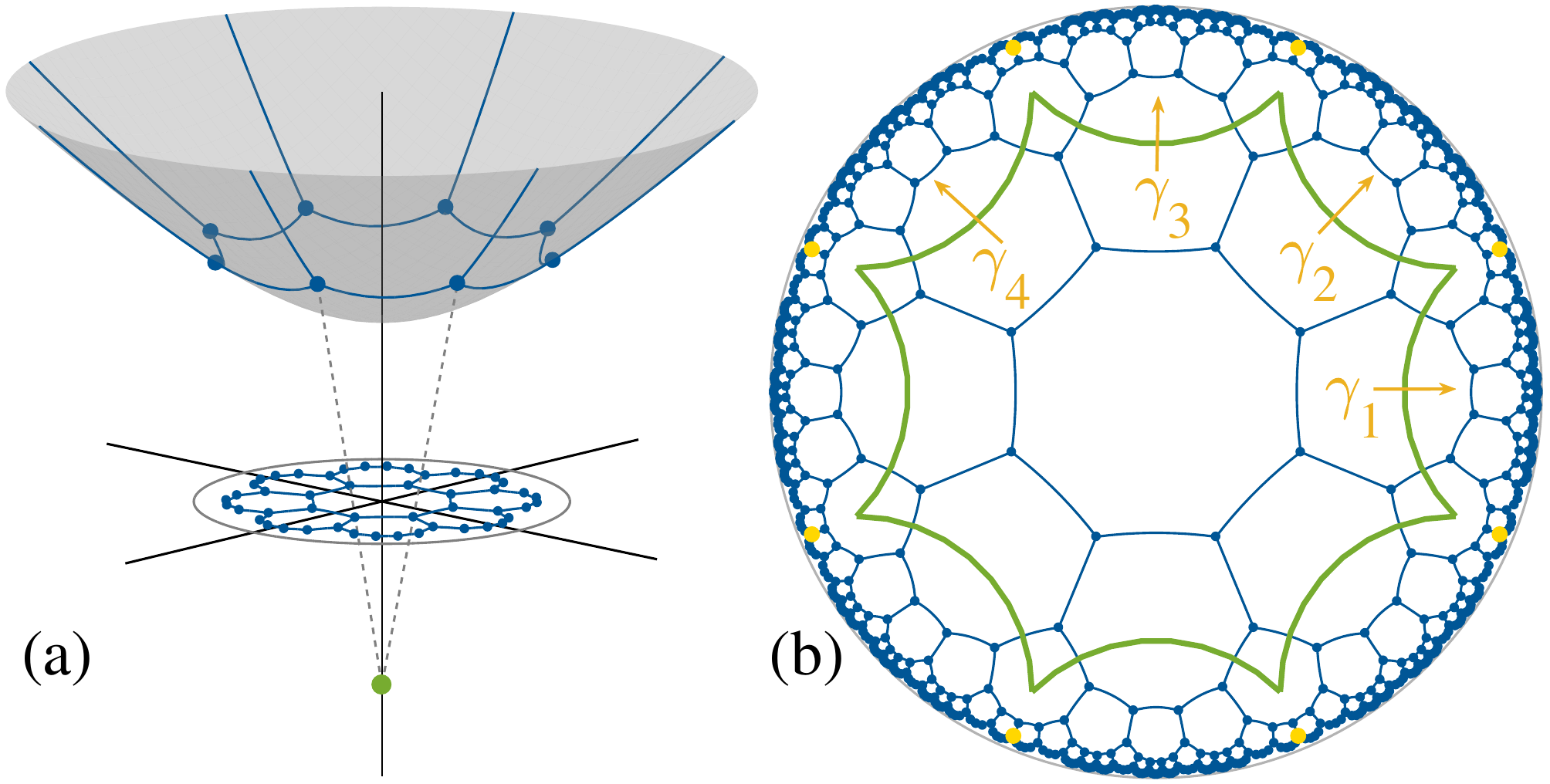}
	\caption{(a) Schematic illustration of the \{8,3\} hyperbolic lattice projected onto the
		Poincar$\acute{e}$ disk.
		(b) The \{8,3\} hyperbolic lattice on the  Poincar$\acute{e}$ disk with the first unit cell enclosed
		by the green curve and $\gamma_1,\gamma_2,\gamma_3$ and $\gamma_4$ representing four
		translational operators.
		The corner states are highlighted by yellow circles.
		From Ref.~\cite{Xu2023PRB}.
	}
	\label{FigHyperbolic1}
\end{figure}

Recently, there has been a surge of interest in hyperbolic lattices~\cite{Houck2019nat,Houck2019CMP,Park2020PRL,Gorshkov2020PRA,
		Rouxinol2021arxiv,Rayan2021scia,Matsuki2021JOP,Gorshkov2022PRL,Rayan2022PNAS,
		Thomale2022PRB,Boettcher2022PRL,
		Zhang2022NC,Zhou2022PRB,
		Lenggenhager2022NC,Mao2022PRL,Mosseri2022PRB,Bzdusek2022PRB,Urwyler2022PRL,
			Boettcher2023NC,lenggenhager2023arxiv}, which belong to the
category of 2D non-crystalline lattices.
Compared with a crystalline system, any rotational symmetry is allowed
in a hyperbolic lattice, providing another platform to realize HOTPs
without crystalline counterparts. The arbitrary rotational symmetry comes from the fact that a hyperbolic
lattice with constant negative curvature can be tessellated by regular $p$-gons with any integer
$p$. The hyperbolic lattice in fact respects the hyperbolic translational symmetry. In other words,
the lattice can be obtained by applying translational operations to the
first unit cell on the hyperbolic plane. For example, in the hyperbolic \{8,3\} lattice, the translational operators
are generated by four generators $\gamma_1,\gamma_2,\gamma_3$ and $\gamma_4$
[see Fig. \ref{FigHyperbolic1}(a)]. These translational operations form a non-Abelian group.
One can also project the hyperbolic lattice onto a Poincar$\acute{e}$ disk as shown in
Fig. \ref{FigHyperbolic1}. Since the lattice sites on the disk break the traditional translational
symmetry, we classify them as non-crystalline lattices.

It is found that hyperbolic lattices can also give rise to HOTPs without crystalline
counterparts due to the rotational symmetry absent in crystalline systems~\cite{Xu2023PRB,ZRLiu2023PRB}.
Here, we follow Ref.~\cite{Xu2023PRB} to present how the new phases arise.
The authors construct two tight-binding models. For the first model, the Hamiltonian
within the first unit cell reads
\begin{eqnarray}
	H_0=\sum_{\alpha,\beta} \Big\{ \sum_i
	&& m \, \hat{c}^{\dagger}_{1i\alpha} (\tau_z\sigma_0)_{\alpha\beta} \hat{c}_{1i\beta} \nonumber \\
	&& +\sum_{\langle i,j \rangle}
	\hat{c}^{\dagger}_{1i\alpha} [T(\theta_{ij})]_{\alpha\beta} \hat{c}_{1j\beta} \Big\}
\end{eqnarray}
where $\hat{c}^{\dagger}_{ri\alpha} $ represents the creation operator for the $\alpha$th orbital at site $i$ in the $r$th unit cell on a hyperbolic lattice,
and $T(\theta_{ij})$ is given by Eq.~(\ref{Eq:2DAmorHopping}).
The hopping between sites in the first unit cell and sites in neighboring unit cells (the set $S_1$)
is given by
\begin{equation}
H_1=\sum_{r\in S_1}\sum_{\alpha,\beta}\sum_{\langle i,j\rangle}
\hat{c}^{\dagger}_{ri\alpha} [T(\tilde{\theta}_{(ri),(1j)})]_{\alpha\beta} \hat{c}_{1j\beta} +\mathrm{H.c.}
\end{equation}
with modified polar angle $\tilde{\theta}_{(ri),(1j)}$.
The entire Hamiltonian can be derived through the application of hyperbolic translational operations,
ensuring that it respects the hyperbolic translational symmetry.

\begin{figure*}[t]
	\includegraphics[width=\linewidth]{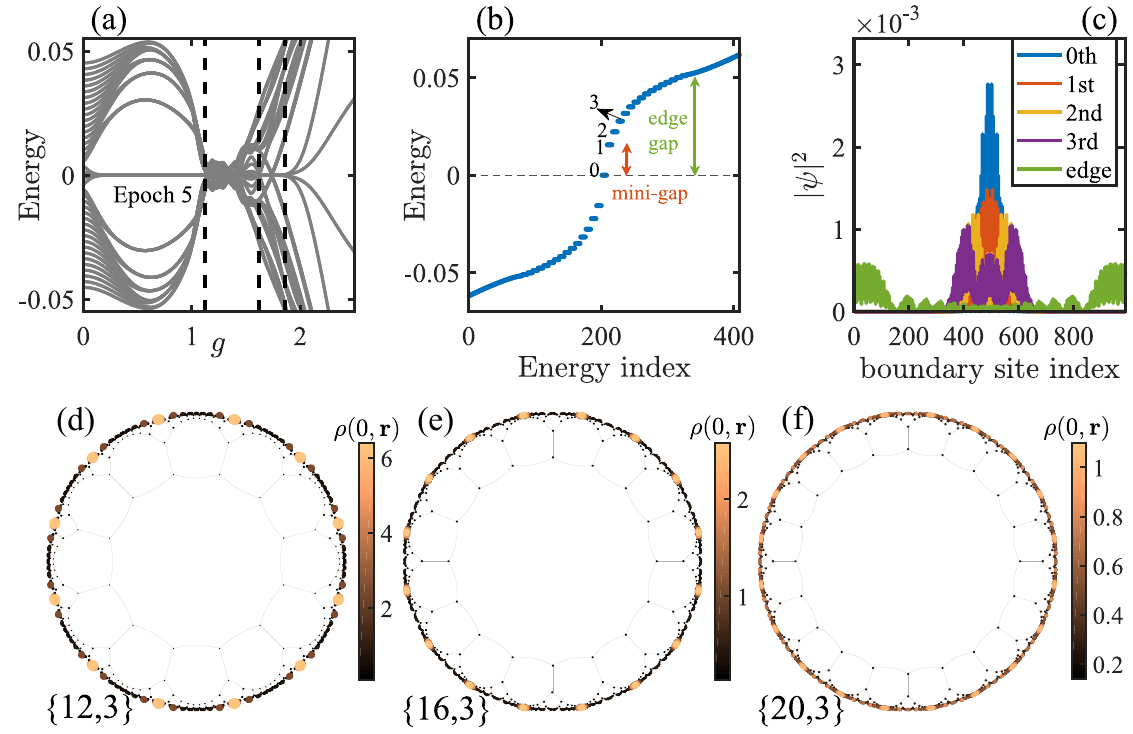}
	\caption{(a) The energy spectrum of the Hamiltonian in Eq.~(\ref{Eq:Hyp2}) with respect to the system parameter $g$
		on a hyperbolic \{8,3\} lattice at epoch 5 with 2888 sites, illustrating the presence of the zero-energy states.
		(b) The eigenenergies when $g = 0.6$ at epoch 6. The minigap represents the first nonzero positive energy and
		the edge gap represents the energy where the degeneracy suddenly changes from the eight-fold to the double one.
		Inside the edge gap, all states are eight-fold degenerate. (c) The spatial distributions of the wave functions in the vicinity
		of zero energy over the 1/8 sector on the boundary of the Poincar$\acute{e}$ disk, suggesting that the eight-fold
		degenerate states are spatially localized around a corner while the states whose energy is slightly above the edge gap
		are more like the edge state.
		The zero-energy local DOS of hyperbolic (d) \{12, 3\}	lattice, (e) \{16, 3\} lattice and (f) \{20, 3\} lattice.
		Reproduced from Ref.~\cite{Xu2023PRB}.
	}
	\label{FigHyperbolic2}
\end{figure*}

The momentum-space Hamiltonian $H(k_1,k_2,k_3,k_4)$ in a 4D Brillouin zone
with $k_j\in [0,2\pi]$ ($j=1,2,3,4$) is constructed based on the hyperbolic band theory~\cite{Rayan2021scia}.
One can keep two momenta $k_{\bar{i}},k_{\bar{j}}$ out of the 4D momentum space $(k_1,k_2,k_3,k_4)$
as parameters and apply the Fourier transformation to the other two momenta, yielding a
real-space Hamiltonian on a square lattice.
In this way, the quadrupole moment $q_{ij}(k_{\bar{i}},k_{\bar{j}})$ can be calculated
and the averaged quadrupole moment over $k_i,k_j$ is defined as
$\overline{q}_{ij}=1/(2\pi)^2\int dk_{\bar{i}}dk_{\bar{j} }q_{ij}(k_{\bar{i}},k_{\bar{j}})$.
It is proved that $\overline{Q}_{12}=\overline{Q}_{23}=\overline{Q}_{34}=\overline{Q}_{41}$
and $\overline{Q}_{13}=\overline{Q}_{24}$ due to the $C_8M$ symmetry.
It is found that the average quadrupole moment $\overline{q}_{12}$ is quantized to $0.5$
when $g$ is not very large, suggesting the existence of HOTP.
However, the real-space calculations do not suggest the existence of robust corner modes as
the energy spectrum differs significantly
for distinct system sizes.

We note that the authors in Ref.~\cite{lenggenhager2023arxiv} calculate the bulk DOS in the
BBH model on a $\{6, 4\}$ hyperbolic lattice that preserves the hyperbolic translational symmetry
by a supercell method and find that the DOS exhibits a gap closing at a critical point.
However, the bulk-edge correspondence has not been explored.

To find a higher-order phase, the requirement of the hyperbolic translational
symmetry is relaxed and the following Hamiltonian on the hyperbolic lattice is studied~\cite{Xu2023PRB},
\begin{equation} \label{Eq:Hyp2}
H=\sum_{\alpha,\beta} \Big\{\sum_i
m \, \hat{c}^{\dagger}_{i\alpha} (\tau_z\sigma_0)_{\alpha\beta} \hat{c}_{i\beta} +\sum_{\langle i,j \rangle}
\hat{c}^{\dagger}_{i\alpha} [T(\theta_{ij})]_{\alpha\beta} \hat{c}_{j\beta} \Big\},
\end{equation}
where $\hat{c}^{\dagger}_{i\alpha}$ represents the creation operator for the $\alpha$th orbital at site $i$. This system
exhibits zero-energy modes in the energy spectrum for a hyperbolic lattice under
open boundary conditions [see Fig.~\ref{FigHyperbolic2}(a)]. The topological properties are
characterized by the corner charge which approaches the quantized value of $0.5$.
Interestingly, in contrast to the quasicrystal case, there are two energy gaps:
the minigap determined by
the first nonzero eigenenergy and the edge energy gap determined by the energy
where the degeneracy suddenly changes from the eight-fold to the double one
[see Fig.~\ref{FigHyperbolic2}(b)]. Inside the edge gap, there are infinite number of eight-fold
degenerate corner modes besides the zero-energy ones [see the spatial distribution
of the states in Fig.~\ref{FigHyperbolic2}(c)]. Although the minigap
may decrease to zero in the thermodynamic limit, finite-size analysis suggests that
the zero-energy corner modes persist.
In addition, HOTIs with $12$, $16$ or $20$ corner modes have
been found as shown in Fig.~\ref{FigHyperbolic2}(d)-(f).

\subsection{Fractal lattices}

\begin{figure}[t]
\includegraphics[width=\linewidth]{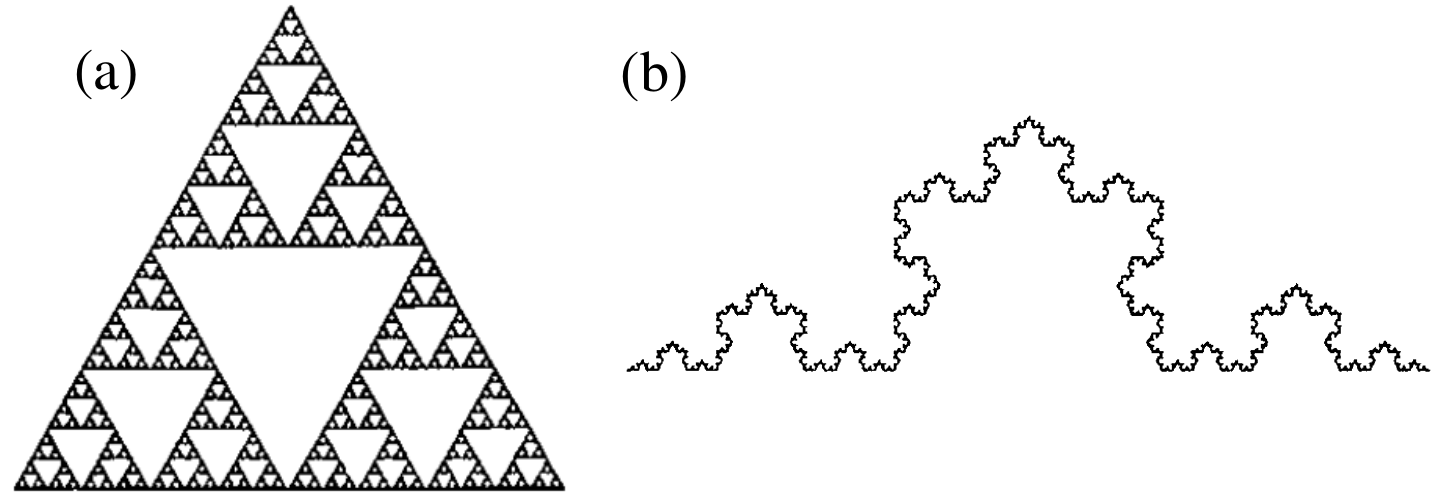}
\caption{
Two famous examples of fractals, (a) Sierpi\'{n}ski triangle and (b) von Koch curve.
Reproduced from Ref.~\cite{falconer2004fractal}.}
\label{figFracEg}
\end{figure}

\begin{figure*}[t]
\includegraphics[width=\linewidth]{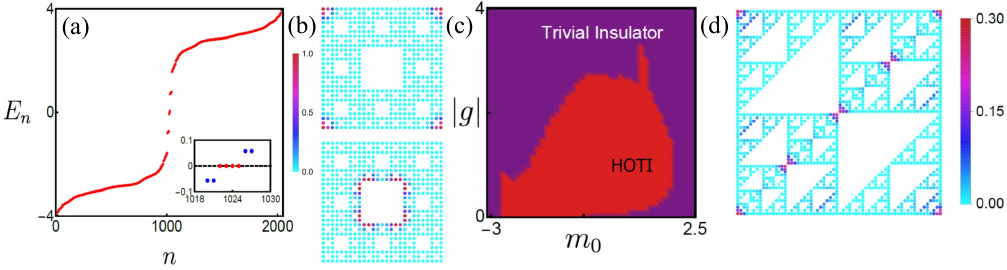}
\caption{
(a) The energy spectrum of the model on the square Sierpi\'{n}ski carpet fractal.
The inset displays the zoom-in view of the states near zero energy.
(b) Upper panel: local DOS of the four zero-energy states colored red in the inset of (a)
which are localized at outer corners.
Lower panel: local DOS of the four states near zero energy colored blue in the inset of
(a).
(c) The phase diagram of the model on the square Sierpi\'{n}ski carpet fractal in the $|g|-m_0$ plane.
(d) The local DOS of the zero-energy states of the model on the glued Sierpi\'{n}ski triangle fractal.
Reproduced from Ref.~\cite{manna2022higher}.}
\label{Fig_Fractal}
\end{figure*}

Fractals are another class of non-crystalline systems with detailed structure at arbitrarily small
scale~\cite{falconer2004fractal}.
The fine structure at small scale is the same as or reminiscent of the whole pattern, which is the
so-called self-similarity property.
Usually a fractal lattice can be generated recursively.
Figure~\ref{figFracEg} shows two well-known examples of fractals, Sierpi\'{n}ski triangle and von Koch curve.
Another salient feature of fractals is the non-integer dimension.
For the Sierpi\'{n}ski triangle, if each line segment is enlarged 2 times, we will get 3 copies of itself,  giving the Hausdorff dimension $D=\frac{\log 3}{\log 2}\approx 1.585$.
Similarly, the dimension of the von Koch curve is $D=\frac{\log 4}{\log 3}\approx 1.262$.
Note that there exist different definitions of dimensions in the context of
fractals~\cite{falconer2004fractal}.
	
Recently, it has been demonstrated that fractal lattices can also harbor
HOTPs~\cite{pai2019topological,manna2022higher,lijk2022higher,zheng2022observation,chenh2023higher,ma2023elastic,nunez2023topological}.
In Ref.~\cite{manna2022higher}, the authors study a Hamiltonian in Eq.~(\ref{Ham2})
on a square Sierpi\'{n}ski carpet fractal with $T_0=(m_0+2t_2)\sigma_0\tau_3$ and
\begin{eqnarray}
T({\bm d})=&&(f(d)/2) \left\{[-it_1(\sigma_3\tau_1\cos \theta+\sigma_0\tau_2\sin \theta)-t_2\sigma_0\tau_3] \right. \nonumber \\
&&\left. +g \cos (2\theta) \sigma_1\tau_1 \right \}.
\end{eqnarray}
As we have discussed previously, the Hamiltonian describes a quantum spin Hall insulator with the
edge states gapped by a mass term.
The energy spectrum exhibits in-gap corner states, as shown in Fig.~\ref{Fig_Fractal}(a) and (b).
The four zero-energy states denoted by red dots in the inset of Fig.~\ref{Fig_Fractal}(a) are localized at the outer
corners [see the upper panel of Fig.~\ref{Fig_Fractal}(b)] and the four states near zero energy colored blue in
the inset of Fig.~\ref{Fig_Fractal}(a) are localized around the inner corners [see the lower panel of Fig.~\ref{Fig_Fractal}(b)].
Note that the peaks of the local DOS of the inner corner states are not located at the exact
inner corner sites (see the lower
panel of Fig.~\ref{Fig_Fractal}(b)), which might be because the coordination number of these sites is larger than 2.
The higher-order topological nature of the system is further illuminated by the quantized quadrupole
moment $Q_{xy}=0.5$ calculated under open boundary conditions, which is independent of the origin choice
in the thermodynamic limit.
The phase diagram is shown in Fig.~\ref{Fig_Fractal}(c)~\cite{manna2022higher}.
The authors of Ref.~\cite{manna2022higher} also study the same Hamiltonian on a glued Sierpi\'{n}ski triangle fractal,
which exhibits zero-energy states located at both the inner and outer corners [see Fig~\ref{Fig_Fractal}(d)].
The number of zero-energy modes can be large and rely on the generation number of the fractal lattice,
while for a specific generation number, the number of corner states is the same under periodic and open
boundary conditions.
The higher-order topological nature of the model on the glued Sierpi\'{n}ski triangle fractal is also characterized
by the quadrupole moment $Q_{xy}=0.5$~\cite{manna2022higher}.
Besides HOTIs, higher-order topological superconductors can also reside on fractal lattices~\cite{manna2022higher}.

\section{Experimental progress}\label{sec5}
In this section, we briefly summarize the progress in experimental realizations of HOTPs
for both crystalline and non-crystalline systems.

\subsection{Solid state materials}

\begin{figure*}[t]
\centering
\includegraphics[width=0.7\linewidth]{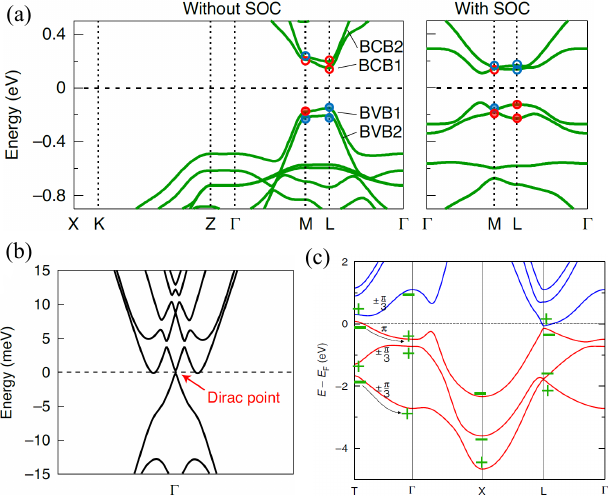}
\caption{
(a) The band structure for Bi$_4$Br$_4$ without (left) and with (right) spin-orbit coupling (SOC).
There is a nonzero gap separating the conduction bands and valence bands.
(b) The energy bands calculated for a nanowire of Bi$_4$Br$_4$,
which exhibit the gapless hinge states crossing at a Dirac point. From Ref.~\cite{Kondo2021NM}.
(c) The band structure of bismuth. There is no gap at the Fermi energy denoted by the dashed line.
From Ref.~\cite{Neupert2018NP}.}
\label{FigMaterBand}
\end{figure*}

HOTPs in both 2Ds~\cite{ezawa2018minimal,Park2019PRL,YangSA2019PRL,liu2019two,chen2020universal,radha2020buckled,YangBJ2020npjQM,chen2021graphyne,liu2021higher,costa2021discovery,xue2021higher,hitomi2021multiorbital,krishtopenko2021higher,mao2021magnetism,zeng2021multiorbital,qian2021second,li2022second,liu2022second,wang2022two,luo2022fragile,hu2022intrinsic,mu2022antiferromagnetic,li2022robust,huang2022higher,qian2022c,mu2022kekule,guo2022quadrupole,zhu2022phononic,mao2023ferroelectric,feng2023tunable,wang2023magnetic,huang2023phononic,han2023cornertronics}
and
3Ds~\cite{Schindler2018SA,XuYF2019PRL,yue2019symmetry,WangZJ2019PRL,fang2020higher,hirayama2020higher,ZhangRX2020Mobius,zeng2023topological}
have been predicted to exist in a bunch of solid state materials.
Several classes of materials host multiple candidates for HOTIs in 2Ds, including hexagonal group-IV hydrides and halides
(XY material with X=C, Si, Ge, Sn and Y=H, F, Cl)~\cite{qian2021second}, hexagonal group-V materials (P, As and
 Sb)~\cite{ezawa2018minimal,qian2021second,radha2020buckled,hitomi2021multiorbital,huang2022higher},
Kekul\'{e} and anti-Kekul\'{e} group-IV materials (C, Si, Ge, Sn)~\cite{qian2021second},
ferroelectric materials (e.g. In$_2$Se$_3$)~\cite{mao2023ferroelectric}, group-VIB transition metal dichalcogenides (MX$_2$
material with M=W, Mo and X=Te, Se, S)~\cite{zeng2021multiorbital,qian2022c} and carbon allotrope
with C$\equiv$C bond (graphyne and
graphdiyne)~\cite{liu2019two,YangSA2019PRL,YangBJ2020npjQM,chen2021graphyne}.
Other typical proposals for non-magnetic 2D HOTIs include twisted bilayer graphene and boron nitride
for a range of twist angles~\cite{Park2019PRL,liu2021higher}, antidot Dirac materials~\cite{xue2021higher},
semiconductor quantum wells~\cite{krishtopenko2021higher} and so on.
There are also many theoretical predictions of magnetic 2D HOTIs, such as bismuth on EuO substrate~\cite{chen2020universal},
NpSb~\cite{mao2021magnetism}, CrSiTe$_3$~\cite{wang2023magnetic}, 2H-RuCl$_2$~\cite{li2022robust},
Janus VSSe~\cite{li2022robust} and CoBr$_2$/Pt$_2$HgSe$_3$/CoBr$_2$ van der Waals heterostructure~\cite{liu2022second}.
FeSe is a possible HOTI with antiferromagnetic order~\cite{luo2022fragile,mu2022antiferromagnetic}.
Besides electronic 2D HOTIs, the phononic conterparts are also predicted in C$_3$N~\cite{huang2023phononic},
graphyne~\cite{zhu2022phononic} and
graphdiyne~\cite{mu2022kekule}.
Most recently, the authors of Ref.~\cite{han2023cornertronics} propose TiSiCO family monolayers as a new class
of 2D HOTIs for which the corner degree of freedom can be controlled by electric and optical fields and might have
potential applications in novel electronic devices.
For 3D materials, EuIn$_2$As$_2$~\cite{XuYF2019PRL} and Bi$_{2-x}$Sm$_x$Se$_3$~\cite{yue2019symmetry}
are predicted to be HOTIs with
chiral hinge states.
Antiperovskites such as Ca$_3$SnO and Ca$_3$PbO~\cite{fang2020higher}, distorted SnTe~\cite{Schindler2018SA},
MoTe$_2$~\cite{WangZJ2019PRL}, WTe$_2$~\cite{WangZJ2019PRL} and La apatite~\cite{hirayama2020higher}
are possibly 3D HOTIs carrying helical hinge states.
Besides, MnBi$_{2n}$Te$_{3n+1}$ is predicted to be a 3D higher-order M{\"o}bius insulator~\cite{ZhangRX2020Mobius}
and $\beta$-CuI is a candidate for higher-order Dirac semimetals~\cite{zeng2023topological}.


For experimental aspects, HOTPs have only been found in a few solid state materials.
Second-order topological corner modes in 2D fractal lattices have been observed in Bi monolayer deposited on InSb substrate recently~\cite{nunez2023topological}.
Bulk bismuth (Bi)~\cite{Neupert2018NP,Mad2021NC}, bismuth bromide (Bi$_4$Br$_4$)
~\cite{Kondo2021NM,Hasan2022NM} and layered $T_d$-WTe$_2$~\cite{Schonen2020NL,
	Xiu2020NSR,Lee2020NM,Choi2023NC} are
experimentally denonstrated to have
time-reversal symmetric 3D higher-order topology with 1D helical hinge states.
The higher-order topology of bismuth are protected by the ${C}_3$ and inversion symmetry and
Bi$_4$Br$_4$ are protected by the ${C}_2$ symmetry.
$T_d$-WTe$_2$ is noncentrosymmetric and hosts nonsymmetry-indicated higher-order topology~\cite{WangZJ2019PRL}.
Among them, bulk $T_d$-WTe$_2$ (WTe$_2$ in the orthorhombic structure) has been identified as a type-II Weyl semimetal
in previous studies~\cite{soluyanov2015type2,DiSante2017WTe2} (note that type-II Weyl points are also called structured
Weyl points~\cite{xu2015structured}),
bismuth is also metallic~\cite{Neupert2018NP}, and only Bi$_4$Br$_4$ is a gapped insulator~\cite{Kondo2021NM}.
Figure~\ref{FigMaterBand} displays the calculated bulk band structures of Bi$_4$Br$_4$ and bismuth.
Bi$_4$Br$_4$ is an insulator with a bulk gap [see Fig.~\ref{FigMaterBand}(a)].
Its higher-order topology gives rise to gapless hinge states in the band spectrum of a Bi$_4$Br$_4$ nanowire [see Fig.~\ref{FigMaterBand}(b)].
As a comparison, bismuth is not an insulator since the energy bands have electron and hole pockets at the Fermi energy [see Fig.~\ref{FigMaterBand}(c)].
Nevertheless, the valence bands and conduction bands are separable by a direct gap across the Brillouin zone.
Therefore, we can still define the bulk topological indices from the symmetry eigenvalues of the valence bands,
corresponding to helical hinge states traversing the bulk gap in the spectrum for a nanowire of bismuth~\cite{Neupert2018NP}.

The hinge states can be observed directly by scanning tunnelling microscopy (STM)
~\cite{Neupert2018NP,Mad2021NC,Hasan2022NM} and angle-resolved
photoemission spectroscopy (ARPES)~\cite{Kondo2021NM} and their transport properties
can be measured by Josephson
interferometry~\cite{Neupert2018NP,Schonen2020NL,Xiu2020NSR,Lee2020NM}.
The spin polarization of the helical hinge states is resolved by the magneto-optic Kerr-rotation
measurement in $T_d$-WTe$_2$~\cite{Choi2023NC}.
Helical hinge states are also observed in Cd$_3$As$_2$ by Josephson interferometry
measurements~\cite{Liao2020PRL},
consistent with the hinge Fermi arc calculated theoretically~\cite{Wieder2020NC}, implying Cd$_3$As$_2$ is a
higher-order topological semimetal.
Spin-momentum locking of the hinge states is evidenced by spin transport measurements
~\cite{Liao2022SB}.
There is also evidence that FeTe$_{0.55}$Se$_{0.45}$ might be a higher-order topological
superconductor with helical hinge zero modes~\cite{Burch2019NL}.

\subsection{Synthetic systems}

In addition to solid state materials, HOTPs are also realized in quantum simulators.
An electronic SOTI in 2Ds is realized by manipulating carbon
monoxide molecules as a breathing kagome
lattice on Cu surface~\cite{Smith2019NM}.
The 2D HOTI is also simulated by a superconducting qubit~\cite{YuDP2021SB}.
Much richer HOTPs are observed in a variety of metamaterials, such as mechanical
\cite{Huber2018Nature,YuDJ2019PRL,WuDJ2020JAP,Hatsugai2020PRB,
	HuangGL2020PRApp,Yang2021CM,JiangJH2021SB,YuDL2022APL,ZhuJ2022IJMS,
	Xia2023JMPS}, microwave~\cite{Bahl2018Nature},
acoustic~\cite{ZhangBL2019NM,Khan2019NM,JiangJH2019NP,ZhangBL2019PRL,Chris2019AM,
	ChenYF2019NC,JiangJH2020NC,Khan2020SA,Khan2020NC,ChenYF2020NC_He,ZhangBL2020NC,
	JiangJH2020PRB,YuDJ2020PRB,JiangJH2020PRBX,Huang2020APL,LiuZY2020PRBY,LiuZY2020PRL,ChengJC2020PRL,MaGC2021PRX,
	Jia2021NM,JiangJH2021NM,ChenYF2021PRLZ,LiuZY2021PRLY,LiuXJ2021APL,WuY2021SB,Huang2021NCC,
	QiuCY2021PRLQ,YangH2021PRB,MaGC2021FP,LiuZY2022SBH,Jia2021PRL,ZhangBL2022NC,
	JiangJH2022PRAWu,lijk2022higher,LiuXJ2022NJP,QiuCY2022PRL,ZhuJ2022PRB,GaoZ2022PRL,
	zheng2022observation,JiangJH2023PRApp,ChenYF2023PRApp,XuChen2023arXiv,QiuCY2023arXiv,
	LiuZY2023CP,Su2023IJMS,Jing2023arXiv}, photonic systems~\cite{ChenYF2018PRB,ChenYF2019PRL,Iwam2019Opt,
	DongJW2019PRL,Hafezi2019NP,
	ChenHY2019OL,Bour2019NP,Khan2020NP,JiangJH2020PRBJ,Chen2020AS,XieXC2020SB,Liew2020PRL,
	Rho2020Nano,Jin2020arXiv,Xu2020LPR,Zhen2020NC,XuXL2020LSA,JiangJH2020LPR,ChenYF2020NC,
	Khan2021NM,Khan2021JPCC,Heinrich2021NP,JinXM2021LSA,ChenZG2021LSA,Xu2021OE,
	ChenZG2021ACS,ZhangBL2022PRBW,LiY2022PRR,Rechtsman2022PRB,Huang2022AS,ChenZG2022arXiv,
	LiuJJ2022arXiv,ChenZG2023elight,Nori2023LPR,XiongQH2023SA,GaoF2023PRL,ChenF2023OL}
and electric circuits~\cite{Thomale2018NP,ZhangXD2019PRB,YanP2020PRR,ZhangS2020LSA,
	LiuZY2020PRB,ZhangXD2021PRL,XuShi2021CP,Bahl2022NC,lizuka2023SR,Kan2023APL},
blessed by their high tunability.
HOTIs with corner states have been extensively simulated in various systems in both 2Ds~\cite{Huber2018Nature,YuDJ2019PRL,WuDJ2020JAP,Hatsugai2020PRB,HuangGL2020PRApp,Yang2021CM,
		JiangJH2021SB,YuDL2022APL,ZhuJ2022IJMS,Xia2023JMPS,Bahl2018Nature,ZhangBL2019NM,Khan2019NM,
		JiangJH2019NP,Chris2019AM,JiangJH2020NC,Khan2020SA,JiangJH2020PRB,LiuZY2020PRBY,ChenYF2021PRLZ,
		LiuZY2021PRLY,LiuXJ2021APL,JiangJH2022PRAWu,LiuXJ2022NJP,ZhuJ2022PRB,JiangJH2023PRApp,
		ChenYF2023PRApp,QiuCY2023arXiv,LiuZY2023CP,Su2023IJMS,ChenYF2018PRB,ChenYF2019PRL,
		Iwam2019Opt,DongJW2019PRL,Hafezi2019NP,ChenHY2019OL,Bour2019NP,Khan2020NP,JiangJH2020PRBJ,
		Chen2020AS,XieXC2020SB,Liew2020PRL,Rho2020Nano,Jin2020arXiv,Xu2020LPR,Zhen2020NC,XuXL2020LSA,
		JiangJH2020LPR,ChenYF2020NC,Khan2021NM,Khan2021JPCC,JinXM2021LSA,Xu2021OE,ChenZG2021ACS,
		LiY2022PRR,ChenZG2023elight,Nori2023LPR,Jing2023arXiv,XiongQH2023SA,GaoF2023PRL,ChenF2023OL,
		Thomale2018NP,YanP2020PRR,LiuZY2020PRL,ChengJC2020PRL,LiuZY2020PRB,Bahl2022NC,lizuka2023SR,Kan2023APL} and 3Ds~\cite{ZhangBL2019PRL,Khan2020NC,ZhangBL2020NC,JiangJH2020PRBX,YangH2021PRB,MaGC2021FP,
		LiuZY2022SBH,GaoZ2022PRL,ZhangXD2019PRB,ZhangS2020LSA,Bahl2022NC}.
In 3Ds, SOTIs with chiral~\cite{ChenYF2020NC_He}, helical~\cite{Jia2021PRL,
	GaoZ2022PRL} and zero-energy hinge states~\cite{QiuCY2022PRL}, higher-order Klein bottle TIs~\cite{XuChen2023arXiv}, second-order
topological semimetals~\cite{Jia2021NM,JiangJH2021NM,QiuCY2021PRLQ,ZhangBL2022PRBW}, and HOTIs with surface,
hinge and corner states coexisting in the same
band gap~\cite{ChenYF2019NC,Khan2020SA,YuDJ2020PRB,Huang2020APL,WuY2021SB} have also been
realized in acoustic~\cite{ChenYF2020NC_He,Jia2021PRL,GaoZ2022PRL,QiuCY2022PRL,Jia2021NM,QiuCY2021PRLQ,
		ChenYF2019NC,Khan2020SA,YuDJ2020PRB,Huang2020APL,WuY2021SB,Huang2022AS,
		ChenZG2022arXiv} and photonic systems~\cite{ZhangBL2022PRBW}.
Nonlinear second-order topological corner states in 2Ds are observed in photonic systems
~\cite{Heinrich2021NP,ChenZG2021LSA}
and Floquet SOTIs are investigated experimentally in an acoustic
lattice~\cite{ZhangBL2022NC}.
By introducing synthetic dimensions, HOTPs in higher dimensions are realized in
photonic and acoustic crystals~\cite{MaGC2021PRX,Huang2021NCC,Rechtsman2022PRB}.
The phase transition from an anomalous quantum Hall insulator to SOTIs induced by disorder is demonstrated experimentally in electric
circuits~\cite{ZhangXD2021PRL}.
Non-0D corner states are observed in acoustic fractal lattices
~\cite{lijk2022higher,zheng2022observation} and photonic quasicrystals with six-fold rotational symmetry~\cite{LiuJJ2022arXiv}.
Quadrupole TIs on quasicrystalline lattices have also been experimentally realized in electric circuits~\cite{XuShi2021CP}.
However, there, quasicrystal sites are arranged in a square sample and only four corner modes are
observed, which is fundamentally equivalent to the crystalline case.
Despite the theoretical predictions of HOTIs in quasicrystals with five, eight, ten or twelve corner modes~\cite{Fulga2019PRL,Xu2020PRL,Cooper2020PRR,Xu2020PRB,Liu2021NanoLett,Zhou2021PRB,LiuFeng2022PRL},
which cannot exist in crystalline systems,
their implementation poses a significant challenge for experiments. The challenge arises from
the complex form of the Hamiltonian, which involves long-range hopping for metamaterial simulations.

\section{Summary and perspectives}\label{sec6}

In this review, we have introduced the recent progress on HOTPs
in both crystalline and non-crystalline systems. Different from conventional first-order topological phases,
the HOTPs are featured by the topological boundary states of lower dimension.
We review several prototypical examples of HOTPs including
2D quadrupole TIs, 3D SOTIs with either chiral or helical hinge states,
and higher-order topological semimetals with hinge Fermi arcs, and
introduce their topological boundary states and corresponding topological invariant.
It has been demonstrated that some HOTIs are protected by the topological invariants
which do not require the presence of crystalline symmetries.
Therefore, these HOTPs can exist in disordered systems or non-crystalline systems that break crystalline symmetries.
Then, we turn to discuss the HOTPs in the presence of quenched disorder.
Similar to first-order topological phases, it has been shown that HOTPs are robust to weak disorder.
We further demonstrate that disorder can induce a HOTI from a trivial insulator,
which is the generalization of topological Anderson insulator to the higher-order case.
On the other hand, structural disorder in irregular lattices can also have effects on HOTPs.
It has been found that there exist SOTIs with chiral or helical hinge states in amorphous systems
and structural disorder can remarkably drive a topological phase transition between a trivial insulator and a SOTI.
In addition, we review the progress on HOTPs in other non-crystalline systems
such as quasicrystals, hyperbolic lattices, and fractal lattices, which can host HOTPs beyond crystalline systems.

We remark that this review has only discussed a few models for HOTPs
and many other aspects of higher-order topology have not been covered here.
In addition to static systems, the HOTPs in nonequilibrium systems have also been extensively investigated.
In Refs.~\cite{Nag2019PRR,Yang2020PRR,mizoguchi2021PRL,okugawa2021mirror,li2021direct,YuDP2021SB,Long2021PRA,Xiongjun2023unified,maslowski2023dynamical}, people have studied the far-from-equilibrium dynamics of various HOTPs after a sudden quench of the Hamiltonian.
Interestingly, the authors in Ref.~\cite{Yang2020PRR} found that a quadrupole TI after the quench will undergo a topological transition from a trivial phase to a quadrupole topological insulating phase during the unitary time evolution.
On the other hand, a number of works studied the Floquet higher-order topology in periodically driven systems~\cite{Peng2019PRL,Gong2019PRB,Rodriguez2019PRB,Seshadri2019PRB,Plekhanov2019PRR,Nag2019PRR,
Huang2020PRL,Hu2020PRL,Ghosh2020PRB,Peng2020PRR,ZhangRX2020arXiv,
Gong2021PRB,Ghosh2021PRB,Nag2021PRB,Vu2021PRB,An2021PRB,Zhou2021Nano,
Ghosh2022PRB,Zhou2022Symmetry,Ghosh2023PRB,Huang2023FOP,Zhou2023arXiv}.
Similarly to the first-order cases, people discovered the Floquet HOTPs with anomalous corner or hinge modes
and the corresponding topological invariants were also proposed~\cite{Huang2023FOP}.
In the review, we have focused on the HOTPs for noninteracting fermions, which can be
effectively described by topological band theory.
Several works have studied higher-order symmetry-protected topological phases in interacting fermion or boson systems
which exhibit gapless corner or hinge modes~\cite{You2018PRB,Kudo2019PRL,Dubinkin2019PRB,You2019PRB,laubscher2019fractional,laubscher2020majorana,Rasmussen2020PRB,Araki2020PRR, You2020PRR,Dubinkin2020arXiv,Hackenbroich2021PRB,You2021PRB,ZhangJH2022PRB,li2022green,You2022PRB,Mondaini2023PRB,
ZhangJH2023PRB,YangShuo2023PRB}.
Besides HOTIs and semimetals we have discussed, higher-order topological superconductors
have also gained much attention for the Majorana modes at corners or hinges,
which provide alternative platforms to realize non-Abelian braiding for topological quantum computation in the future~\cite{ZhangSB2020PRR,ZhangSB2020PRB,bomantara2020measurement,pahomi2020braiding,wu2020non,lapa2021symmetry,pan2022detecting}.

So far, most studies have been focused on HOTPs at zero temperature, while only a few of works have covered HOTPs at finite temperature.
For a topological insulator of noninteracting fermions, the system is in a many-body ground state that all particles occupy the bands below the gap at zero temperature.
At finite temperatures, the system in thermal equilibrium is described by a mixed state rather than a single ground state.
In recent years, people have developed concepts and characterizations for the first-order TIs at finite temperatures,
including the topological invariant for ensembles~\cite{hastings2011topological,viyuela2014uhlmann,budich2015topology,viyuela2015symmetry,bardyn2018probing,gonccalves2019temperature}.
One immediate problem worth studying is generalizing the formalism to HOTPs.
Another interesting problem is about higher-order topology tuned by the temperature.
At finite temperatures, thermal fluctuations play a similar role as disorder which can drive phase transitions~\cite{gonccalves2019temperature}.
On the other hand, various symmetry-breaking orders such as superconducting pairings or magnetizations may have dependence on the temperature,
which can affect the bulk topology at mean-field level accordingly.
In Ref.~\cite{kheirkhah2020PRL}, the authors studied a 2D extended Hubbard model with spin-orbit coupling
and found that the temperature can drive a phase transition between a gapped second-order topological superconductor and a gapless superconductor,
which results from the change of favored pairing type at a critical temperature.

On the topological characterization of HOTIs,
there exists an open question about the properties of the quadrupole moment as a topological invariant.
It has been found that the quadrupole moment can change when either the bulk energy gap or edge energy gap closes
while the quantity itself is evaluated from the bulk wave functions without boundaries.
Moreover, the current operator-based definition of the quadrupole moment may have some inconsistent results for some cases as pointed out in Ref.~\cite{Watanabe2019PRB}.
Another problem worth studying is about HOTPs in a system with gauge flux leading to momentum-space nonsymmorphic symmetry~\cite{chen2022brillouin}.
Recently, multipole TIs with corner modes and HOTIs with hinge modes have been found
in lattice models respecting the glide-reflection symmetry in momentum space~\cite{XuChen2023arXiv,Tao2023Quadrupole,YangYi2023Quadrupole}.
The momentum-space nonsymmorphic symmetries may enrich the family of HOTPs.

\begin{acknowledgments}
We appreciate the collaboration with L.-M. Duan, G. Chen, S. A. Yang, Q.-B. Zeng, Y.-L. Tao, Y.-F. Mao, N. Dai and M. Yan on the topic.
We thank B. Wieder, R.-J. Slager and S. Zhang for helpful discussions.
This work is supported by the National Natural Science Foundation of China (Grant No. 11974201) and Tsinghua University Dushi Program
and Innovation Program for Quantum Science and Technology (Grant No. 2021ZD0301604).
We also acknowledge the support by center of high performance computing, Tsinghua University.
\end{acknowledgments}


\bibliography{reference.bib}

\end{document}